\documentclass[pdflatex,sn-mathphys-num]{sn-jnl}

\usepackage{amsmath,amssymb}
\usepackage[utf8]{inputenc}
\usepackage[section]{placeins}
\usepackage{graphicx}
\usepackage{siunitx}
\usepackage{booktabs}
\usepackage{rotating}
\usepackage{multirow}

\DeclareSIUnit{\wattpeak}{Wp}
\DeclareSIUnit{\kilowattpeak}{kWp}
\DeclareSIUnit{\megawattpeak}{MWp}
\DeclareSIUnit{\gigawattpeak}{GWp}
\DeclareSIUnit{\Wh}{Wh}
\DeclareSIUnit{\kWh}{kWh}
\DeclareSIUnit{\MWh}{MWh}
\DeclareSIUnit{\GWh}{GWh}
\DeclareSIUnit{\TWh}{TWh}

\begin{document}
\setcounter{secnumdepth}{2}

\title[PV net energy beyond the panel]{Photovoltaics beyond the panel: system-level net energy under grid and firming constraints}

\author*[1,2]{\fnm{Hans Peter} \sur{Beck}}\email{hanspeter.beck@unibe.ch}

\affil*[1]{\orgdiv{Albert Einstein Center for Fundamental Physics and Laboratory for High Energy Physics}, \orgname{University of Bern}, \orgaddress{\street{Sidlerstrasse 5}, \city{Bern}, \postcode{3012},  \country{Switzerland}}}
\affil[2]{\orgdiv{Department of Physics}, \orgname{University of Fribourg}, \orgaddress{\street{Chemin du Musée 3}, \city{Fribourg}, \postcode{1700}, \country{Switzerland}}}

\abstract{

Photovoltaics can exhibit favourable installation-level energy return on
investment (EROI), yet electricity must also be delivered when solar and
wind generation are insufficient. This study evaluates how the energy
balance changes when the boundary is expanded to reliable electricity
supply. A fleet-equivalent benchmark model includes component turnover,
grids, storage, renewable overbuild, dispatchable backup, and fuel supply.
It is not a simulation, optimization, or real-system prediction.
Photovoltaics provide the accounting reference in an illustrative Central
European mix of 80\% wind and 20\% photovoltaics, while the firming penalties
arise more generally from weather-dependent generation.

Expanding the boundary is decisive. The unfirmed renewable fleet retains an
EROI of 10.7--16.0, but does not provide continuous electricity. Battery
load-balancing references decline from 7.1--12.5 for \SI{1}{h} to
0.7--2.3 for \SI{24}{h}. Seasonal hydrogen yields 3.4--7.7; adding a
\SI{1}{h} battery reduces the combined range to 2.9--6.8. Pipeline-gas and
LNG backup yield only 1.2--2.6 and 1.0--2.2.

The lifecycle-carbon advantage narrows accordingly. Under present industrial
supply chains, GWP$_{100}$ intensities rise from
38--75~gCO$_2$eq/kWh without firming to 80--289~gCO$_2$eq/kWh for hydrogen
and battery--hydrogen firming, and 121--327~gCO$_2$eq/kWh for gas backup.
These favourable benchmarks set orientation, site, curtailment, and delivery
factors to unity; real deployment lowers EROI and raises emissions. The
results identify firming as a critical energetic constraint:
reliable decarbonization requires firm low-carbon generation with sufficient
net-energy surplus to sustain a resilient industrial society.

}


\maketitle

\section{Introduction}

Photovoltaics (PV) have expanded rapidly over the past two decades, driven by policy support and steep declines in manufacturing costs. They are widely regarded as a major contributor to future low-carbon electricity systems. At the same time, global energy demand and energy-related CO$_2$ emissions have continued to increase, highlighting the importance not only of deploying low-carbon technologies, but also of understanding their system-level energetic performance and infrastructure requirements~\cite{IEA:2023:CO2}.

A critical but often overlooked perspective is the net energy return on investment (EROI): not only at the module or installation level, where PV generally appears favorable, but also at the system level, where storage, backup generation, grid integration, and firming requirements fundamentally alter the energetic balance.

EROI, defined as the ratio of lifetime electricity delivered to the total primary energy required to build, operate, maintain, and replace the system~\cite{Hall2014,Lambert2014}, provides a benchmark for evaluating whether an energy technology can sustain the energetic surplus required by modern industrial societies:
\begin{equation*}
\text{EROI} = \frac{E_{\text{out}}}{E_{\text{in}}}
= \frac{\text{Lifetime electricity output (kWh)}}%
        {\text{Total primary energy invested (kWh)}}
\end{equation*}
Assessments of PV performance employ system boundaries of varying scope. Values often cited for PV systems are typically 15--25 under Central European conditions and up to 40--50 in high-irradiance regions. These figures generally correspond to \emph{as-installed} systems that include the PV module and part of the initial balance-of-system (BOS), such as mounting structures, inverter, and cabling, but exclude broader system-level requirements associated with reliable electricity supply~\cite{Raugei2017,Fthenakis2021:ScSiSustainability}. This is consistent with established LCA guidelines for photovoltaic electricity systems, which treat temporal storage, dispatchable backup generation, and broader grid-level balancing infrastructure as being outside the standard PV system boundary~\cite{IEA2016:PVPS_Task12_LCA}.

Fraunhofer~ISE, for example, reports energy payback times of approximately 0.9--1.1~years for modern PV systems under such assumptions~\cite{Fraunhofer2023:PVReport}, corresponding to installation-level EROI values of roughly 20 for a 20-year lifetime 
and about 30 for a 30-year module lifetime. 

These remain technology-level metrics. Extending the system boundary to include repeated component replacement, storage, curtailment, grid integration, and firming infrastructure substantially introduces additional primary-energy 
investments that become part of the system-level energy balance. In the following, this expanded system-level quantity is denoted EROI$_{\text{sys}}$ to distinguish it from installation-level metrics commonly reported in the literature.

At high shares of intermittent generation, reliable electricity supply requires additional system-level measures beyond the PV installation itself. These include grid integration, balancing infrastructure, storage, dispatchable backup generation, synthetic inertia, and other firming mechanisms needed to maintain stable operation during periods of low renewable output. Such requirements introduce additional embodied-energy investments that are generally excluded from installation-level PV assessments.

To represent large-scale deployment consistently, the analysis adopts a fleet-equivalent lifecycle accounting framework. Here, a ``fleet'' denotes the aggregate PV capacity deployed across a region, consisting of many individual installations that enter service at different times and therefore coexist simultaneously at different stages of their life cycles. Components are continuously replaced according to their characteristic service lifetimes. Short-lived elements such as inverters, storage systems, and power electronics therefore contribute through their average turnover rates within the fleet, while longer-lived infrastructure contributes proportionally to its service life.

Under this formulation, component replacement factors need not be integers. They represent the average cumulative turnover required to maintain the regional fleet over time rather than the replacement history of any individual installation. This ensures that cumulative electricity generation and embodied-energy reinvestment are treated consistently across technologies with different component lifetimes. The resulting EROI$_{\text{sys}}$ therefore reflects the energetic consequences of maintaining the generation fleet over time rather than the replacement schedule of any individual installation. This fleet-equivalent embodied-energy inventory subsequently serves as the reference against which the additional energetic investments associated with system-level firming are evaluated.

Within this framework, $E_{\text{out}}$ denotes the usable electricity
delivered per installed photovoltaic capacity, expressed in MWh/kWp.
Output-side reductions associated with orientation, site conditions,
curtailment, and electricity delivery are treated separately from the
primary-energy investments represented by $E_{\text{in}}$.
The latter include photovoltaic modules, balance-of-system components,
storage and firming infrastructure, replacement cycles, and grid-related
equipment. Conversion losses are represented through the
architecture-specific energy requirements needed to maintain useful
electricity delivery.

The purpose of the present framework is not to reproduce a fully
optimized future electricity system, but to provide a
boundary-consistent benchmark for examining how progressively expanded
reliability requirements affect system-level net energy return.
Different technical pathways may redistribute these energetic costs
across infrastructure, conversion losses, renewable overbuild, and
fuel consumption, but they do not remove the underlying energetic
consequences of weather-dependent generation.

In the terminology used here, load balancing denotes the shifting of
electricity over short and intermediate timescales in response to
routine variations in renewable generation and demand. Firming denotes
the broader requirement to maintain reliable electricity supply during
extended periods of insufficient renewable generation, including
long-duration and seasonal deficits. Short-duration storage can provide
valuable balancing and grid-support services, but repeated cycling does
not create the energy inventory required for seasonal adequacy.

The storage, balancing, and firming architectures examined below are
therefore illustrative benchmark cases rather than optimized system
designs. Their purpose is to quantify the energetic consequences of
prescribed infrastructure and reliability requirements, not to define
preferred storage durations or predict a particular electricity-system
configuration. The dominant challenge for deeply decarbonized
electricity systems in Central Europe remains the provision of reliable
energy during prolonged periods of weak wind generation and low solar
irradiation.

The present analysis focuses on photovoltaics as the primary case
study, but the energetic constraints associated with storage,
balancing, and firming infrastructure arise more generally from
weather-dependent generation itself. They therefore apply, with
quantitative differences, to other forms of variable renewable
generation such as wind power. The corresponding system-level energy
investments are quantified later through normalized primary-energy
contributions relative to the photovoltaic fleet.

Accordingly, the analysis proceeds in three stages: first, installation-level PV energetics are established and compared with published literature; second, these are extended to a fleet-equivalent representation that accounts for continuous component replacement; finally, representative low-carbon electricity-system architectures are constructed by incorporating the additional primary-energy investments required for storage, balancing, and firming.

\paragraph{Interpretation of EROI and societal relevance}
\label{sec:eroi_societal-relevance}

EROI quantifies the fraction of gross energy production that remains available to society after reinvestment into the energy system itself. As EROI declines, an increasing share of total energy production must be devoted to constructing, operating, maintaining, and replacing the energy infrastructure, leaving less surplus energy available for other societal functions.

Values below unity correspond to net energy sinks, while modest surpluses permit only limited economic and technological complexity. Previous studies suggest that EROI values of approximately 7--10 are required to sustain modern industrial societies, whereas higher values provide greater resilience against economic, environmental, and infrastructure stresses~\cite{Hall2014, Lambert2014,Weissbach2013}. Higher net-energy surpluses also support the long-term maintenance of energy-intensive services and infrastructure such as healthcare, water supply, education, transportation, scientific research, and environmental remediation.

In this context, high-EROI low-carbon energy systems are relevant not only for climate mitigation, but also for sustaining the broader societal capabilities associated with the United Nations Sustainable Development Goals (SDGs)~\cite{UN2015:SDGs}, which depend fundamentally on the long-term availability of affordable, reliable, and scalable energy supplies.

Indicative EROI thresholds commonly discussed in the literature are summarized in Table~\ref{tab:societal-thresholds}. For clarity, the table also reports the corresponding energy surplus, defined as the fraction of gross energy production remaining after reinvestment into the energy system itself.

\begin{table}[h!]
\centering
\caption{Indicative societal thresholds for EROI, based on previous studies~\cite{Hall2014,Lambert2014,Weissbach2013}. Thresholds are approximate and vary across studies, but all point to the need for high net energy to sustain complex societies. Energy surplus is shown as the fraction of gross output remaining after reinvestment.}
\label{tab:societal-thresholds}
\begin{tabular}{lll}
\toprule
EROI (approx.) & Energy surplus & Societal implications \\
\midrule
$<$\phantom{0}1   & negative & Energy sink \\
\phantom{$<$0}1--3   & \phantom{0}0--67\%  & Bare subsistence, minimal surplus \\
\phantom{$<$}3--7   & 67--86\% & Simple societies, limited specialization \\
\phantom{$<$0}7--10  & 86--90\% & Industrial society, but fragile \\
\phantom{$<$}10--20 & 90--95\% & Sustainable abundance: resilience, welfare, \\
                                   &                & culture, science, and progress toward SDGs \\
\bottomrule
\end{tabular}
\end{table}

\section{Installation-level embodied-energy inventories}
\label{sec:installation_inventory} 

Installation-level photovoltaic performance reported in the literature primarily reflects the balance between embodied manufacturing energy and lifetime electricity generation at a given site. Under Central European conditions, representative installation-level EROI values are typically about 15--25 for rooftop systems, roughly 17--25 for utility-scale installations, and around 20--30 for alpine PV benefiting from enhanced irradiation and seasonal albedo effects~\cite{Raugei2017,Bhandari2015,Fthenakis2021:ScSiSustainability,JRC2018:Task5PVSystems}.

These values generally describe PV systems operating without storage and without broader system-level integration requirements. They therefore characterize the energetic performance of individual PV installations and provide the installation-level reference from which the progressively expanded system boundaries developed in this work are constructed.

The representative installation-level embodied-energy inventories summarized in
Table~\ref{tab:installation_inventory} are based on detailed LCA
inventories and harmonized reviews for rooftop and utility-scale
systems \cite{Raugei2017,Bhandari2015,Leccisi2016:Energies,
Fthenakis2021:ScSiSustainability}. For alpine PV, comparable
harmonized inventories are not yet widely available. The values
adopted here therefore represent engineering estimates derived from
the corresponding rooftop and utility-scale inventories together with
additional infrastructure requirements associated with remote and
elevated terrain, including heavier mounting structures, longer cable
runs, grid connection, access infrastructure, and seasonal snow
management. The resulting estimates are intended to be representative
of alpine photovoltaic deployment under Central European conditions
rather than precise site-specific values.

Recent LCA studies report cumulative energy demand (CED), expressed as primary energy, for crystalline silicon PV systems (modules plus BOS) in the range of approximately 10,000--20,000~MJ/kWp \cite{Bhandari2015}, with more recent assessments reporting values around 13,600--14,800~MJ/kWp \cite{Fthenakis2021:ScSiSustainability}. These correspond to approximately 2.8--5.6~MWh/kWp, with central values around 3.5--4.0~MWh/kWp.

Such CED values generally reflect broader upstream supply-chain
boundaries than installation-level EPBT and EROI assessments and
therefore yield larger embodied-energy estimates. The inventories
adopted in Table~\ref{tab:installation_inventory} are chosen to remain
consistent with the installation-level EROI ranges commonly reported in
the literature while preserving a transparent component-level
decomposition. Accordingly, they correspond to the installation-level
system boundaries typically used in EPBT and EROI assessments rather
than to the broader primary-energy accounting of full CED analyses.
They represent installation-level embodied-energy estimates under
favorable assumptions and provide the baseline for the fleet-level and
system-level extensions developed in subsequent sections.

Module manufacturing constitutes the dominant contribution to the
inventories summarized in Table~\ref{tab:installation_inventory},
while BOS components, including mounting structures, inverters,
cabling, and installation, account for a significant but smaller
share of the total embodied energy.

\begin{table}[ht]
\centering
\caption{Typical installation-level embodied-energy contributions ($E_{\text{in}}$) for rooftop, utility-scale, and alpine PV systems in Central Europe, expressed per installed capacity (MWh/kWp).}
\label{tab:installation_inventory}

\renewcommand{\arraystretch}{1.1}
\begin{tabular}{lccc}
\hline
Component & Rooftop PV & Utility-scale PV & Alpine PV \\
 & (MWh/kWp) & (MWh/kWp) & (MWh/kWp) \\
\hline
Module manufacturing        & 1.20       & 1.20          & 1.20 \\
Mounting structures         & 0.14       & 0.20--0.25    & 0.30--0.40 \\
Inverter manufacturing      & 0.20       & 0.18--0.20    & 0.20 \\
Cabling/connectors          & 0.05       & 0.08          & 0.12 \\
Transport/installation      & 0.10       & 0.15          & 0.20 \\
Grid connection             & --         & 0.20--0.30    & 0.30--0.40 \\
Access roads/snow clearance & --         & --            & 0.10--0.20 \\
\hline
\textbf{Total $E_{\text{in}}$}
 & \textbf{1.7}
 & \textbf{2.0--2.2}
 & \textbf{2.4--2.7}
 \\
\hline
\end{tabular}
\end{table}

The total installation-level embodied energy lies in the range
\SIrange{1.7}{2.7}{MWh/kWp} depending on system type. These values are
consistent with installation-level EROI estimates reported in the
literature and should be interpreted as installation-level baseline
inputs under favorable assumptions.

\section{Installation-level EROI}
\label{sec:installation_eroi}

Installation-level EROI characterizes the energetic performance of an
individual photovoltaic installation over its operating lifetime. It is
defined by the relationship between the lifetime electricity output
$E_{\text{gross}}$ and the embodied-energy input $E_{\text{in}}$
required for the initial installation.

For consistency with the assumptions adopted throughout this study,
photovoltaic systems are assumed to operate under optimal orientation
and tilt conditions, without partial shading or significant soiling
losses. Under these conditions, representative annual yields in Central
Europe are typically in the range of
\SIrange{1.05}{1.15}{MWh/kWp} for rooftop PV,
\SIrange{1.10}{1.20}{MWh/kWp} for utility-scale installations, and
approximately \SIrange{1.25}{1.35}{MWh/kWp} for alpine PV. These values
are consistent with European PV performance data reported in lifecycle
assessment inventories and system-performance studies
\cite{JRC2018:Task5PVSystems,Fthenakis2021:ScSiSustainability}.

Assuming a representative PV module lifetime of 30 years, these annual
yields correspond to the gross lifetime electricity outputs summarized
in Table~\ref{tab:eout}. Combined with the installation-level
embodied-energy inventories reported in
Table~\ref{tab:installation_inventory}, they yield the representative
installation-level EROI values shown in Table~\ref{tab:eout}.

The resulting reference installation-level EROI values reflect the embodied 
energy required for the initial photovoltaic installation, including modules, 
mounting structures, electrical balance-of-system components, transport, 
and installation. They deliberately exclude component turnover, operational 
and maintenance requirements, delivery losses, curtailment, storage systems, and other system-level effects.

Subsequent sections progressively extend this installation-level
baseline toward fleet-level and system-level EROI$_{\text{sys}}$ by
introducing these additional requirements.

\begin{table}[ht]
\centering
\caption{Reference gross lifetime electricity output ($E_{\text{gross}}$), installation embodied-energy input ($E_{\text{in}}$), and resulting installation-level EROI values for photovoltaic systems in Central Europe without storage.}
\label{tab:eout}

\begin{tabular}{lccc}
\toprule
System type & $E_{\text{gross}}$ & $E_{\text{in}}$ & Installation-level EROI \\
& (MWh/kWp) & (MWh/kWp) & \\
\midrule
Rooftop PV       & 31.5--34.5 & 1.7        & 18.6--20.4 \\
Utility-scale PV & 33.0--36.0 & 2.0--2.2  & 15.1--17.9 \\
Alpine PV        & 37.5--40.5 & 2.4--2.7  & 13.8--16.7 \\
\bottomrule
\end{tabular}
\end{table}

Although alpine PV benefits from higher irradiation and seasonal
albedo effects, these gains are partly offset in the present framework
by the additional infrastructure required for deployment in remote and
elevated terrain, including heavier mounting structures, longer
transmission connections, grid connection, and site-access
infrastructure.

\section{Grid integration and stability}
\label{sec:grid}

PV expansion affects the electricity grid not only at the project
level but increasingly at the system level as deployment scales.
Beyond conventional BOS components such as mounting structures, cabling, and inverters, large-scale deployment requires feeder reinforcements, additional transformers, substations, and in some cases upstream transmission infrastructure~\cite{JRC2018:Task5PVSystems,UNECE2022:LCA_ElectricityOptions}. These reinforcements are material- and energy-intensive, involving excavation, metals, power electronics, and civil works.

International assessments indicate that representative grid
reinforcement measures add roughly 10--15\% to the embodied energy of
PV systems under moderate deployment levels
\cite{UNECE2022:LCA_ElectricityOptions}. The required reinforcement depends on network topology, local
conditions, and renewable penetration. As PV penetration increases,
integration burdens generally grow nonlinearly. At low penetration
levels, existing feeders and substations are often sufficient,
whereas higher shares increasingly require upstream reinforcement,
balancing infrastructure, congestion management, and accommodation of
surplus generation.

IEA scenarios emphasize that large-scale solar and wind deployment requires substantial expansion of both transmission and distribution infrastructure to avoid congestion and reliability issues~\cite{IEA2021:NetZero,IEA2023:ElectricityMarketReport}.

Interconnection through large transmission networks can reduce short-term local variability by redistributing electricity geographically. 
However, the grid itself neither generates nor stores energy. Even
when supported by the rotational inertia of conventional synchronous
generators or by synthetic inertia in inverter-dominated systems,
interconnection alone cannot eliminate extended periods of low
renewable output. In continental-scale weather systems, solar irradiation and wind conditions remain strongly correlated across large parts of Central Europe, particularly during persistent winter anticyclonic conditions associated with Dunkelflaute events. Interconnection can therefore mitigate local fluctuations but cannot generally replace the requirement for firm generation, storage, or other balancing resources.

\subsection{Curtailment}
\label{sec:curtailment}

Curtailment directly reduces the usable electricity delivered to the
grid. Because the embodied-energy investment required to construct the
generating infrastructure remains unchanged, curtailed electricity
lowers the effective energy return obtained from the system.
Consequently, increasing curtailment reduces the system-level energy
return on investment, EROI$_{\text{sys}}$.

Germany curtailed approximately 1.4~TWh of solar generation in 2024,
corresponding to roughly 3--4\% of total PV output and nearly double
the value of the previous year~\cite{BNetzA2024:Curtailment}. The
International Energy Agency reports curtailment levels of 5--15\% in
high-penetration regions such as parts of China, Chile, and California,
with further increases expected as renewable shares continue to
grow~\cite{IEA2024:Renewables}. Plant-level data from the United States
show similar behaviour, with curtailment of approximately 2--3\%
nationally and 5--10\% in regions with high solar penetration such as
California and Texas~\cite{Millstein2021:JouleCurtailment}.

Curtailment and storage requirements are closely related. Storage can
partially reduce curtailment by shifting excess generation from periods
of surplus production to periods of higher demand. However, finite
storage capacity, charging-rate limitations, prolonged oversupply
periods, and seasonal imbalances generally prevent complete elimination
of curtailment in high-penetration renewable systems.

Curtailment can also differ among PV configurations. Utility-scale
installations are generally more exposed to system-level curtailment
because they inject large quantities of electricity into the network at
common locations and times. Distributed rooftop systems are often less
affected because generation is partly consumed locally and many
jurisdictions provide priority grid access for small installations.

Alpine PV may likewise experience lower curtailment than conventional
utility-scale installations where flexible hydropower resources are
available nearby. In regions with reservoir or pumped-storage
hydropower, surplus solar generation can often be accommodated by
temporarily reducing hydroelectric output or increasing pumping
activity. Such flexibility can substantially reduce curtailment,
although the achievable benefit depends on local infrastructure and
operating conditions.

\subsection{Synthetic inertia}

Large shares of inverter-based renewables such as PV and wind alter
the dynamical stability properties of the electrical grid.
Conventional synchronous generators inherently provide rotational
inertia through their spinning masses, thereby helping to stabilize
grid frequency following disturbances. PV inverters, by contrast, do
not naturally provide inertia because they are electronically coupled
to the grid.

Grid-forming inverters can emulate inertia and provide fast frequency
response through advanced control algorithms. Such systems can
contribute to frequency stability and support grid operation in
electricity systems with high shares of inverter-based generation.
However, these capabilities generally require additional control
systems, reserve-power headroom, DC buffers, and more sophisticated
converter architectures than conventional grid-following inverters.

ENTSO-E notes that such capabilities require additional converter
dimensioning and control infrastructure~\cite{ENTSOE2020}, while IEEE
analyses indicate that a substantial fraction of inverter-based
resources may need to operate in grid-forming mode to maintain system
stability in low-inertia power systems~\cite{Matevosyan2019:GFM}.
Demonstration projects reviewed by IEA PVPS further show that PV
inverters, with or without battery support, can already provide
synthetic inertia and fast frequency-response services in practical
electricity systems~\cite{IEA2024:PVFrequencyServices}.

Recent large-scale grid disturbances have highlighted the growing
importance of such services. The Iberian blackout of 2025 triggered
extensive discussion regarding the provision of frequency support,
voltage control, fault-ride-through capability, and grid-forming
services in systems with high shares of inverter-based generation.
While the causes of such events are generally multifactorial, they
illustrate that maintaining system stability requires dedicated
technical measures beyond electricity generation itself.

Unlike curtailment, whose impact can be expressed directly as lost
electricity output, the energetic cost of synthetic inertia arises
through additional converter hardware, control systems, reserve
capacity, and operational constraints. The deployment of such systems
requires additional embodied-energy investment associated with their
manufacture, installation, and integration into the electricity
network. Long-term operation further entails ongoing maintenance,
periodic replacement of components reaching end of life, and other
operational support activities. In addition, every additional system
component introduces finite conversion, control, standby, or
transmission losses, thereby reducing the fraction of generated
electricity ultimately available for useful consumption. Quantifying
these contributions therefore requires explicit assumptions regarding
system architecture and is incorporated later into the fleet-equivalent 
energy accounting and the subsequent system-level EROI methodology
in Sections~\ref{sec:fleet}~and~\ref{sec:EROImethodology}.

\section{The grid is still the backbone}

It is a common belief that a rooftop solar array, especially when
paired with a household battery, makes a household largely independent
from the grid. In practice, this is not the case. Such systems can
cover part of the daytime load and shift a limited fraction of
generation to evening hours, but they cannot bridge extended periods
of low solar output or seasonal deficits. Reliable supply therefore
continues to depend on the public electricity grid, particularly
during winter conditions.

From a system perspective, distributed PV changes power-flow patterns
but does not eliminate the need for the underlying network. Periods
of high solar irradiation lead to simultaneous injection from many
distributed systems, requiring adjustments in dispatchable generation
and, in some cases, curtailment or export to external markets. During
periods of low solar output, however, electricity demand must still be
met through other generation sources and the existing transmission and
distribution infrastructure.

This operational pattern also affects the economic structure of the
system. Periods of high PV generation tend to coincide with low
wholesale electricity prices, reducing revenues for dispatchable
generation precisely when solar output is abundant. At these times,
dispatchable generators must either reduce output or sell electricity
into wholesale markets at low, and occasionally negative, prices.
Nevertheless, the same dispatchable capacity must remain available to
ensure supply during periods of low renewable output. As a result, a
growing share of system costs is associated with maintaining
infrastructure that is only partially utilized in energy terms, but
remains essential for reliability.

As a consequence, the electricity grid continues to function as a
full-capacity system providing firm supply while accommodating
variable and partially utilized generation assets. Dispatchable
generation capacity, transmission infrastructure, and distribution
networks must remain available even when renewable output is high.
From an energetic perspective, maintaining infrastructure that is
utilized less intensively increases the effective energy investment
per unit of delivered electricity. This effect links system-level
EROI$_{\text{sys}}$ directly to infrastructure utilization and
constitutes one of the principal differences between installation-level
and system-level assessments.

\section{Impact of orientation and latitude on yield}
\label{sec:impact}

Unlike dispatchable generation technologies, PV systems produce
electricity only when solar irradiation is available. The delivered
electricity output therefore depends strongly on module orientation,
tilt angle, latitude, and local environmental conditions.

For Central Europe, an optimally oriented south-facing PV installation
typically produces approximately \SI{1150}{kWh/kWp} annually
\cite{Fraunhofer2023:PVReport}. Alternative deployment configurations
generally reduce this output. East–west orientations spread
generation more evenly across morning and evening hours but typically
reduce annual yield by about 10--15%
\cite{Keiner2024:AssessingYieldDisparities,
IEA:PVPS2015:Task13:LongTermPerformance}. Flat-roof installations typically incur losses of approximately
15--20\%, while vertically mounted south-facing
fa\c{c}ade-integrated systems may experience yield reductions of up
to 40\% relative to optimally tilted south-facing modules
\cite{IEA:PVPS2015:Task13:LongTermPerformance,
Schram2025:FacadePV}. Less favorable fa\c{c}ade orientations can
result in substantially larger losses. Representative yield-adjustment factors are
summarized in Table~\ref{tab:yield_factors}.

\begin{table}[ht]
\centering
\caption{Representative yield-adjustment factors relative to an
optimally oriented south-facing PV installation in Central Europe.
Annual yields are shown relative to a benchmark yield of
\SI{1150}{kWh/kWp/yr}. Values for fa\c{c}ade systems are indicative
and depend strongly on latitude, local horizon, and shading.}
\label{tab:yield_factors}

\begin{tabular}{lcc}
\toprule
Configuration &
$f_{\text{orientation}}$ &
Annual yield \\
&
&
(kWh/kWp/yr) \\
\midrule
South-facing, optimal tilt          & 1.00        & 1150 \\
East--west roof orientation         & 0.85--0.90 & 980--1040 \\
Flat roof                           & 0.80--0.85 & 920--980 \\
Vertical south-facing fa\c{c}ade    & 0.60--0.70 & 690--805 \\
Vertical east/west fa\c{c}ade       & 0.45--0.60 & 520--690 \\
Vertical north-facing fa\c{c}ade    & 0.20--0.40 & 230--460 \\
\bottomrule
\end{tabular}
\end{table}

Latitude further influences PV performance through seasonal
variations in solar irradiation. In Central Europe, winter
electricity production is commonly less than one third of summer
production, with the highest-yield month producing up to five times
more electricity than the lowest-yield month
\cite{PVGIS:Status2024,Huld2012:PVGIS}. By contrast, desert regions
such as California or Saudi Arabia can achieve annual yields of
\SIrange{1800}{2000}{kWh/kWp}, approximately twice those typical for
Central Europe.

Alpine installations can partially mitigate seasonal variability
through higher irradiance, reduced fog, and enhanced winter
performance. However, these benefits are accompanied by additional
infrastructure requirements associated with remote terrain, access,
and grid connection, as discussed previously.

Because embodied-energy inputs remain largely unchanged, reductions in
electricity yield translate directly into corresponding reductions in
energy return on investment. The adjustment factors summarized in
Table~\ref{tab:yield_factors} are incorporated later through the
output-side modifiers used in the fleet-level EROI framework.

\section{Fleet-equivalent embodied-energy inventory and EROI}
\label{sec:fleet}

Installation-level EROI values characterize the energetic performance
of individual photovoltaic installations. Long-term operation of a
large regional photovoltaic fleet, however, requires ongoing
replacement of shorter-lived components, operational support, grid
integration measures, and additional infrastructure beyond the initial
installation inventory.

The present section extends the installation-level inventories of
Table~\ref{tab:installation_inventory} to a fleet-equivalent
embodied-energy inventory. All values are expressed per installed
photovoltaic capacity (MWh/kWp) and represent steady-state fleet
operation. Storage systems, dispatchable backup generation, and other
firming pathways are excluded at this stage and are treated separately
in subsequent sections.

Table~\ref{tab:component} summarizes the resulting fleet-equivalent
embodied-energy inventory that defines the reference embodied-energy
investment for the subsequent system-level analysis.
Relative to the installation-level inventories
of Table~\ref{tab:installation_inventory}, the values incorporate
component turnover, operational and maintenance requirements, grid
reinforcement, and synthetic-inertia provisions required for
large-scale deployment of inverter-dominated generation fleets.

\begin{table}[htbp]
\centering
\caption{Fleet-equivalent embodied-energy inputs of photovoltaic systems in Central Europe expressed per installed capacity (MWh/kWp). Relative to the installation-level inventories of Table~\ref{tab:installation_inventory}, the values incorporate component turnover, operational and maintenance requirements, grid reinforcement, and synthetic-inertia provisions. Firming infrastructure is not included.}
\label{tab:component}

\renewcommand{\arraystretch}{1.1}

\begin{tabular}{lccc}
\toprule
Component & Rooftop PV & Utility-scale PV & Alpine PV \\
& (MWh/kWp) & (MWh/kWp) & (MWh/kWp) \\
\midrule

\multicolumn{4}{l}{\emph{Installation inventory (Table~\ref{tab:installation_inventory})}} \\
Installation inventory &
1.69 &
2.01--2.18 &
2.42--2.72 \\

\addlinespace

\multicolumn{4}{l}{\emph{Fleet-level additions}} \\
Inverter turnover        & 0.20 & 0.17--0.25 & 0.20--0.30 \\
O\&M                     & 0.05 & 0.05--0.08 & 0.05--0.08 \\
Grid reinforcement       & 0.20 & 0.25--0.40 & 0.25--0.40 \\
Synthetic inertia        & 0.02 & 0.02 & 0.02 \\

\midrule

\textbf{Total $E_{\text{in}}$} &
\textbf{2.16} &
\textbf{2.50--2.93} &
\textbf{2.94--3.52} \\

\bottomrule
\end{tabular}
\end{table}

The fleet-equivalent embodied-energy inventories summarized in
Table~\ref{tab:component} define the reference embodied-energy
investment of the photovoltaic fleet. Combined with the representative
lifetime electricity outputs introduced in
Section~\ref{sec:installation_eroi}, they yield the fleet-level EROI
values summarized in Table~\ref{tab:fleet_eroi}. Compared with the
installation-level EROI values of Table~\ref{tab:eout}, the
fleet-level EROI additionally accounts for component turnover,
operation and maintenance, grid reinforcement, and synthetic-inertia
provisions associated with the large-scale deployment of
photovoltaic generation.

For consistency with the benchmark framework adopted throughout this
study, fleet-level EROI values are evaluated for optimally oriented
systems under favorable site conditions. Orientation losses,
site-specific yield reductions, curtailment, storage systems, and
other firming measures are excluded at this stage and are introduced
separately in subsequent sections.

The fleet-level EROI is therefore defined as $\mathrm{EROI}_{\mathrm{fleet}}=\frac{E_{\mathrm{gross}}}{E_{\mathrm{in,fleet}}}$
where $E_{\mathrm{gross}}$ denotes the gross lifetime electricity
output and $E_{\mathrm{in,fleet}}$ the fleet-equivalent embodied
energy input of Table~\ref{tab:component}.

\begin{table}[ht]
\centering
\caption{Fleet-level EROI values for photovoltaic systems in Central
Europe derived from the lifetime electricity outputs of
Table~\ref{tab:eout} and the fleet-equivalent embodied-energy
inventories of Table~\ref{tab:component}. The values include
component turnover, operational and maintenance requirements, grid
reinforcement, and synthetic-inertia provisions, but exclude
orientation losses, curtailment, storage systems, and other firming
measures.}
\label{tab:fleet_eroi}

\begin{tabular}{lccc}
\toprule
System type &
$E_{\text{gross}}$ &
$E_{\text{in,fleet}}$ &
Fleet-level EROI \\
&
(MWh/kWp) &
(MWh/kWp) &
\\
\midrule
Rooftop PV       & 31.5--34.5 & 2.16        & 14.6--16.0 \\
Utility-scale PV & 33.0--36.0 & 2.50--2.93 & 11.3--14.4 \\
Alpine PV        & 37.5--40.5 & 2.94--3.52 & 10.7--13.8 \\
\bottomrule
\end{tabular}
\end{table}

The fleet-level EROI values quantify the energetic performance of the 
photovoltaic fleet itself after accounting for continuous component replacement 
and long-term operation. They therefore constitute the reference fleet-level EROI 
benchmark prior to introducing the additional energetic investments associated 
with system-level firming.

The fleet-level EROI values are systematically lower than the
corresponding installation-level values of Table~\ref{tab:eout}, reflecting the additional embodied-energy
requirements associated with long-term fleet operation. The reduction is most
pronounced for utility-scale and alpine PV, where grid integration
requirements and infrastructure needs contribute more strongly to the
overall embodied-energy inventory.

The fleet-equivalent embodied-energy inventory established here
provides the reference energy-accounting framework for the
photovoltaic generation fleet itself. The remaining sections
progressively evaluate the additional primary-energy investments
required to transform this generating fleet into a reliable
electricity system.

\section{Firming pathways}
\label{sec:firming}

The previous sections established the fleet-equivalent
embodied-energy inventory of the photovoltaic fleet and the main
output-side factors governing usable electricity delivery. The present
section examines the principal pathways by which intermittent
electricity generation can be transformed into reliable electricity
supply. Each pathway modifies the system-level energy balance through
a combination of additional primary-energy investments, conversion
losses, renewable overbuild, or dispatchable backup capacity.

The first and most natural mitigation of photovoltaic intermittency is
not storage, but the addition of wind generation. In Central Europe,
photovoltaic systems typically achieve capacity factors of
approximately \mbox{10--15\%}, while wind generation commonly reaches
\mbox{25--45\%} depending on site conditions and the relative contribution of
onshore and offshore installations. Wind and solar generation exhibit
partial complementarity on both daily and seasonal timescales:
photovoltaic output is concentrated around midday and during summer,
whereas wind generation tends to be stronger during winter and is not
constrained by the day--night cycle. 

A representative benchmark used throughout this section is a
renewable system in which approximately 20\% of annual electricity is
supplied by PV and 80\% by wind. This ratio broadly reflects the
relative capacity factors of photovoltaic and wind generation in
Central Europe and serves as a convenient reference for discussing
firming requirements. It should be regarded as illustrative rather
than prescriptive.

This complementarity substantially reduces variability and lowers
storage requirements relative to a photovoltaic-only system.
Nevertheless, the correlation is incomplete. Analyses of multi-decadal
European weather records show that prolonged periods of simultaneously
low wind and solar generation remain a characteristic feature of
renewable electricity systems, even when geographical aggregation is
considered~\cite{Kittel2026:Dunkelflaute}. Such renewable-energy
droughts (Dunkelflauten) can persist for several weeks and therefore
represent a critical design constraint for highly renewable power
systems. Reliable electricity supply consequently requires some
combination of storage, renewable overbuild, dispatchable generation,
imports, or demand-side flexibility.

The scale of this challenge extends well beyond short-term balancing.
Multi-hour storage can mitigate fluctuations associated with the
day--night cycle and short weather events, but cannot address
prolonged periods of low renewable generation or seasonal mismatches
between electricity supply and demand. National and regional studies
therefore identify firming requirements on the scale of multiple
terawatt-hours. 

Switzerland provides a contemporary example of the magnitude of these
firming requirements. A recent assessment by the Swiss Federal Electricity
Commission (ElCom) estimates winter net-import requirements ranging
from approximately \SIrange{2}{8}{TWh} in 2030 and up to \SI{13}{TWh} in 2035,
depending on assumptions regarding demand growth, domestic generation
expansion, and import availability
\cite{ElCom2025:Winterproduktionsfaehigkeit}. These values correspond
to approximately 3--14\% and up to about 22--25\%, of
present-day Swiss annual electricity consumption. The same study
emphasizes that security of supply depends critically on both import
availability and adverse weather conditions, highlighting the close
connection between adequacy and firming requirements in future
electricity systems.

A Royal Society assessment based on 37 years of hourly weather data
estimates seasonal storage requirements exceeding \SI{100}{TWh} for a
highly renewable UK electricity system
\cite{RoyalSociety2023:LargeScaleStorage}, equivalent to
approximately one third of present-day UK annual electricity demand.

Different firming pathways address these challenges through markedly
different combinations of additional infrastructure, conversion
losses, storage capacity, renewable overbuild, and dispatchable
generation. As a result, they impose substantially different
energetic penalties and lead to markedly different values of EROI$_{\text{sys}}$.

\subsection{Battery storage}
\label{sec:battery_storage}

Compared with seasonal storage options, batteries exhibit relatively
high electric-to-electric round-trip efficiencies, typically of order
80--90\%. They are therefore well suited for shifting electricity over
short and intermediate timescales. The favourable benchmark
calculations presented below adopt the upper-limit value
$\eta_{\mathrm{batt}}=0.90$.

The principal energetic limitation of battery storage is not its
conversion efficiency, but the embodied energy of the installed
capacity. Literature values for lithium-ion battery manufacturing
indicate primary-energy requirements of approximately
\SIrange{0.3}{0.7}{MWh} per kWh of installed battery capacity
\cite{Ellingsen2014,Degen2024}. Because battery lifetimes are
substantially shorter than the assumed 30-year photovoltaic-module
lifetime, replacement becomes a recurring fleet-level energy
investment rather than a one-time installation cost. Battery turnover
must therefore be included in the fleet-equivalent framework adopted
in this study.

Battery energy capacity and cumulative battery throughput must be
distinguished. A battery can cycle repeatedly during a year and thereby
shift an annual quantity of electricity substantially larger than its
installed energy capacity. This throughput does not accumulate into a
long-duration energy reserve. During an extended renewable-energy
deficit, the battery can discharge only the energy stored at the onset
of the event and cannot recharge until a new electricity surplus becomes
available. Repeated short-duration cycling therefore cannot be counted
as an equivalent reduction of seasonal storage requirements.

This limitation does not diminish the system-level value of batteries.
They can absorb short-lived renewable surpluses, reduce curtailment,
shift electricity from periods of high generation to periods of high
demand, limit ramps and peak loads, and provide rapid-response grid
services. They may also reduce the number and severity of operating
transients imposed on dispatchable generators or seasonal-storage
systems. Quantifying these interactions, however, requires a
chronological electricity-system model and lies outside the
fleet-equivalent benchmark developed here.

The energetic cost of battery load balancing is instead evaluated from
the installed energy capacity and its replacement over the analysis
lifetime. Because both quantities increase with the prescribed storage
duration, the fleet-equivalent embodied-energy investment rises
approximately in proportion to battery capacity. It can consequently
become comparable to, or exceed, the embodied energy of the renewable
generation fleet even though the battery itself operates with high
round-trip efficiency.

Battery storage is therefore treated in this work as a load-balancing
and grid-support technology, rather than as a substitute for the energy
inventory required during prolonged or seasonal renewable-generation
deficits. The representative battery durations and their associated
fleet-equivalent energy investments are introduced in
Section~\ref{sec:battery_infrastructure} and evaluated at system level
in Section~\ref{sec:battery_balancing}.

\subsection{Pumped-hydropower storage}

Pumped-hydropower storage (PHS) is widely regarded as the most mature
large-scale electricity-storage technology. Its combination of high
round-trip efficiency, long service life, and comparatively low
embodied-energy requirements makes it one of the most energetically
favorable storage options currently available.

The principal limitation of PHS is not efficiency but geography.
Suitable sites require large elevation differences, substantial
reservoir volumes, and appropriate geological conditions. As a
result, the technically and environmentally accessible storage
capacity is limited even in countries with extensive hydropower
resources.

Switzerland provides a useful illustration of both the strengths and
limitations of hydropower-based firming. The country possesses one of
Europe’s most flexible hydroelectric systems, comprising run-of-river
plants, reservoir hydropower, and pumped-storage facilities.
Hydropower currently produces approximately \SI{37}{TWh/y}, nearly
half of Swiss electricity generation, with roughly equal
contributions from run-of-river and reservoir plants
\cite{SFOE2025:HydropowerOverview}. Switzerland’s ability to maintain
such a high share of renewable electricity depends on the combined
contribution of these complementary hydropower resources rather than
on storage alone.

Pumped-storage facilities constitute only a small fraction of total hydropower generation but provide valuable operational flexibility. The discussion below focuses on the three largest Swiss pumped-storage schemes, which constitute the country’s principal large-scale pumped-storage assets and provide the only GW-scale pumped-storage resources available for system-level balancing. Numerous smaller facilities also exist, but their individual storage
volumes and power ratings are substantially lower than those of the three schemes considered here. According to the Swiss Federal Office of Energy, the usable stored energy of these three schemes is approximately
\SI{22.9}{GWh} for Linth–Limmern (Muttsee),
\SI{17.8}{GWh} for Nant de Drance (Vieux-Emosson), and
\SI{93.3}{GWh} for Hongrin/Veytaux
\cite{SFOE2025:EnergieinhalteStauseen}. Together, these facilities
provide approximately \SI{0.13}{TWh} of dispatchable stored energy.

This figure refers only to the energy that can be actively shifted
through pumped-storage operation and should not be confused with the
much larger energy content of Switzerland’s reservoir-hydropower
system. Reservoir lakes contain approximately \SI{8.0}{TWh} of recoverable hydroelectric energy under normal operating conditions, obtained by summing the energy contents of the major Swiss storage-reservoir systems reported by the Swiss Federal Office of Energy \cite{SFOE2025:EnergieinhalteStauseen}, and provide an important source
of seasonal flexibility by allowing water inflows accumulated during
summer and autumn to be utilized during periods of higher winter
demand.

Nevertheless, even this exceptional hydroelectric infrastructure does not eliminate concerns regarding winter adequacy. A recent assessment by the Swiss Federal Electricity Commission (ElCom) estimates winter net-import requirements ranging from approximately \SIrange{2}{8}{TWh} in 2030 and up to \SI{13}{TWh} in 2035, depending on assumptions regarding demand growth, domestic generation expansion, and import availability \cite{ElCom2025:Winterproduktionsfaehigkeit}. The combined stored energy of the three largest Swiss pumped-storage facilities (\SI{0.13}{TWh}) corresponds to only about 1–6\% of these projected winter firming requirements. While pumped-storage plants can cycle repeatedly and therefore shift substantially larger amounts of energy over the course of a year, their principal contribution in Switzerland is to support reservoir management by conserving water during periods of surplus electricity production and thereby increasing reservoir inventories available for winter electricity generation. The comparison nevertheless illustrates the disparity between the instantaneous energy inventory of even large pumped-storage facilities and the multi-terawatt-hour energy reserves required for seasonal adequacy. The study further emphasizes that security of supply depends critically on import availability and adverse weather conditions.

More fundamentally, Switzerland already represents one of the most favorable
hydroelectric systems in Europe, possessing extensive reservoir storage,
major pumped-storage infrastructure, and a large fleet of run-of-river plants.
Despite these advantages, projected winter supply deficits remain substantial.
Consequently, the limitations identified here should be interpreted as an
upper bound on the contribution that hydropower-based firming can provide
in most Central European electricity systems, which generally possess
significantly smaller hydropower resources.

Opportunities for further large-scale expansion of pumped-storage
capacity are also constrained by geography, environmental
considerations, competing water uses, and the fact that many of the
most favorable sites have already been developed. While additional
projects remain possible, they would increase storage capacity only
incrementally and would not bridge the multi-terawatt-hour gap
between existing storage volumes and projected winter firming
requirements.

A similar conclusion emerges at larger scales, particularly in
countries that lack Switzerland’s extensive hydropower and reservoir
resources. Studies of highly renewable electricity systems in
Germany, the United Kingdom, and Europe more broadly identify storage
requirements of many terawatt-hours despite geographical aggregation
and transmission interconnection. The Royal Society, for example,
estimates seasonal storage requirements exceeding \SI{100}{TWh} for a
highly renewable UK electricity system based on 37 years of hourly
weather data \cite{RoyalSociety2023:LargeScaleStorage}. Such values
exceed the combined storage volumes of existing pumped-storage
systems by orders of magnitude and illustrate the scale of the
challenge faced by countries without substantial reservoir-hydropower
resources. Even hydro-rich regions such as Switzerland cannot rely on pumped storage alone to provide the energy reserves required for winter adequacy, while most
Central European countries possess far more limited hydropower-based
firming potential.

From the perspective of system-level EROI, pumped-hydropower storage remains one of the most attractive firming options because its high round-trip efficiency minimizes delivery losses and its long service life limits embodied-energy turnover. Its principal limitation is not energetic performance but deployable storage volume. Geographic constraints severely restrict the number of suitable sites, while even hydro-rich regions possess storage volumes that are limited when compared with the multi-terawatt-hour energy reserves associated with winter adequacy. Pumped hydropower can therefore make an important contribution to short-term balancing and reservoir management but cannot by itself provide the seasonal energy reserves required for highly renewable electricity systems.

\subsection{Hydrogen storage}
\label{sec:hydrogen_storage}

Unlike batteries and pumped-hydropower storage, hydrogen can, in
principle, provide very large energy-storage capacities without a
proportional increase in storage-vessel infrastructure. Hydrogen is
therefore one of the few proposed energy-storage technologies capable
of providing the multi-terawatt-hour energy reserves associated with
seasonal balancing in highly renewable electricity systems. Its
principal attraction lies in the possibility of storing large
quantities of surplus renewable electricity over extended periods and
recovering that energy during prolonged intervals of low wind and
solar generation.

The present discussion is restricted to hydrogen as a storage medium
for renewable electricity. Hydrogen may also serve as an industrial
feedstock, transport fuel, or chemical energy carrier, and these
applications may be justified by considerations that differ from those
governing electricity storage. The analysis presented here addresses
only the use of hydrogen as an electric-to-electric storage pathway for
firming renewable electricity generation.

Hydrogen storage involves multiple conversion stages. Renewable
electricity is first converted into hydrogen through electrolysis,
after which the hydrogen is compressed, transported, and stored.
Electricity is subsequently recovered through fuel cells or gas
turbines. Unlike batteries or pumped-hydropower storage, whose
principal limitations are storage duration and deployable volume,
the dominant challenge for hydrogen storage is the cumulative
efficiency loss associated with these successive conversion steps.

Electric-to-electric hydrogen storage based on electrolysis and
subsequent electrical reconversion commonly achieves a round-trip
efficiency of approximately 40\%~\cite{DOE2022:HydrogenStorage}.
A recent integrated supply-chain review reports overall efficiencies
of approximately 30--40\% for gaseous-hydrogen pathways with pipeline
distances of \SI{1000}{km} to \SI{3000}{km}, with still lower values
for longer transport distances
~\cite{Restrepo2025:HydrogenSupplyChains}. Although the end-use
boundary considered in that review differs from electrical
reconversion, the results illustrate the additional losses introduced
by compression, storage, distribution, and long-distance transport.
The present benchmark therefore adopts a central electric-to-electric
round-trip efficiency of
$\eta_{\mathrm{rt}}=0.35$.

The storage requirement considered here corresponds to a deliberately
stringent firming case in which photovoltaic and wind generation,
together with hydrogen storage, are assumed to provide firm electricity
without reliance on fossil-fuelled backup, nuclear generation, or
imports from neighbouring systems that themselves depend on
dispatchable generation. Imports are considered only insofar as they
represent geographically remote wind or photovoltaic generation
delivered through long-distance transmission. This assumption reflects
the objective of testing whether variable renewable generation can
provide autonomous firm electricity under deep-decarbonization
constraints.

Long-distance transmission can reduce storage requirements by
aggregating geographically diverse renewable resources and thereby
smoothing short-term fluctuations in generation. However, the
meteorological conditions responsible for prolonged periods of low
renewable output are often correlated over continental scales
~\cite{Kittel2026:Dunkelflaute}. Consequently, even geographically
aggregated renewable systems may require substantial seasonal energy
reserves~\cite{RoyalSociety2023:LargeScaleStorage}. In the present
PV--wind--hydrogen case, external firming from fossil, nuclear, or
other dispatchable sources is deliberately excluded. Geographical
aggregation can therefore reduce, but not eliminate, the need for
seasonal storage.

Seasonal storage requirements arise from two distinct but related
mechanisms. The first is the systematic seasonal mismatch between
renewable generation and electricity demand. In Central Europe,
photovoltaic production is highest during summer, whereas electricity
demand tends to increase during winter due to reduced daylight,
increased lighting requirements, and the anticipated electrification
of heating through heat pumps. Periods of surplus generation and
periods of peak demand therefore occur at opposite times of the year,
creating an intrinsic requirement for seasonal energy transfer.

The second mechanism consists of prolonged intervals of unusually low
renewable output, commonly referred to as Dunkelflauten, during which
both wind and solar generation may remain significantly below average
for days or weeks. Wind generation partially mitigates the seasonal
mismatch because average wind resources are generally stronger during
the colder months. However, adequacy requirements are determined not
by average conditions but by periods of exceptionally low renewable
output. Analyses of multi-decadal European weather records demonstrate
that prolonged renewable-energy droughts can occur even in
geographically aggregated wind and solar systems
~\cite{Kittel2026:Dunkelflaute}. Seasonal storage must therefore
compensate not only for structural seasonal imbalances but also for
extended periods of reduced renewable generation.

The magnitude of the required seasonal reserve may be expressed as

\[
s
=
\frac{E_{\text{seasonal storage}}}
     {E_{\text{annual demand}}},
\]

where $s$ denotes the fraction of annual electricity demand that must
be supplied through seasonal storage. This quantity cannot be inferred
uniquely from any single weather event or national case, but several
reference points constrain its plausible scale.

At the lower end, analyses of multi-decadal European weather records
show that renewable-energy droughts can persist for several weeks even
in geographically aggregated wind and solar systems. The most severe
event identified by Kittel et al. occurred during the winter of
1996/97 and lasted approximately 55 days
~\cite{Kittel2026:Dunkelflaute}. Although renewable generation remained
significantly above zero throughout the event, such extended periods
of reduced wind and solar availability imply that seasonal reserves
corresponding to only a few percent of annual demand are unlikely to
provide a robust adequacy basis.

A useful real-world benchmark is provided by Switzerland, whose
storage reservoirs contain approximately \SI{8}{TWh} of recoverable
hydroelectric energy, corresponding to roughly 13\% of annual Swiss
electricity consumption, in a system where hydropower already supplies
nearly half of annual electricity generation. Although hydroelectric
reservoir storage is not directly equivalent to hydrogen storage, this
benchmark illustrates the scale of seasonal reserves available in one
of Europe's most favorable hydroelectric systems and suggests that
countries lacking comparable hydropower resources are unlikely to
require smaller seasonal reserves. As a simple scaling argument,
systems that do not benefit from Switzerland's approximately 50\%
hydropower contribution would require correspondingly larger seasonal
reserves. This consideration suggests storage requirements of order
$2\times13\%$, i.e. approximately one quarter of annual electricity
demand.

At the upper end, the Royal Society estimates seasonal storage
requirements exceeding \SI{100}{TWh} for a highly renewable UK
electricity system, corresponding to roughly one third of annual
electricity demand~\cite{RoyalSociety2023:LargeScaleStorage}. These
reference points indicate that seasonal storage requirements in
autonomous PV--wind systems plausibly lie in the range
$s\simeq10\text{--}30\%$, depending on geography, renewable mix,
demand flexibility, transmission availability, and the acceptable
level of adequacy risk.

The reference calculations are performed for
$s=15\%,\ 20\%,\ 25\%$, while $s=30\%$ is treated as a conservative
high-security case.

The renewable overbuild required to support seasonal hydrogen storage
is estimated from the electricity that must be routed through the
hydrogen pathway in order to deliver the required seasonal reserve
back to the grid. It may be approximated as
\[
f_{\mathrm{overbuild}}
=
\frac{s\,c_{\mathrm{reserve}}}
     {\eta_{\mathrm{rt}}\,f_{\mathrm{refill}}},
\]
where $\eta_{\mathrm{rt}}$ is the electric-to-electric round-trip
efficiency of the hydrogen pathway, $f_{\mathrm{refill}}$ denotes
the effective fraction of annual renewable generation available for
replenishing seasonal hydrogen stores, and
$c_{\mathrm{reserve}}$ is a contingency factor accounting for
operational reserve requirements, forecast uncertainty, interannual
variability, and recurrent low-renewable periods.

The round-trip efficiency appears in this expression because it
determines how much additional renewable generation must be built in
order to compensate the conversion losses of the hydrogen cycle. The
resulting overbuild is therefore treated as an input-side embodied-
energy contribution associated with the additional renewable fleet,
rather than as a separate output-side delivery factor.

The refill parameter should not be interpreted as a fixed calendar
window. Hydrogen production can occur whenever renewable generation
exceeds contemporaneous demand, including summer photovoltaic
surpluses, windy spring and autumn periods, and occasional winter
intervals with high wind output. For Central European conditions, a
representative value of $f_{\mathrm{refill}}\simeq0.50$ is adopted.

The resulting renewable-overbuild requirements are summarized in
Table~\ref{tab:h2_overbuild}.

\begin{table}[htbp]
\centering
\caption{Representative assumptions and resulting renewable-overbuild
requirements for seasonal hydrogen storage. The central values
$\eta_{\mathrm{rt}}=0.35$, $f_{\mathrm{refill}}=0.50$, and
$c_{\mathrm{reserve}}=1.25$ are applied to the seasonal-storage
fractions discussed in the text.}
\label{tab:h2_overbuild}

\begin{tabular}{ccccc}
\toprule
Seasonal storage fraction $s$ &
$\eta_{\mathrm{rt}}$ &
$f_{\mathrm{refill}}$ &
$c_{\mathrm{reserve}}$ &
$f_{\mathrm{overbuild}}$ \\
\midrule
15\% & 0.35 & 0.50 & 1.25 & 1.07 \\
20\% & 0.35 & 0.50 & 1.25 & 1.43 \\
25\% & 0.35 & 0.50 & 1.25 & 1.79 \\
30\% & 0.35 & 0.50 & 1.25 & 2.14 \\
\bottomrule
\end{tabular}
\end{table}

Even at $s=15\%$, the additional renewable-generation fleet is
slightly larger than the reference fleet itself. Across the central
benchmark range $s=15\text{--}25\%$, the overbuild factor increases
from 1.07 to 1.79, reaching 2.14 in the high-security case.

The dominant driver of this overbuild is the relatively low
round-trip efficiency of the hydrogen cycle, which requires multiple
units of renewable electricity to be generated for each unit
subsequently recovered from storage. For the central benchmark range,
the additional renewable-generation fleet is therefore larger than
the fleet needed to satisfy annual electricity demand itself.

Additional embodied-energy investments arise from the construction,
maintenance, and replacement of the infrastructure required to support
the hydrogen pathway, including electrolysers, storage facilities,
reconversion systems, compressors, and associated grid connections.
These contributions are distinct from the additional renewable-
generation fleet represented by $f_{\mathrm{overbuild}}$. Within the
present benchmark, they are assumed to remain secondary to the
renewable-generation overbuild and are therefore neglected in the
calculations.

Hydrogen-based seasonal firming therefore substantially increases the
energetic burden of the renewable system primarily through the
additional renewable-generation capacity required to compensate
conversion losses and maintain seasonal reserves. While hydrogen is
one of the few storage technologies capable of providing
multi-terawatt-hour seasonal reserves, this capability comes at the
cost of a large renewable overbuild before the embodied energy of the
hydrogen infrastructure itself is even included.

\paragraph{Large-scale hydrogen storage}

The storage capacities considered here are far beyond the range that
can be provided economically using conventional pressurized storage
tanks. While compressed-gas tanks and liquefied-hydrogen systems are
well established for industrial applications, the material, energy,
and cost requirements of such approaches increase rapidly with storage
volume and become prohibitive at the multi-terawatt-hour scales
associated with seasonal electricity storage
~\cite{Bussar2016,Caglayan2020:HydrogenStorageReview}.

Consequently, most analyses of large-scale hydrogen storage assume the
use of underground geological storage, particularly solution-mined
salt caverns
~\cite{Bussar2016,Caglayan2020:HydrogenStorageReview,
IEA2023:HydrogenStorage}. Salt caverns can store very large quantities
of hydrogen with comparatively low standing losses and represent the
most mature option with operational experience relevant to large-scale
geological hydrogen storage
~\cite{Caglayan2020:HydrogenStorageReview,IEA2023:HydrogenStorage}.
Other geological options, including depleted hydrocarbon reservoirs
and saline aquifers, have been proposed and may ultimately provide
additional storage capacity, but their large-scale deployment remains
less mature~\cite{Heinemann2021:EuropeanSaltCaverns}.

However, suitable geological formations are geographically constrained
and are not uniformly available across Europe
~\cite{Caglayan2020:HydrogenStorageReview,
Heinemann2021:EuropeanSaltCaverns}. Consequently, although hydrogen can
provide very large storage volumes, the locations at which geological
storage can be deployed are considerably more limited. Large-scale
hydrogen systems may therefore require substantial pipeline
infrastructure linking renewable-generation regions, storage sites,
and demand centres. Recent European planning studies envisage a
dedicated hydrogen transmission network of approximately
\SI{58000}{km} by 2040~\cite{EuropeanHydrogenBackbone2024}. The
resulting increase in hydrogen handling and transport also increases
the importance of controlling leakage throughout the hydrogen supply
chain
~\cite{Caglayan2020:HydrogenStorageReview,
Heinemann2021:EuropeanSaltCaverns,IEA2023:HydrogenStorage}.

\paragraph{Hydrogen leakage}

Hydrogen storage introduces an additional environmental consideration
through hydrogen leakage. Owing to hydrogen's small molecular size,
high diffusivity, and low viscosity, leakage can occur throughout the
supply chain, including production, storage, transport, and
reconversion systems~\cite{IEA2023:HydrogenStorage}.

A recent integrated review reports constructed leakage estimates of
approximately 1--8\% for complete gaseous-hydrogen supply chains, 
with average values of approximately 4.5\%~\cite{Restrepo2025:HydrogenSupplyChains}. 
These values are assembled from literature estimates for individual 
supply-chain stages rather than direct measurements of complete 
operating systems and therefore remain uncertain. The full interval 
should consequently be interpretedas a scenario envelope rather than 
a statistically established range.The present benchmark adopts a representative
leakage range of 2--7\%,centred on the reported average while excluding the 
most optimistic and pessimistic combinations of component-level assumptions.

Although hydrogen is not itself a greenhouse gas, it influences
atmospheric chemistry by affecting concentrations of methane, ozone,
and stratospheric water vapour. Multi-model atmospheric-chemistry
calculations estimate an effective global-warming potential for
hydrogen of $11.6\pm2.8$ over a 100-year horizon and
$37.3\pm15.1$ over a 20-year horizon
~\cite{Sand2023:HydrogenGWP}.

While these effects do not enter directly into the EROI calculation,
they represent an additional environmental consideration beyond the
energetic penalties associated with hydrogen storage and renewable
overbuild.

\subsection{Synthetic fuels}

Synthetic fuels are often proposed as alternatives to direct hydrogen
storage because they can be transported and stored using infrastructure
that is similar to, or compatible with, existing fuel systems.
However, from an energetic perspective they are best regarded as
extensions of the hydrogen pathway. Their production requires hydrogen
as an intermediate feedstock and therefore inherits the conversion
losses associated with hydrogen production before introducing
additional synthesis, storage, transport, and reconversion stages.

These pathways can provide energy carriers compatible with existing
infrastructure and are particularly relevant for sectors that are
difficult to electrify directly. However, each additional processing
step increases energy consumption, infrastructure requirements, and
renewable-overbuild needs. Consequently, synthetic fuels generally
incur larger energetic penalties than direct hydrogen storage and are
therefore expected to exhibit correspondingly lower system-level EROI
values when used as electricity-storage media.

Ammonia (NH$_3$) is produced from hydrogen and nitrogen through the
Haber–Bosch process. Compared with molecular hydrogen, ammonia offers
several advantages for storage and transport: it can be liquefied
under substantially milder conditions, possesses a higher volumetric
energy density than compressed hydrogen, and benefits from mature
industrial production, handling, and distribution infrastructure
\cite{Lan2012:AmmoniaHydrogenStorage,ValeraMedina2018:AmmoniaReview,IEA2021:Ammonia}. These characteristics can mitigate some of the
storage and transport challenges associated with molecular hydrogen.

However, ammonia does not eliminate the fundamental energetic
penalties associated with hydrogen production. Rather, it should be
viewed as a hydrogen carrier that exchanges some of hydrogen’s storage
and transport difficulties for additional conversion steps and their
associated energy losses.

For electricity storage, however, ammonia introduces additional
conversion requirements. Green ammonia production requires
electrolytic hydrogen, nitrogen separation, and Haber--Bosch synthesis;
recovery of electricity then requires either ammonia cracking followed
by hydrogen use in fuel cells or turbines, or direct ammonia combustion
with associated combustion and emissions-control challenges. Studies
of renewable ammonia production also show that variable renewable
input requires additional process flexibility, including hydrogen
buffering, plant oversizing, and curtailment management~\cite{NayakLuke2020:GreenAmmonia}. 
Consequently, ammonia may improve storage and transport logistics
relative to molecular hydrogen, but it does not provide an
energetically superior pathway for bulk electricity storage.
Compared with direct hydrogen storage in suitable geological
reservoirs, the additional conversion stages further reduce the
fraction of renewable electricity ultimately recovered as useful
electrical power.

Unlike ammonia, synthetic methane (CH$_4$) and synthetic liquid
hydrocarbons require not only hydrogen but also a sustainable carbon
source. Because the carbon released during fuel use must ultimately be
balanced by an equivalent carbon input, large-scale deployment depends
on biogenic carbon, carbon capture from industrial sources, or direct
air capture (DAC). Biomass availability is inherently limited,
industrial point sources are expected to decline under deep
decarbonization, and DAC requires substantial additional energy input.

Methanation, Fischer–Tropsch synthesis, and subsequent refining
stages therefore introduce further conversion losses, infrastructure
requirements, and energy consumption beyond those already associated
with hydrogen production~\cite{IPCC:2021:WG3}.

From an energetic perspective, ammonia, synthetic methane, synthetic
liquid hydrocarbons, and related synthetic fuels inherit the losses of
the hydrogen pathway while introducing additional conversion stages.
Their overall round-trip efficiencies are therefore lower than those
of direct hydrogen storage, implying larger renewable-overbuild
requirements, greater EROI penalties, and correspondingly lower
system-level EROI values.

Accordingly, synthetic fuels may be justified where chemical energy
carriers are intrinsically required, particularly in aviation,
shipping, and selected industrial processes. Their use as bulk
electricity-storage media is less attractive because each additional
conversion stage further reduces the fraction of renewable electricity
that can ultimately be recovered as useful electrical power.

\subsection{Alternative firming and storage options}

Beyond batteries, pumped-hydropower storage, and chemical-storage
pathways, a variety of additional technologies have been proposed to
improve the reliability of electricity systems with high shares of
variable renewable generation. These include gravity-based storage,
compressed-air energy storage, thermal-energy storage, and various
forms of demand-side flexibility. While such approaches can provide
valuable balancing services, they address only part of the broader
challenge of maintaining reliable electricity supply during prolonged
periods of low renewable generation.

Gravity-based storage systems operate on the same physical principle
as pumped hydropower, storing energy through the elevation of a mass
in a gravitational field. Their scale limitation follows directly from
the relation $E=mgh$. Storing \SI{1}{TWh} at a height difference of
\SI{500}{m} would require lifting approximately
$7\times10^{11}\,\mathrm{kg}$, or about
\SI{7e8}{tonnes}, of material. Even at such a substantial elevation
difference, multi-terawatt-hour storage would require the repeated
movement of billions of tonnes of solid mass. Gravity-based storage
may therefore provide useful short- or medium-duration balancing, but
it is unlikely to constitute a practical solution for seasonal firming
at the scale required by highly renewable electricity systems.

Compressed-air energy storage similarly stores energy in a physical
medium for later recovery. Both technologies may provide valuable
short- to medium-duration balancing, but their practical deployment
remains constrained by geography, available storage volumes, and
site-specific requirements. They are therefore unlikely to provide the
multi-terawatt-hour reserves associated with seasonal balancing.

Thermal-energy storage can provide valuable flexibility where electricity
is used for heating. Examples include hot-water tanks, district-heating
storage, and seasonal thermal-energy storage systems. Such technologies
can reduce peak electricity demand and shift heat production to periods
of surplus generation. Thermal storage may also improve heat-pump
performance by providing a more favourable heat source during cold
periods, thereby reducing electricity consumption when heating demand
is highest.

However, thermal storage stores thermal energy rather than electrical
energy and therefore cannot substitute for firm electrical generation.
Although heat pumps can deliver several units of heat per unit of
electricity consumed under favourable conditions, their performance
typically declines during periods of high heating demand and low
ambient temperatures. Consequently, the large quantity of heat
currently supplied by fossil fuels must still be replaced by
substantial amounts of electricity. Thermal storage can mitigate this
requirement, but it does not eliminate the need for reliable
electricity supplies during winter periods of low renewable generation.

Demand-side flexibility represents another important source of system
balancing. Flexible charging of electric vehicles, industrial load
management, smart appliances, adaptive operation of heat pumps, and
vehicle-to-grid (V2G) concepts can shift electricity demand in time
and thereby reduce storage requirements.

The potential contribution of V2G can be illustrated through a simple
scaling argument. A fleet of 10 million electric vehicles with
\SI{80}{kWh} batteries would contain a total storage capacity of
\SI{0.8}{TWh}. If half of the vehicles were connected to the grid and
half of their battery capacity were made available for balancing, the
accessible storage volume would be approximately \SI{0.2}{TWh}. 
While substantial for short-duration balancing, this remains small
compared with the multi-terawatt-hour energy reserves associated with
seasonal firming.

Demand-side measures can therefore improve system flexibility and
reduce balancing requirements, but their effectiveness depends on user
participation, infrastructure availability, and the duration of
low-renewable periods. They appear more relevant to hourly and daily
balancing than to long-duration or seasonal energy storage.

Taken together, these technologies can improve system flexibility,
reduce balancing requirements, and in some cases provide valuable
short- or medium-duration storage. However, they primarily address the
timing of energy consumption and delivery rather than the fundamental
requirement for large quantities of firm energy during prolonged
periods of low renewable generation. Thermal storage stores heat
rather than electricity, demand-side flexibility shifts consumption
without creating additional energy, and gravity-based or vehicle-based
storage remains limited in scale when compared with seasonal storage
requirements.

Consequently, while such measures can reduce the magnitude of the
firming challenge, they do not eliminate it. Highly renewable
electricity systems therefore continue to require access to firm
energy reserves measured on the scale of multiple terawatt-hours in
order to maintain security of supply during extended periods of low
renewable output.

\subsection{Dispatchable backup generation}

An alternative to large-scale storage is the retention of dispatchable
generation capacity capable of supplying electricity whenever wind and
solar output are insufficient. Unlike storage-based firming pathways,
dispatchable generation does not require electricity to be stored and
later recovered. It therefore avoids the conversion losses, renewable
overbuild, and storage infrastructure associated with batteries,
hydrogen, and synthetic fuels.

Reservoir hydropower represents the most favorable form of dispatchable
generation where suitable resources are available. Reservoir storage
allows electricity generation to be shifted in time with comparatively
high efficiency and long infrastructure lifetimes. In countries with
substantial hydroelectric resources, hydropower can provide both
balancing services and firm generation capacity. However, deployment is
strongly constrained by geography, and opportunities for major
expansion are limited in many regions.

Where large-scale hydropower is unavailable, dispatchable generation is
typically provided by thermal power plants. Natural-gas turbines and
combined-cycle gas plants are particularly attractive because of their
high operational flexibility and ability to respond rapidly to changes
in renewable generation. As a result, gas-fired generation is widely
regarded as the principal backup technology for electricity systems
with high shares of intermittent renewables.

From an energetic perspective, however, gas backup does not eliminate
the need for a dispatchable generation fleet. Electricity systems must
be designed to satisfy peak load rather than annual energy demand, and
extended periods of low renewable output may require dispatchable
generation to supply a large fraction of system load. Consequently,
even if photovoltaic and wind generation provide most annual
electricity production, sufficient thermal capacity must remain
available to meet peak demand during prolonged Dunkelflaute events.
The system must therefore maintain not only the renewable-generation
fleet but also a largely complete dispatchable-generation fleet.

From a system-level EROI perspective, this implies that a largely
redundant backup infrastructure must be constructed, maintained,
operated, and periodically replaced despite operating at relatively
low annual utilization factors. Renewable generation may substantially
reduce fuel consumption, but it does not eliminate the energetic cost
of maintaining the dispatchable capacity required to guarantee
security of supply. Storage pathways primarily increase energy inputs
through storage infrastructure and conversion losses. Gas backup avoids
these storage penalties, but instead requires the construction,
maintenance, operation, and periodic replacement of dispatchable
generation assets that may operate only intermittently. The energetic
cost of maintaining this parallel infrastructure therefore becomes
part of the system-level energy investment.

In practice, photovoltaic and wind generation combined with gas backup
function primarily as fuel-saving systems rather than as complete
substitutes for dispatchable generation. Annual gas consumption may be
substantially reduced, but the firm generation capacity itself cannot
be reduced proportionally because it remains essential for reliability
during Dunkelflaute events and other prolonged periods of low renewable
generation. Consequently, the EROI benefit obtained from reducing fuel
consumption is partially offset by the continued requirement to
construct, maintain, and periodically replace a dispatchable
generation fleet sized to meet peak system demand.

Nuclear power occupies a different position within the firming
landscape. Although fully dispatchable, nuclear generation is generally
deployed as a primary source of electricity rather than as a backup
technology. Modern nuclear fleets are capable of demand-following
operation and can provide a substantial fraction of system flexibility
while maintaining high capacity factors. In systems with significant
nuclear generation, a large fraction of electricity demand can therefore
be supplied directly by the nuclear fleet, reducing the need for
storage, renewable overbuild, and dispatchable fossil backup. 

The following section develops normalized primary-energy
contributions for the representative firming architectures considered
in the subsequent system-level EROI analysis.

\section{System-level EROI methodology}
\label{sec:EROImethodology}

The preceding sections examined the principal pathways by which 
intermittent photovoltaic generation can be transformed into reliable 
electricity supply. Batteries, pumped-hydropower storage, hydrogen, 
synthetic fuels, dispatchable backup generation, and alternative firming 
measures each introduce additional primary-energy requirements through 
embodied-energy investments, conversion losses, renewable overbuild, 
or the maintenance of redundant infrastructure. The engineering quantities 
established in the preceding sections are now translated into delivery factors 
and normalized primary-energy investments, thereby extending the fleet-equivalent
framework to the system level.

The fleet-level photovoltaic EROI values derived earlier in this study
represent photovoltaic generation before the introduction of firming
requirements. They therefore provide a useful reference, but they do
not describe the energetic performance of a system capable of
delivering reliable electricity under all operating conditions.

\subsection{System-level methodology}
\label{sec:system_losses}

To distinguish between gross electricity generation and electricity
ultimately available for useful consumption, the delivered lifetime
electricity output is written as
\[
E_{\text{out,delivered}}
=
E_{\text{gross}} \cdot
f_{\text{orientation}} \cdot
f_{\text{site}} \cdot
f_{\text{curtailment}} \cdot
f_{\text{delivery}},
\]
where the individual factors represent reductions associated with
module orientation, site-specific conditions, curtailment, and
delivery losses occurring after electricity has been generated.

The orientation and site factors were discussed in
Section~\ref{sec:impact}, while curtailment was introduced in
Section~\ref{sec:curtailment}. The delivery factor accounts for
unavoidable losses incurred during electricity delivery, including
transmission losses, transformer losses, power-electronics conversion
losses, standby losses, and similar grid-related processes. It does
not include storage round-trip efficiencies. The energetic
consequences of storage losses are instead represented explicitly
through the additional primary-energy investments required for storage
infrastructure and, where applicable, the additional renewable
generation needed to compensate storage losses.

The resulting delivered electricity output forms the basis for the
system-level EROI$_{\text{sys}}$ calculations presented later in this
study.

\subsection{System architecture and methodology}

The following calculations are based on an illustrative renewable
electricity system consisting of 80\% wind and 20\% photovoltaic
generation on an annual electricity-output basis. This mix is not
intended to represent an optimized national electricity system, but
serves as a simple Central European benchmark for evaluating the
energetic implications of different firming strategies.

The purpose of this reference system is not to determine the absolute 
EROI of a complete wind–photovoltaic electricity system. Rather, it 
provides a consistent benchmark for quantifying the additional primary-energy 
investments required to transform variable renewable generation 
into reliable electricity supply. These additional primary-energy investments 
are normalized to the embodied-energy investment of the
photovoltaic fleet and subsequently combined in
Section~\ref{sec:EROI_sys} to construct representative low-carbon
electricity-system architectures.

Unless stated otherwise, all calculations assume favourable
photovoltaic siting conditions with
$f_{\mathrm{site}}=1$ and
$f_{\mathrm{orientation}}=1$, such that only the energetic
consequences of the firming strategies themselves are evaluated.
Curtailment is treated explicitly where relevant for the
corresponding firming architecture.

The fleet-level photovoltaic EROI values reported in
Table~\ref{tab:fleet_eroi} provide the reference embodied-energy
inventory against which the additional primary-energy requirements
associated with the various firming technologies are evaluated.

The engineering quantities introduced in the preceding sections are
now translated into these normalized contributions for each
representative firming pathway.

For each firming technology, the corresponding primary-energy
investment is normalized to the embodied-energy investment of the
photovoltaic fleet according to
\[
f_i
=
\frac{E_i}{E_{\mathrm{PV}}},
\]
where $E_i$ denotes the primary-energy investment associated with the
firming technology under consideration. In the following subsections,
$i$ refers in turn to battery storage, seasonal hydrogen storage,
dispatchable backup generation, and the primary-energy input required
for fuel supply.

The dimensionless quantities $f_i$ therefore directly express the energetic importance 
of each firming measure relative to the embodied-energy investment required to construct the photovoltaic fleet
and provide a compact parameterization of
the corresponding primary-energy investments associated with the
different firming strategies. Because all quantities are normalized to the 
same reference embodied-energy investment, they can be added directly to 
construct the denominator of the system-level EROI expression developed in 
Section~\ref{sec:EROI_sys}.

\subsection{Battery infrastructure}
\label{sec:battery_infrastructure}

Electrochemical battery storage introduces an embodied-energy
investment associated with manufacturing the installed storage capacity
and replacing it over the lifetime of the renewable-generation fleet.
Finite round-trip efficiency additionally limits the electricity
recovered from each charge--discharge cycle. The present benchmark
quantifies the infrastructure investment associated with prescribed
battery capacities. Consistent with the physical distinction established
in Section~\ref{sec:battery_storage}, battery cycling is not credited as
a reduction in long-duration or seasonal firming requirements.

To expose how the infrastructure investment varies with installed
battery capacity, representative load-balancing durations of
\SIlist{0.25;0.5;1;2;4;24}{h} are considered. This sequence spans
sub-hourly, hourly, multi-hour, and daily-scale storage. The selected
durations define engineering benchmark cases rather than optimized
storage capacities for a specific electricity system.

For a battery power rating normalized to the installed photovoltaic
capacity, a storage duration $t_{\mathrm{batt}}$ corresponds to an
installed battery energy capacity of
\[
C_{\mathrm{batt}}
=
t_{\mathrm{batt}}\,
\mathrm{kWh/kWp}.
\]

The corresponding fleet-equivalent battery embodied-energy investment
is
\[
E_{\mathrm{batt}}
=
C_{\mathrm{batt}}\,
e_{\mathrm{batt}}\,
\frac{\tau_{\mathrm{PV}}}
     {\tau_{\mathrm{batt}}},
\]
where $e_{\mathrm{batt}}$ denotes the battery manufacturing energy per
unit storage capacity,
$\tau_{\mathrm{PV}}=\SI{30}{yr}$ is the photovoltaic-fleet lifetime,
and $\tau_{\mathrm{batt}}$ is the battery service lifetime. Consistent
with Section~\ref{sec:battery_storage}, the favourable benchmark assumes
$e_{\mathrm{batt}}=\SI{0.3}{MWh/kWh}$ and
$\tau_{\mathrm{batt}}=\SI{15}{yr}$, whereas the conservative benchmark
assumes $e_{\mathrm{batt}}=\SI{0.7}{MWh/kWh}$ and
$\tau_{\mathrm{batt}}=\SI{12}{yr}$.

Following the methodology introduced in
Section~\ref{sec:EROImethodology}, the normalized battery contribution
becomes
\[
f_{\mathrm{batt}}
=
\frac{E_{\mathrm{batt}}}
     {E_{\mathrm{PV}}},
\]
where $E_{\mathrm{PV}}$ denotes the fleet-equivalent embodied-energy
investment of the photovoltaic fleet given in
Table~\ref{tab:component}.

\begin{table}[htbp]
\centering
\caption{Fleet-equivalent battery embodied-energy investments and
normalized battery contributions for representative load-balancing
durations. The normalized contribution is defined as
$f_{\mathrm{batt}}=E_{\mathrm{batt}}/E_{\mathrm{PV}}$, with
$E_{\mathrm{PV}}$ taken from Table~\ref{tab:component}. The lower and
upper limits combine the favourable and conservative battery assumptions
with the corresponding photovoltaic embodied-energy inventories.}
\label{tab:battery_factors}

\begin{tabular}{lccccc}
\toprule
Battery &
$E_{\mathrm{batt}}$ fav. &
$E_{\mathrm{batt}}$ cons. &
Rooftop PV &
Utility-scale PV &
Alpine PV \\
duration &
(MWh/kWp) &
(MWh/kWp) &
$f_{\mathrm{batt}}$ &
$f_{\mathrm{batt}}$ &
$f_{\mathrm{batt}}$ \\
\midrule

\SI{0.25}{h} & 0.15 & 0.44 & 0.07--0.20 & 0.05--0.18 & 0.04--0.15 \\
\SI{0.5}{h}  & 0.30 & 0.88 & 0.14--0.41 & 0.10--0.35 & 0.09--0.30 \\
\SI{1}{h}    & 0.60 & 1.75 & 0.28--0.81 & 0.20--0.70 & 0.17--0.60 \\
\SI{2}{h}    & 1.20 & 3.50 & 0.56--1.62 & 0.41--1.40 & 0.34--1.19 \\
\SI{4}{h}    & 2.40 & 7.00 & 1.11--3.24 & 0.82--2.80 & 0.68--2.38 \\
\SI{24}{h}   & 14.40 & 42.00 & 6.67--19.44 & 4.91--16.80 & 4.09--14.29 \\
\bottomrule
\end{tabular}
\end{table}

Table~\ref{tab:battery_factors} shows that the embodied-energy
investment associated with battery storage scales linearly with
installed storage duration. Consequently, the normalized battery
contribution $f_{\mathrm{batt}}$ likewise increases linearly with
battery capacity. Even comparatively short-duration systems require a
significant additional embodied-energy investment, while multi-hour and
especially daily-scale systems can exceed the embodied-energy
investment of the photovoltaic fleet itself.

Consistent with the favourable benchmark assumptions adopted here, no
dedicated renewable-generation overbuild is assigned to battery load
balancing. Batteries are assumed to recover naturally occurring
renewable surpluses that would otherwise be curtailed. Their finite
round-trip efficiency reduces the useful electricity recovered from
those surpluses. No numerical credit against seasonal hydrogen or
dispatchable backup is assigned.

The normalized contribution $f_{\mathrm{batt}}$ is consequently the
sole battery-specific energy investment carried into the system-level
EROI benchmark of Section~\ref{sec:EROI_sys}. It quantifies the energetic
cost of prescribing a given battery capacity without claiming that the
same capacity provides long-duration or seasonal adequacy.

\subsection{Seasonal hydrogen infrastructure}
\label{sec:H2_infrastructure}

Seasonal hydrogen storage differs fundamentally from battery storage.
For batteries, the dominant energetic contribution arises from the
embodied energy of the storage system itself. For seasonal hydrogen,
the present benchmark assumes that the dominant contribution arises
from the additional renewable generation required to compensate
hydrogen-cycle losses and maintain the seasonal reserve. The embodied-
energy investments associated with electrolysers, compressors, storage
caverns, reconversion equipment, and transport infrastructure are
therefore neglected. This deliberately favorable boundary provides an
upper benchmark for the system-level EROI of the hydrogen-storage
pathway.

The principal parameter describing seasonal firming is the seasonal
storage fraction
\[
s
=
\frac{E_{\mathrm{seasonal}}}
     {E_{\mathrm{annual}}},
\]
where $E_{\mathrm{seasonal}}$ denotes the annual electricity delivered
from seasonal hydrogen storage and $E_{\mathrm{annual}}$ the annual
electricity demand. Thus, $s$ represents the fraction of annual
electricity demand that must ultimately be supplied through the
hydrogen-storage pathway.

Delivering this seasonal energy requires substantially larger amounts
of renewable electricity because only a fraction of the electricity
used to charge the hydrogen store is recovered after electrolysis,
compression, transport, storage, and reconversion. The dominant
physical parameter governing this process is therefore the complete
electric-to-electric round-trip efficiency
$\eta_{\mathrm{rt}}$. As discussed in
Section~\ref{sec:hydrogen_storage}, representative complete-pathway
efficiencies lie approximately in the range 30--40\%. The present
benchmark adopts the central value
$\eta_{\mathrm{rt}}=0.35$.

In addition to round-trip efficiency, two further quantities influence
the renewable-generation requirement. The refill fraction
$f_{\mathrm{refill}}$ denotes the effective fraction of annual
renewable generation available for replenishing the seasonal hydrogen
store. Hydrogen production can occur whenever renewable generation
exceeds contemporaneous demand, including summer photovoltaic
surpluses, windy spring and autumn periods, and occasional winter
intervals with high wind output. For representative Central European
conditions, a value
$f_{\mathrm{refill}}=0.50$
is adopted.

Finally, a contingency factor
$c_{\mathrm{reserve}}$
accounts for operational reserve requirements, forecast uncertainty,
interannual meteorological variability, and recurrent periods of low
renewable generation. This factor reflects the requirement that
seasonal hydrogen stores provide adequate security of supply under
adverse operating conditions rather than merely under average
meteorological conditions. A value
$c_{\mathrm{reserve}}=1.25$
is adopted.

The representative calculations consider seasonal-storage fractions
of $s=15\%,\,20\%,\,25\%$, and $30\%$, spanning seasonal-reserve
requirements from moderate to conservative security-of-supply cases.
Together with $\eta_{\mathrm{rt}}=0.35$,
$f_{\mathrm{refill}}=0.50$, and
$c_{\mathrm{reserve}}=1.25$, these values provide the basis for
deriving the normalized renewable-overbuild contributions of the
representative electricity-system architectures evaluated in
Section~\ref{sec:EROI_sys}.

The corresponding overbuild is evaluated explicitly only after the
representative generation architectures have been specified. Because
the embodied-energy requirements of the hydrogen infrastructure are
omitted, the resulting EROI values should be interpreted as optimistic
upper benchmarks rather than complete lifecycle estimates of deployed
seasonal hydrogen systems.

When synthetic fuels are employed as intermediates for electricity
storage, they inherit all energetic penalties associated with the
hydrogen pathway while introducing additional conversion and
processing stages. Consequently, synthetic-fuel pathways necessarily
require at least as much renewable overbuild as direct hydrogen
storage. The hydrogen pathway therefore defines a lower bound on the
energetic penalty, and an upper bound on the achievable system-level
EROI, of synthetic-fuel-based electricity-storage systems.

\subsection{Dispatchable backup infrastructure}
\label{sec:disp_infrastructure}

Unlike storage-based firming pathways, dispatchable backup generation
does not store electricity produced during periods of surplus
renewable generation. Instead, electricity is generated directly
whenever photovoltaic and wind output are insufficient to satisfy
system demand. The energetic implications therefore differ
fundamentally from those associated with batteries or seasonal
hydrogen storage.

Maintaining dispatchable backup introduces two distinct energetic
contributions. The first is the embodied-energy investment required to
construct, maintain, and periodically replace the dispatchable
generating fleet. The second is the operational primary-energy input
associated with the fuel consumed whenever dispatchable generation is
required. Unlike storage-based firming pathways, dispatchable backup
therefore combines a comparatively modest infrastructure investment
with a potentially large continuing operational energy requirement.

The present subsection considers only the infrastructure contribution.
Within the notation adopted throughout this work, the normalized
embodied-energy investment associated with dispatchable backup is
defined as
\[
f_{\mathrm{disp}}
=
\frac{E_{\mathrm{in,disp}}}
     {E_{\mathrm{in,PV}}},
\]
where $E_{\mathrm{in,disp}}$ denotes the embodied-energy investment
required to construct and maintain the dispatchable generating fleet,
normalized to the embodied-energy investment of the photovoltaic
fleet. The operational primary-energy contribution associated with
fuel consumption depends on the required firm-energy fraction and is
derived separately in
Section~\ref{sec:dispatchable_backup}.

The embodied-energy contribution differs fundamentally from that of
storage systems. Dispatchable backup avoids the construction of
large-scale storage infrastructure and the conversion losses
associated with repeated charge--discharge cycles. However, the
dispatchable fleet must remain capable of supplying essentially the
full system demand during prolonged periods of low renewable
generation. Consequently, its installed capacity is determined
primarily by peak demand rather than annual electricity production.
Although renewable generation substantially reduces fuel consumption,
the embodied-energy investment required to maintain the dispatchable
fleet remains largely independent of its annual utilization.

The present analysis considers the embodied-energy investment
associated with constructing dispatchable gas-fired generating
capacity. Additional upstream energy expenditures associated with
natural-gas extraction, processing, compression, transport, and,
where applicable, liquefaction, shipping, and regasification scale
primarily with fuel consumption rather than installed generating
capacity. These operational energy inputs are therefore excluded from
$f_{\mathrm{disp}}$ and incorporated later in the system-level EROI
analysis. For LNG supply chains, these upstream energy requirements
may be substantially higher than for pipeline gas because
liquefaction, shipping, and regasification introduce additional
energy-intensive stages~\cite{Howarth2024:LNG}.

Published life-cycle assessments of gas-fired power plants
predominantly report results normalized to lifetime electricity
generation rather than to installed capacity
\cite{UNECE2022:LCA_ElectricityOptions,NETL2014:GasLCA}. The latter
quantity is, however, required for the present fleet-equivalent
framework, where dispatchable capacity is maintained irrespective of
its annual utilization. The dispatchable embodied-energy contribution
is therefore estimated from published construction-material
inventories rather than from per-kilowatt-hour life-cycle results.

Representative inventories for a modern combined-cycle gas-turbine
plant comprise approximately \SI{9.8e4}{kg/MW} of concrete,
\SI{3.1e4}{kg/MW} of steel, together with smaller quantities of iron,
aluminium, and other construction materials
\cite{Spath2000:NGCC}. Using representative cradle-to-gate
embodied-energy coefficients from the ICE database
\cite{HammondJones2011:ICE}, namely approximately
\SI{1.9}{MJ/kg} for reinforced concrete and
\SIrange{25}{35}{MJ/kg} for structural steel, with
\SI{30}{MJ/kg} adopted for the central estimate, yields a
material-only lower-bound embodied-energy contribution of
approximately \SI{0.3}{MWh/kW}.

This lower bound accounts only for the principal construction
materials. A realistic dispatchable generating fleet also requires
gas and steam turbines, heat-recovery steam generators, generators,
transformers, switchgear, cabling, auxiliary systems, construction
energy, and maintenance or refurbishment of shorter-lived components.
Rather than attempting to annualize these components individually,
given their widely differing service lives, the present calculations
adopt the representative range
\[
E_{\mathrm{in,disp}}
=
0.3\text{--}0.6\,\mathrm{MWh/kWp},
\]
where the lower bound is anchored by the published material
inventories and the upper bound provides an engineering allowance for
balance-of-plant equipment, maintenance, refurbishment, construction
energy, and uncertainties in the published inventories.

The corresponding normalized embodied-energy contribution depends on
the photovoltaic configuration through the embodied-energy investment
of the photovoltaic fleet. Representative values are summarized in
Table~\ref{tab:dispatchable_embodied_energy}.

\begin{table}[htbp]
\centering
\caption{Representative normalized embodied-energy contributions for
maintaining full-power dispatchable backup capacity. Operational
primary-energy inputs associated with fuel consumption are excluded
and treated separately in
Section~\ref{sec:dispatchable_backup}.}
\label{tab:dispatchable_embodied_energy}

\begin{tabular}{lcc}
\toprule
PV configuration &
$E_{\mathrm{in,disp}}$ &
$f_{\mathrm{disp}}$ \\
&
(MWh/kWp) & \\
\midrule

Rooftop PV       & 0.3--0.6 & 0.14--0.28 \\
Utility-scale PV & 0.3--0.6 & 0.10--0.24 \\
Alpine PV        & 0.3--0.6 & 0.09--0.20 \\

\bottomrule
\end{tabular}
\end{table}

Compared with seasonal hydrogen storage, dispatchable backup avoids
the large renewable-generation overbuild required to compensate
hydrogen-cycle losses. Its embodied-energy infrastructure requirement
is comparatively modest because no large-scale storage system has to
be constructed. This advantage is, however, offset by the continuing
primary-energy input required whenever dispatchable generation is
used. The combined effect of infrastructure and fuel consumption on
system-level EROI is evaluated in
Section~\ref{sec:dispatchable_backup}.

The present analysis treats dispatchable backup as part of the
electricity system supplying firm demand. Whether this capacity is
located within the system itself or provided through imports from
neighbouring regions does not fundamentally alter the energetic
accounting. Reliable imports require equivalent dispatchable
generation capacity to exist elsewhere and remain available during
periods of low renewable output. The embodied-energy investment
associated with this capacity is therefore not eliminated but merely
transferred to the exporting system.

Over much larger geographical scales, long-distance interconnections
can exploit reduced weather correlations, different time zones, and
complementary renewable resources, thereby reducing both storage and
dispatchable-backup requirements. The present analysis is restricted
to Central Europe, where meteorological conditions remain strongly
correlated over large regions and such benefits are correspondingly
more limited. Even in highly interconnected systems, transmission
redistributes firm generation but does not eliminate the need for
dispatchable capacity. Moreover, extensive interconnections require
substantial transmission infrastructure and create strategic
dependencies on external transmission corridors and exporting
regions. These considerations lie beyond the scope of the present
EROI analysis but represent additional practical constraints on
highly interconnected renewable-electricity systems.

\section{System-level EROI of representative low-carbon electricity systems}
\label{sec:EROI_sys}

The preceding section quantified the normalized primary-energy
contributions associated with the principal firming technologies
considered in this work. The present section combines these energetic
building blocks into representative electricity-system architectures
in order to evaluate the resulting system-level EROI of
photovoltaics under different firming strategies.

The representative system architectures considered here are intended
to illustrate the energetic consequences of different firming
strategies in a deeply decarbonized electricity system. To keep the
comparison transparent, only intermittent renewable generation is
considered, represented by an illustrative mix of approximately
80\% wind and 20\% photovoltaics, consistent with previous studies of
highly renewable electricity systems
\cite{RoyalSociety2023:LargeScaleStorage}. Such a wind-dominated mix
is also broadly consistent with the higher annual capacity factors of
wind relative to photovoltaics in Central Europe and serves here as a
representative operating point rather than an optimized solution.

Conventional fossil generation is excluded, except where explicitly
introduced as dispatchable backup. Hydroelectricity is omitted because
its availability is strongly constrained by geography and cannot be
expanded substantially in most regions. Nuclear power is likewise
excluded, reflecting electricity-transition scenarios in which future
electricity generation is assumed to rely predominantly on renewable
sources. The resulting reference architectures are therefore not
intended to predict future electricity systems, but to provide
internally consistent benchmarks for evaluating the energetic
implications of different firming strategies.

The present analysis nevertheless remains centred on photovoltaics.
The energetic penalties associated with overbuild, curtailment,
storage, dispatchable backup, and transmission arise from the
intermittent nature of renewable generation rather than from
photovoltaics specifically. Similar energetic penalties therefore
apply to wind-dominated systems, although their quantitative values
depend on the adopted generation mix and regional meteorological
conditions. The assumed 80/20 wind--photovoltaic mix is therefore not
introduced to analyse wind power itself, but to provide representative
system-level firming requirements for evaluating the EROI of
photovoltaics within a representative low-carbon electricity system.

In existing electricity systems, renewable overproduction does not necessarily
result in curtailment. Dispatchable generators may reduce their output, electricity 
may be exported to neighbouring systems, or demand-side flexibility may 
absorb part of the surplus. 

The representative electricity-system architectures considered here instead 
correspond to fully decarbonized electricity-transition scenarios in which 
 balancing by conventional dispatchable generation is deliberately 
excluded, except where such generation is explicitly introduced 
as a separate firming option. Whenever instantaneous renewable generation exceeds 
demand, the surplus must therefore either be stored or curtailed. 
Renewable-generation overbuild, storage, and curtailment are therefore 
inseparable aspects of the representative benchmark systems analysed in this work.

Alpine PV is retained as a separate photovoltaic configuration because
it is frequently proposed as a means of improving winter electricity
production in Switzerland and other mountain regions. However, it is
not treated as a separate hydro-dominated electricity-system
architecture. Even in Switzerland, where hydropower and reservoir
storage provide exceptional flexibility, projected winter electricity
deficits remain substantial
\cite{ElCom2025:Winterproduktionsfaehigkeit}. Alpine PV therefore
tests whether its higher winter yield can materially improve the
system-level EROI of photovoltaics once the additional embodied-energy
investments for mounting structures, site access, grid connection,
and the remaining system-level firming requirements are taken into
account.

\subsection{System-level EROI framework}

The purpose of the present framework is to identify the dominant
energetic requirements of representative low-carbon electricity
architectures. It is a transparent benchmark calculation rather than
a chronological simulation or optimization of a real electricity
system. Its results should therefore be interpreted primarily as
identifying the energetic regime in which an architecture operates,
rather than as precise predictions of a particular electricity
system.

For the representative electricity-system architectures considered
in this study, the complete system-level EROI is defined as
\[
\mathrm{EROI}_{\mathrm{sys}}
=
\frac{E_{\mathrm{out,sys}}}
     {E_{\mathrm{PV}}
      +E_{\mathrm{wind}}
      +E_{\mathrm{batt}}
      +E_{\mathrm{H2}}
      +E_{\mathrm{disp}}
      +E_{\mathrm{fuel}}
      +E_{\mathrm{overbuild}}},
\]
where $E_{\mathrm{out,sys}}$ denotes the total delivered electrical
energy of the system. The denominator comprises the lifetime
primary-energy requirements associated with constructing,
maintaining, and operating the complete generation and firming
infrastructure.

Here, $E_{\mathrm{PV}}$ and $E_{\mathrm{wind}}$ denote the
embodied-energy investments associated with the photovoltaic and
wind fleets, respectively. The remaining terms represent the
additional primary-energy investments associated with battery
storage, hydrogen infrastructure, dispatchable backup generation,
fuel supply, and the renewable generation capacity required to
compensate storage and conversion losses. In particular,
$E_{\mathrm{overbuild}}$ denotes the primary-energy investment in
constructing, maintaining, and replacing this additional generation
capacity. It does not denote the electrical energy passing through
the storage or conversion pathway.

The present work focuses on the energetic performance of
photovoltaics. Rather than calculating the EROI of the complete
wind--photovoltaic electricity system, the analysis therefore
considers the photovoltaic-attributed system-level EROI,
\[
\mathrm{EROI}_{\mathrm{PV,sys}}
=
\frac{E_{\mathrm{out,PV}}}
     {E_{\mathrm{PV}}
      +E_{\mathrm{batt,PV}}
      +E_{\mathrm{H2,PV}}
      +E_{\mathrm{disp,PV}}
      +E_{\mathrm{fuel,PV}}
      +E_{\mathrm{overbuild,PV}}},
\]
where the subscript ``PV'' denotes the share of the electricity
system attributed to the photovoltaic fleet.

The representative 80/20 wind--photovoltaic annual-generation mix
introduced above serves to define the benchmark firming
requirements. These requirements are allocated to wind and
photovoltaics in proportion to their annual electricity generation.
Under this accounting convention, both the useful electrical output
and the associated primary-energy investments scale with the
photovoltaic generation fraction $f_{\mathrm{PV}}$. The common factor
$f_{\mathrm{PV}}$ therefore cancels throughout the normalized
photovoltaic-attributed formulation.

This proportional allocation is a deliberate simplification. It does
not imply that wind and photovoltaic generation have identical
temporal profiles or make identical physical contributions to
curtailment, storage demand, and residual load. Resolving such
differences would require chronological system simulation and lies
outside the scope of the benchmark model.

Following the methodology established in
Section~\ref{sec:EROImethodology}, all additional primary-energy
contributions are normalized to the embodied-energy investment of
the photovoltaic fleet,
\[
f_i
=
\frac{E_i}{E_{\mathrm{PV}}},
\]
where $i$ denotes the corresponding firming technology.

Renewable overbuild associated with storage and conversion losses is
treated separately. The corresponding normalized contribution for a
given firming pathway is written as
\[
f_{\mathrm{overbuild},i}
=
\frac{E_{\mathrm{overbuild},i}}
     {E_{\mathrm{PV}}},
\]
and the total renewable-overbuild contribution becomes
\[
f_{\mathrm{overbuild}}
=
\sum_i
f_{\mathrm{overbuild},i}.
\]

To preserve the transparent benchmark character of the model, the
required overbuild fraction is assigned uniformly to wind and
photovoltaic generation and is independent of the particular
generation configuration. The same fractional overbuild is therefore
applied to onshore and offshore wind and to rooftop, utility-scale,
and Alpine photovoltaic installations. This is an accounting
convention rather than a claim that these technologies would require
identical overbuild in a chronologically resolved electricity system.

The energetic cost of the additional capacity nevertheless retains
the characteristics of the underlying generation technology. A given
fractional overbuild therefore produces a larger primary-energy
investment when applied to a generation configuration with a lower
fleet EROI. This separation allows the storage and conversion pathway
to determine the overbuild fraction, while the generation technology
determines the energetic cost of providing the additional capacity.
In the representative architectures considered here, a dedicated
renewable-overbuild contribution is assigned only to the seasonal
hydrogen pathway. Under the favourable benchmark assumptions adopted
for battery load balancing, no dedicated renewable-generation overbuild 
is assigned to battery operation.

The denominator of the photovoltaic-attributed system-level EROI can
therefore be written as
\[
E_{\mathrm{PV}}
\left(
1
+
f_{\mathrm{batt}}
+
f_{\mathrm{H2}}
+
f_{\mathrm{disp}}
+
f_{\mathrm{fuel}}
+
f_{\mathrm{overbuild}}
\right).
\]

The delivered lifetime electricity output is obtained from the
fleet-level output through the delivery factors introduced in
Section~\ref{sec:system_losses},
\[
E_{\mathrm{out,PV}}
=
E_{\mathrm{gross}}
\,
f_{\mathrm{orientation}}
\,
f_{\mathrm{site}}
\,
f_{\mathrm{curtailment}}
\,
f_{\mathrm{delivery}}.
\]

Since
\[
\mathrm{EROI}_{\mathrm{fleet}}
=
\frac{E_{\mathrm{gross}}}
     {E_{\mathrm{PV}}},
\]
the photovoltaic-attributed system-level EROI becomes
\[
\mathrm{EROI}_{\mathrm{PV,sys}}
=
\frac{\mathrm{EROI}_{\mathrm{fleet}}
      \,
      f_{\mathrm{orientation}}
      \,
      f_{\mathrm{site}}
      \,
      f_{\mathrm{curtailment}}
      \,
      f_{\mathrm{delivery}}}
     {1
      +
      f_{\mathrm{batt}}
      +
      f_{\mathrm{H2}}
      +
      f_{\mathrm{disp}}
      +
      f_{\mathrm{fuel}}
      +
      f_{\mathrm{overbuild}}}.
\]

Normalizing the firming requirements to the embodied-energy
investment of the photovoltaic fleet provides a direct measure of
their energetic importance relative to the generating fleet. The
resulting $f_i$ values are PV-attributed quantities, but they also
indicate, to first order, the magnitude of the firming burden imposed
by weather-dependent renewable generation more generally.

As a limiting case, consider wind and photovoltaic generation to have
identical embodied-energy intensities per unit of delivered lifetime
electricity. Under the proportional-allocation convention, the
explicit dependence of the attributed EROI on the assumed
wind--photovoltaic generation mix would then disappear, because the
generation-specific embodied-energy investment would scale directly
with the electricity produced.

In practice, wind and photovoltaic generation do not have identical
embodied-energy intensities. Nevertheless, the standard-boundary EROI
values for onshore wind, offshore wind, and
photovoltaics differ by factors of order unity rather than by orders
of magnitude~\cite{Castro2020}. This observation supports their
treatment as comparable generation technologies in a first-order
benchmark, but it does not establish exact energetic equivalence or
uniquely determine the allocation of shared firming requirements.

Differences between the embodied-energy intensities of wind and
photovoltaic generation can remain relevant, particularly for
architectures requiring little firming. They are, however, secondary
to the largest energetic penalties associated with extensive
storage, renewable overbuild, conversion losses, and continuous fuel
consumption. The benchmark is therefore intended to identify the
trends produced by these requirements and to determine whether an
architecture retains a robust energy surplus or enters a low-EROI
regime. Its principal conclusions do not depend on numerical
precision within a given energetic regime.

The generation mix nevertheless remains important because it
determines the magnitude of the firming requirements through its
influence on curtailment, storage demand, dispatchable backup, and
seasonal balancing. The present framework does not resolve these
effects chronologically. Instead, it exposes their energetic
consequences systematically and allows the principal firming pathways
to be compared under a common set of transparent benchmark
assumptions.

\subsection{Battery load balancing}
\label{sec:battery_balancing}

The battery-only benchmark isolates the energetic consequence of adding
a prescribed load-balancing capacity to the renewable-generation fleet.
As established in Sections~\ref{sec:battery_storage} and
\ref{sec:battery_infrastructure}, batteries can provide valuable
short-duration shifting and grid-support services, but the installed
capacity is not credited as a substitute for long-duration or seasonal
firming. The configurations evaluated here are therefore reference
load-balancing cases and do not independently provide firm electricity.

The benchmark sequence
\SIlist{0.25;0.5;1;2;4;24}{h} spans sub-hourly, hourly,
multi-hour, and daily-scale storage. These durations define prescribed
battery capacities rather than optimized electricity-system designs.
Their fleet-equivalent embodied-energy contributions
$f_{\mathrm{batt}}$ are taken directly from
Table~\ref{tab:battery_factors}.

No dedicated renewable-generation overbuild is assigned to battery
operation. Moreover, the benchmark does not prescribe a fraction of
annual electricity that must pass through the battery. Battery
round-trip efficiency therefore cannot be applied as a single output
factor. The calculation instead isolates the embodied-energy penalty of
the installed battery infrastructure. Any operational losses associated
with electricity routed through the battery would reduce delivered
output further and are not included in the values reported below.

The photovoltaic-attributed system-level EROI becomes
\[
\mathrm{EROI}_{\mathrm{PV,sys}}
=
\frac{\mathrm{EROI}_{\mathrm{fleet}}
      \,
      f_{\mathrm{orientation}}
      \,
      f_{\mathrm{site}}
      \,
      f_{\mathrm{curtailment}}
      \,
      f_{\mathrm{delivery}}}
     {1+f_{\mathrm{batt}}}.
\]

The benchmark calculations assume
$f_{\mathrm{orientation}}
=f_{\mathrm{site}}
=f_{\mathrm{curtailment}}
=f_{\mathrm{delivery}}
=1$,
such that only the fleet-equivalent embodied-energy investment of the
battery reduces the resulting system-level EROI. The corresponding
values are summarized in
Table~\ref{tab:eroi_battery_balancing}.

\begin{table}[htbp]
\centering
\caption{System-level photovoltaic EROI for representative battery
load-balancing benchmark cases. The calculation adds the
fleet-equivalent embodied energy of the prescribed battery capacity but
does not include operational output losses or assign a reduction in
long-duration firming requirements.}
\label{tab:eroi_battery_balancing}

\begin{tabular}{lccc}
\toprule
Battery duration &
Rooftop PV &
Utility-scale PV &
Alpine PV \\
\midrule

\SI{0.25}{h} & 12.1--14.9 & 9.8--13.6 & 9.5--13.1 \\
\SI{0.5}{h}  & 10.4--14.0 & 8.7--12.9 & 8.5--12.5 \\
\SI{1}{h}    & 8.1--12.5 & 7.1--11.6 & 7.1--11.4 \\
\SI{2}{h}    & 5.6--10.3 & 5.1--9.7 & 5.3--9.8 \\
\SI{4}{h}    & 3.4--7.6 & 3.3--7.3 & 3.6--7.6 \\
\SI{24}{h}   & 0.7--2.1 & 0.7--2.1 & 0.8--2.3 \\
\bottomrule
\end{tabular}
\end{table}

The system-level EROI decreases monotonically with prescribed battery
duration because the fleet-equivalent embodied-energy investment rises
in direct proportion to installed capacity. Short-duration batteries
retain substantially higher EROI than multi-hour and daily-scale cases,
but this comparison does not identify an energetically optimal storage
duration. It quantifies the cost of successively larger battery
installations under otherwise unchanged benchmark assumptions.

The \SI{24}{h} case approaches, or falls below, energetic break-even
over much of the investigated parameter space. It nevertheless provides
only daily-scale load balancing and does not resolve prolonged or
seasonal renewable-generation deficits. The battery-only cases must
therefore not be interpreted as complete firm-electricity systems.

\subsection{Seasonal hydrogen storage}
\label{sec:h2_storage}

Seasonal hydrogen storage addresses a fundamentally different role
from battery load balancing. Whereas batteries shift renewable
electricity over hours, hydrogen storage provides long-duration energy
reserves capable of maintaining electricity supply during extended
periods of insufficient renewable generation. Within the system-level
EROI framework adopted here, the dominant energetic penalty therefore
arises from the renewable-generation overbuild required to sustain the
seasonal hydrogen cycle. As discussed in
Section~\ref{sec:H2_infrastructure}, the embodied-energy investments
associated with electrolysers, hydrogen storage, compressors,
transport, and reconversion equipment are assumed to be secondary and
are therefore neglected in the present benchmark calculations.

Seasonal hydrogen storage is supplied by the representative 80\% wind
/ 20\% photovoltaic generation mix introduced at the beginning of
this section. Unlike battery load balancing, seasonal hydrogen storage
relies on dedicated renewable-generation overbuild to create the
surplus electricity required for hydrogen production. This surplus is
accumulated over extended periods and subsequently used to maintain
electricity supply during prolonged intervals of insufficient
renewable generation.

Following the methodology established in
Section~\ref{sec:H2_infrastructure}, the normalized renewable-generation
overbuild associated with seasonal hydrogen storage is represented by
the quantity $f_{\mathrm{overbuild,H2}}$, which depends on the seasonal
storage fraction $s$, the complete-pathway hydrogen round-trip
efficiency $\eta_{\mathrm{rt}}$, the refill fraction
$f_{\mathrm{refill}}$, and the reserve factor
$c_{\mathrm{reserve}}$.

The physical energy losses associated with hydrogen leakage are
contained within the complete-pathway round-trip efficiency and are
therefore not applied again as a separate EROI penalty. The adopted
2--7\% leakage range is used separately when evaluating the indirect
climate contribution of hydrogen emissions.

Although expressed relative to the embodied-energy investment of the
photovoltaic fleet, $f_{\mathrm{overbuild,H2}}$ represents the
additional renewable generation required from the combined
80\% wind / 20\% photovoltaic electricity system. As discussed in
Section~\ref{sec:EROI_sys}, this normalization provides a good
approximation because the embodied-energy intensities of wind and
photovoltaic generation are similar when expressed per unit of
delivered lifetime electricity~\cite{Castro2020}.

Using the system-level EROI framework developed above, the
photovoltaic-attributed system-level EROI therefore becomes

\[
\mathrm{EROI}_{\mathrm{PV,sys}}
=
\frac{\mathrm{EROI}_{\mathrm{fleet}}
      \,
      f_{\mathrm{orientation}}
      \,
      f_{\mathrm{site}}
      \,
      f_{\mathrm{curtailment}}
      \,
      f_{\mathrm{delivery}}}
     {1+f_{\mathrm{overbuild,H2}}},
\]

where the benchmark calculations assume
$f_{\mathrm{orientation}}
=f_{\mathrm{site}}
=f_{\mathrm{curtailment}}
=f_{\mathrm{delivery}}=1$,
such that the system-level EROI is reduced solely by the
renewable-generation overbuild required for seasonal hydrogen storage.

Four representative seasonal-storage fractions of
$s=15\%,\,20\%,\,25\%$, and $30\%$ are considered. The corresponding
renewable-generation overbuild factors
$f_{\mathrm{overbuild,H2}}$ are taken directly from
Section~\ref{sec:H2_infrastructure}. The resulting benchmark
photovoltaic system-level EROI values are summarized in
Table~\ref{tab:eroi_h2_storage}.

\begin{table}[htbp]
\centering
\caption{System-level photovoltaic EROI for representative seasonal
hydrogen-storage benchmark cases. Here, $s$ denotes the seasonal
storage fraction and $f_{\mathrm{overbuild,H2}}$ the corresponding
normalized renewable-generation overbuild factor, calculated using
$\eta_{\mathrm{rt}}=0.35$, $f_{\mathrm{refill}}=0.50$, and
$c_{\mathrm{reserve}}=1.25$. Embodied-energy investments associated
with hydrogen infrastructure are neglected.}
\label{tab:eroi_h2_storage}

\begin{tabular}{ccccc}
\toprule
$s$ &
$f_{\mathrm{overbuild,H2}}$ &
Rooftop PV &
Utility-scale PV &
Alpine PV \\
\midrule
15\% & 1.07 & 7.0--7.7 & 5.4--7.0 & 5.1--6.7 \\
20\% & 1.43 & 6.0--6.6 & 4.6--5.9 & 4.4--5.7 \\
25\% & 1.79 & 5.2--5.7 & 4.0--5.2 & 3.8--4.9 \\
30\% & 2.14 & 4.6--5.1 & 3.6--4.6 & 3.4--4.4 \\
\bottomrule
\end{tabular}
\end{table}

The reduction in EROI is controlled by the factor
\[
\frac{1}{1+f_{\mathrm{overbuild,H2}}}.
\]
For the benchmark assumptions adopted here,
$\eta_{\mathrm{rt}}=0.35$, $f_{\mathrm{refill}}=0.50$, and
$c_{\mathrm{reserve}}=1.25$, the renewable-overbuild factor becomes
\[
f_{\mathrm{overbuild,H2}}
\simeq
7.14\,s.
\]
Consequently, even a seasonal-storage fraction of $s=15\%$ requires
$f_{\mathrm{overbuild,H2}}\simeq1.07$, corresponding to an additional
renewable-generation fleet slightly larger than the reference fleet.
The resulting denominator exceeds two, reducing the system-level EROI
to less than one half of the fleet-level value.

This reduction is not a consequence of the hydrogen round-trip
efficiency alone. It arises from the combined effect of low
electric-to-electric efficiency, the limited fraction of annual
renewable generation available for refilling seasonal stores, and the
reserve margin required for security of supply. Larger seasonal-storage
fractions further increase the renewable-overbuild term and reduce the
system-level EROI approximately in proportion to
$1/(1+7.14s)$.

The resulting benchmark EROI values are already reduced to the lower
single-digit range despite neglecting the embodied-energy investments
associated with hydrogen infrastructure. Furthermore, the benchmark
calculations assume
$f_{\mathrm{orientation}}
=f_{\mathrm{site}}
=f_{\mathrm{curtailment}}
=f_{\mathrm{delivery}}=1$,
such that additional reductions associated with suboptimal module
orientation, site-specific effects, curtailment, and electricity
delivery are not included. The values presented here should therefore
be regarded as optimistic upper bounds on the system-level EROI
achievable by representative hydrogen-based seasonal-storage systems
under the adopted benchmark assumptions.

\subsection{Combined battery and seasonal hydrogen storage}
\label{sec:combined_storage}

Battery load balancing and seasonal hydrogen storage perform different
functions within a renewable electricity system. Batteries can shift
electricity over short timescales and provide rapid grid-support
services, whereas hydrogen provides the energy inventory required
during prolonged renewable-generation deficits. The technologies may
therefore be operationally complementary, but their energetic
contributions remain additive within the present benchmark.

In particular, repeated battery cycling does not reduce the seasonal
energy requirement. During an extended deficit, a battery can discharge
only the energy stored at the beginning of the event and cannot recharge
until renewable generation again exceeds demand. The prescribed
seasonal-storage fraction $s_0$ is therefore retained unchanged when a
battery is added. No effective seasonal fraction derived from cumulative
battery throughput is introduced.

The combined system includes both the normalized fleet-equivalent
battery investment $f_{\mathrm{batt}}$ from
Section~\ref{sec:battery_infrastructure} and the renewable-generation
overbuild required for seasonal hydrogen storage. Using the relationship
derived in Section~\ref{sec:H2_infrastructure},
\[
f_{\mathrm{overbuild,H2}}(s_0)
=
\frac{c_{\mathrm{reserve}}}
     {\eta_{\mathrm{rt}}f_{\mathrm{refill}}}
\,s_0
\simeq
7.14\,s_0.
\]

The photovoltaic-attributed system-level EROI of the combined
architecture becomes
\[
\mathrm{EROI}_{\mathrm{PV,sys}}
=
\frac{\mathrm{EROI}_{\mathrm{fleet}}
      \,
      f_{\mathrm{orientation}}
      \,
      f_{\mathrm{site}}
      \,
      f_{\mathrm{curtailment}}
      \,
      f_{\mathrm{delivery}}}
     {1
      +f_{\mathrm{batt}}
      +f_{\mathrm{overbuild,H2}}(s_0)}.
\]

The battery does not reduce
$f_{\mathrm{overbuild,H2}}(s_0)$ in this expression. Any operational
interaction between the two technologies may affect curtailment,
equipment utilization, and dispatch, but it does not convert
short-duration battery capacity into seasonal energy inventory.

Table~\ref{tab:eroi_batt_h2} combines each of the four
seasonal-storage fractions with representative battery durations of
\SIlist{0.25;1;4}{h}. These cases illustrate short, intermediate, and
multi-hour load balancing and are not identified as optimized storage
capacities. As established in
Section~\ref{sec:battery_balancing}, larger battery capacities increase
$f_{\mathrm{batt}}$ and therefore reduce the combined-system EROI
monotonically.

\begin{table}[htbp]
\centering
\caption{System-level photovoltaic EROI for representative combined
battery--hydrogen benchmark systems. Battery durations of
\SIlist{0.25;1;4}{h} are combined with the full seasonal-storage
fraction $s_0$. The calculations use $\eta_{\mathrm{rt}}=0.35$,
$f_{\mathrm{refill}}=0.50$, and $c_{\mathrm{reserve}}=1.25$.
Operational battery losses and the embodied energy of hydrogen
infrastructure are neglected.}
\label{tab:eroi_batt_h2}

\begin{tabular}{lccc}
\toprule
$s_0$ &
Rooftop PV &
Utility-scale PV &
Alpine PV \\
\midrule

\multicolumn{4}{l}{\textit{\SI{0.25}{h} battery}} \\
\addlinespace[2pt]
15\% & 6.4--7.5 & 5.1--6.8 & 4.9--6.5 \\
20\% & 5.5--6.4 & 4.4--5.8 & 4.2--5.6 \\
25\% & 4.9--5.6 & 3.8--5.1 & 3.7--4.9 \\
30\% & 4.4--5.0 & 3.4--4.5 & 3.3--4.3 \\

\addlinespace[4pt]
\multicolumn{4}{l}{\textit{\SI{1}{h} battery}} \\
\addlinespace[2pt]
15\% & 5.1--6.8 & 4.2--6.2 & 4.1--6.1 \\
20\% & 4.5--5.9 & 3.7--5.4 & 3.6--5.2 \\
25\% & 4.1--5.2 & 3.3--4.8 & 3.2--4.6 \\
30\% & 3.7--4.7 & 3.0--4.3 & 2.9--4.1 \\

\addlinespace[4pt]
\multicolumn{4}{l}{\textit{\SI{4}{h} battery}} \\
\addlinespace[2pt]
15\% & 2.7--5.0 & 2.5--4.8 & 2.6--4.8 \\
20\% & 2.6--4.5 & 2.3--4.2 & 2.4--4.2 \\
25\% & 2.4--4.1 & 2.2--3.8 & 2.2--3.8 \\
30\% & 2.3--3.8 & 2.0--3.5 & 2.1--3.5 \\
\bottomrule
\end{tabular}
\end{table}

For every seasonal-storage fraction, adding the battery lowers EROI
relative to the corresponding hydrogen-only architecture because the
battery introduces an additional embodied-energy investment without
reducing the prescribed seasonal requirement. This result does not
imply that batteries lack system value. Their value lies in
load balancing, curtailment management, and rapid grid support, while
the energy required to manufacture and replace them remains an
additional system-level investment.

Even the \SI{0.25}{h} combined systems remain within the single-digit
EROI regime, while the \SI{4}{h} cases fall substantially lower. The
calculations further assume
$f_{\mathrm{orientation}}
=f_{\mathrm{site}}
=f_{\mathrm{curtailment}}
=f_{\mathrm{delivery}}=1$,
exclude operational battery losses, and neglect embodied-energy
investments associated with substantial parts of the hydrogen
infrastructure. The reported values should therefore be interpreted as
optimistic benchmark results.

\subsection{Dispatchable backup generation}
\label{sec:dispatchable_backup}

Dispatchable backup generation represents a fundamentally different
firming pathway from the storage-based architectures considered in the
preceding subsections. Rather than storing surplus renewable
electricity for later use, dispatchable generation produces
electricity directly whenever the combined photovoltaic and wind fleet
is unable to satisfy system demand. For a given seasonal-storage fraction $s$,
the dispatchable pathway is assumed to provide the same firm-electricity output 
as the seasonal hydrogen architecture, $E_{\mathrm{firm}}=s\,E_{\mathrm{gross}}$,
thereby providing an energetically equivalent level of seasonal
adequacy.

Unlike seasonal hydrogen storage, however, dispatchable backup must
also remain capable of supplying essentially the full system demand
during prolonged periods of low renewable generation. Consequently, 
the required dispatchable generating capacity is determined primarily by 
peak demand rather than annual electricity production.
Its embodied-energy investment therefore remains largely independent
of annual utilization, whereas the operational primary-energy input
depends directly on the amount of electricity generated.

Following Section~\ref{sec:disp_infrastructure}, the system-level EROI 
contains two contributions: the embodied-energy investment required to 
maintain the dispatchable generating fleet and the operational primary-energy 
input associated with fuel consumption.

The operational primary-energy input includes not only the fuel converted 
within the power plant but also the energy consumed throughout the fuel-supply 
chain, including extraction, processing, compression, pipeline transport, and, 
for LNG, liquefaction, shipping, and regasification. These upstream requirements are
represented by an effective fuel-supply efficiency,
\[
\eta_{\mathrm{supply}}
=
\frac{E_{\mathrm{fuel,plant}}}
     {E_{\mathrm{extracted}}},
\]
where $E_{\mathrm{fuel,plant}}$ denotes the chemical energy of the
fuel delivered to the generating plant and
$E_{\mathrm{extracted}}$ the total primary energy extracted or
otherwise consumed along the corresponding supply chain.

Using the mass and energy flows reported
in~\cite{Howarth2024:LNG}, the derivation accounts for the energy
required for extraction, processing, compression, and transport,
together with chemical-energy losses through methane leakage,
venting, and flaring. For LNG, the additional energy requirements
associated with liquefaction, shipping, and regasification
are likewise included. The energy consumed by supply-chain
processes is estimated from the corresponding indirect
carbon-dioxide emissions, using the natural-gas combustion factor
of \SI{55}{gCO2/MJ} as the conversion between combustion
emissions and fuel-energy input~\cite{Howarth2024:LNG}.

For pipeline gas, the reported indirect emissions of
\SI{12.6}{gCO2/MJ}, together with upstream and downstream methane
losses amounting to approximately 3.1\% of delivered
gas~\cite{Howarth2024:LNG}, yield an effective fuel-supply
efficiency of approximately $\eta_{\mathrm{supply,pipe}} \simeq 0.80$.

For LNG supplied over representative transport distances, the
additional energy consumed in liquefaction and shipping,
together with upstream and downstream losses, gives
$\eta_{\mathrm{supply,LNG}} \simeq 0.65$.

The electrical conversion efficiency of the dispatchable
generating fleet is represented by
$\eta_{\mathrm{disp}}$. Modern combined-cycle gas-turbine plants
can exceed 60\% efficiency under full-load design conditions,
whereas dispatchable backup operation involves repeated starts,
ramping, and part-load operation that reduce the fleet-average
efficiency. The benchmark calculations therefore adopt
$\eta_{\mathrm{disp}} = 0.60$ for the optimistic case and
$\eta_{\mathrm{disp}} = 0.50$ for the conservative case.

Together with the adopted benchmark values, the overall
extracted-fuel-to-electricity efficiencies become
\[
\eta_{\mathrm{disp}}\eta_{\mathrm{supply}}
=
\begin{cases}
0.40\text{--}0.48, & \text{pipeline gas},\\
0.33\text{--}0.39, & \text{LNG}.
\end{cases}
\]

The corresponding operational primary-energy input associated with fuel
consumption over the fleet lifetime is therefore
\[
E_{\mathrm{fuel}}
=
\frac{E_{\mathrm{firm}}}
     {\eta_{\mathrm{disp}}
      \eta_{\mathrm{supply}}},
\]
which, normalized to the embodied-energy investment of the
photovoltaic fleet, becomes
\[
f_{\mathrm{fuel}}
=
\frac{E_{\mathrm{fuel}}}
     {E_{\mathrm{in,PV}}}
=
\frac{s\,\mathrm{EROI}_{\mathrm{fleet}}}
     {\eta_{\mathrm{disp}}
      \eta_{\mathrm{supply}}}.
\]

The normalized embodied-energy contribution associated with
maintaining the dispatchable generating fleet is denoted by
$f_{\mathrm{disp}}$ and is taken from
Section~\ref{sec:disp_infrastructure}. The corresponding
photovoltaic-attributed system-level EROI therefore becomes
\[
\mathrm{EROI}_{\mathrm{PV,sys}}
=
\frac{\mathrm{EROI}_{\mathrm{fleet}}
      \,
      f_{\mathrm{orientation}}
      \,
      f_{\mathrm{site}}
      \,
      f_{\mathrm{curtailment}}
      \,
      f_{\mathrm{delivery}}}
     {1
      +
      f_{\mathrm{disp}}
      +
      f_{\mathrm{fuel}}}.
\]

As in the preceding benchmark calculations, the benchmark assumptions
\[
f_{\mathrm{orientation}}
=
f_{\mathrm{site}}
=
f_{\mathrm{curtailment}}
=
f_{\mathrm{delivery}}
=
1
\]
are adopted throughout this subsection.

Representative benchmark system-level photovoltaic EROI values for
dispatchable backup supplied by pipeline gas and LNG are summarized in
Tables~\ref{tab:eroi_dispatchable_pipeline}
and~\ref{tab:eroi_dispatchable_lng}.

\begin{table}[htbp]
\centering
\caption{System-level photovoltaic EROI for dispatchable backup
generation supplied by pipeline gas. Benchmark calculations assume
the same seasonal firm-electricity fraction $s$ as in the corresponding
hydrogen-storage cases. The ranges combine
$\eta_{\mathrm{disp}}=0.50$--$0.60$,
$\eta_{\mathrm{supply}}=0.80$, and the
dispatchable-infrastructure embodied-energy contribution
$f_{\mathrm{disp}}$ derived in
Section~\ref{sec:disp_infrastructure}.}
\label{tab:eroi_dispatchable_pipeline}

\begin{tabular}{lccc}
\toprule
$s$ &
Rooftop PV &
Utility-scale PV &
Alpine PV \\
\midrule

15\% & 2.2--2.6 & 2.1--2.6 & 2.1--2.5 \\
20\% & 1.7--2.0 & 1.6--2.0 & 1.6--2.0 \\
25\% & 1.4--1.7 & 1.4--1.7 & 1.4--1.7 \\
30\% & 1.2--1.4 & 1.2--1.4 & 1.2--1.4 \\

\bottomrule
\end{tabular}
\end{table}

\begin{table}[htbp]
\centering
\caption{System-level photovoltaic EROI for dispatchable backup
generation supplied by LNG. Benchmark calculations assume the same
seasonal firm-electricity fraction $s$ as in the corresponding
hydrogen-storage cases. The ranges combine
$\eta_{\mathrm{disp}}=0.50$--$0.60$,
$\eta_{\mathrm{supply}}=0.65$, and the
dispatchable-infrastructure embodied-energy contribution
$f_{\mathrm{disp}}$ derived in
Section~\ref{sec:disp_infrastructure}.}
\label{tab:eroi_dispatchable_lng}

\begin{tabular}{lccc}
\toprule
$s$ &
Rooftop PV &
Utility-scale PV &
Alpine PV \\
\midrule

15\% & 1.8--2.2 & 1.8--2.2 & 1.8--2.2 \\
20\% & 1.4--1.7 & 1.4--1.7 & 1.4--1.7 \\
25\% & 1.2--1.4 & 1.1--1.4 & 1.1--1.4 \\
30\% & 1.0--1.2 & 1.0--1.2 & 1.0--1.2 \\

\bottomrule
\end{tabular}
\end{table}

The benchmark results demonstrate that dispatchable backup imposes a substantial
operational-energy penalty even at comparatively modest firm-energy
fractions. At $s=15\%$, the central system-level EROI is already approximately two
for all photovoltaic configurations. Increasing the
dispatchable contribution further reduces the EROI toward unity,
reaching approximately 1.4--1.7 for pipeline gas and 1.1--1.4 for LNG
at $s=25\%$.

Pipeline gas consistently performs better than LNG because it avoids
the additional energy requirements associated with liquefaction,
shipping, and regasification. However, this difference is secondary to
the dominant effect of the fuel term itself. Once operational
primary-energy consumption becomes the largest contribution to the
denominator, the system-level EROI becomes nearly independent
of the photovoltaic fleet configuration.

Although dispatchable backup avoids the renewable-generation overbuild
required for seasonal hydrogen storage, it replaces this penalty with
continuous primary-fuel consumption. The energetic penalty therefore
changes in origin, but the resulting system-level EROI remains very
low and approaches unity as the required firm-energy fraction
increases.

This behaviour follows directly from the structure of the
system-level EROI expression. Once the operational fuel contribution
$f_{\mathrm{fuel}}$ dominates over the embodied-energy terms, the
denominator becomes approximately proportional to
$\mathrm{EROI}_{\mathrm{fleet}}$. The system-level EROI therefore
approaches the asymptotic form
\[
\mathrm{EROI}_{\mathrm{PV,sys}}
\approx
\frac{\eta_{\mathrm{disp}}
      \eta_{\mathrm{supply}}}
     {s},
\]
becoming increasingly insensitive to the fleet-level photovoltaic EROI.
Consequently, further improvements in photovoltaic manufacturing,
installation, or lifetime yield only marginal increases in the
system-level EROI once dispatchable fuel consumption dominates the
energetic balance. The limiting system-level EROI is governed almost entirely by the
overall efficiency of converting extracted primary fuel into
dispatchable electricity and by the required seasonal firm-energy
fraction.

The comparatively low system-level EROI values obtained here should
not be confused with published EROI values for natural-gas extraction
or gas-fired electricity generation. Conventional EROI analyses treat
natural gas as the primary energy resource and evaluate the energetic
investment required to discover, extract, process, transport, and
convert that resource into electricity. In the present analysis,
however, the objective is fundamentally different. The renewable
photovoltaic--wind fleet is regarded as the primary electricity
system, while dispatchable gas generation represents an additional
energetic input required to maintain firm supply during periods of
insufficient renewable generation. Consequently, the primary energy 
contained in the fuel itself constitutes part of the energetic investment 
required to operate the overall renewable electricity system and therefore 
appears explicitly in the denominator of the system-level EROI.

This difference in system boundaries explains why the present
system-level EROI values are substantially lower than published EROI
values for conventional gas-fired electricity generation.

This distinction has important implications beyond EROI.
Because the dominant energetic penalty arises from the continuous
consumption of primary fuel, the corresponding carbon emissions remain
directly coupled to the operational primary-energy input. 

\subsection{Sensitivity of system-level EROI to photovoltaic
orientation}
\label{sec:orientation_sensitivity}

The benchmark calculations above assume optimally oriented
photovoltaic installations and therefore adopt
$f_{\mathrm{orientation}}=1$. The orientation factors introduced in
Section~\ref{sec:impact} enter the numerator of the system-level EROI
expression multiplicatively. For otherwise unchanged assumptions,
the photovoltaic-attributed system-level EROI therefore becomes
\[
\mathrm{EROI}_{\mathrm{PV,sys}}(\theta)
=
f_{\mathrm{orientation}}(\theta)
\,
\mathrm{EROI}_{\mathrm{PV,sys}}^{\mathrm{optimal}}.
\]

The orientation penalty consequently remains proportional to the
reduction in annual photovoltaic yield irrespective of the firming
pathway. This effect is particularly important when the optimally
oriented benchmark already occupies the single-digit EROI regime.
A further output reduction then removes part of an already limited
net-energy margin.

Table~\ref{tab:eroi_orientation_sensitivity} illustrates this effect
for fully firmed seasonal-hydrogen and combined battery--hydrogen
architectures. Seasonal-storage fractions of 15\% and 25\% represent
moderate and more demanding firming requirements. The combined cases
contain the representative \SI{1}{h} load-balancing battery used in
the summary comparison. All values are derived from the full-precision
rooftop-PV endpoints of Tables~\ref{tab:eroi_h2_storage} and
\ref{tab:eroi_batt_h2}. Site, curtailment, and delivery factors remain
at unity so that only the effect of orientation is exposed.

\begin{table*}[htbp]
\centering
\caption{Sensitivity of photovoltaic-attributed system-level EROI to
module orientation for representative complete firm-electricity
systems using the rooftop-PV accounting configuration. Orientation
factors are taken from
Table~\ref{tab:yield_factors}. Seasonal hydrogen is shown for
$s=15\%$ and $25\%$; the combined systems retain the full seasonal
fraction $s_0$ and add a \SI{1}{h} load-balancing battery. All other
output factors remain at unity.}
\label{tab:eroi_orientation_sensitivity}

\renewcommand{\arraystretch}{1.10}
\setlength{\tabcolsep}{5.5pt}
\small

\begin{tabular}{lccccc}
\toprule
PV orientation &
$f_{\mathrm{orientation}}$ &
\multicolumn{2}{c}{Seasonal hydrogen} &
\multicolumn{2}{c}{Battery--hydrogen, \SI{1}{h}} \\
\cmidrule(lr){3-4}
\cmidrule(l){5-6}
&
& $s=15\%$ & $s=25\%$
& $s_0=15\%$ & $s_0=25\%$ \\
\midrule

South-facing, optimal tilt
& 1.00
& 7.0--7.7 & 5.2--5.7
& 5.1--6.8 & 4.1--5.2 \\

East--west roof
& 0.85--0.90
& 6.0--6.9 & 4.4--5.2
& 4.3--6.1 & 3.4--4.7 \\

Flat roof
& 0.80--0.85
& 5.6--6.6 & 4.2--4.9
& 4.0--5.8 & 3.2--4.4 \\

Vertical south-facing fa\c{c}ade
& 0.60--0.70
& 4.2--5.4 & 3.1--4.0
& 3.0--4.8 & 2.4--3.6 \\

Vertical east--west fa\c{c}ade
& 0.45--0.60
& 3.2--4.6 & 2.4--3.4
& 2.3--4.1 & 1.8--3.1 \\

Vertical north-facing fa\c{c}ade
& 0.20--0.40
& 1.4--3.1 & 1.0--2.3
& 1.0--2.7 & 0.8--2.1 \\

\bottomrule
\end{tabular}
\end{table*}

Even the comparatively modest yield penalties of east--west and flat
roofs reduce the EROI of the fully firmed systems appreciably.
Vertical deployment has a much stronger effect. For a south-facing
fa\c{c}ade, the combined system with $s_0=25\%$ falls to approximately
2.4--3.6, while a vertical east--west fa\c{c}ade reduces it to
approximately 1.8--3.1. North-facing deployment can bring the same
architecture to, or below, energetic break-even despite the favourable
site, curtailment, and delivery assumptions retained here.

These results do not imply that unconventional photovoltaic
installations provide no local value. Fa\c{c}ades, transport corridors,
and other built surfaces may offer land-use or grid-location benefits.
They do show, however, that access to an otherwise unused surface does
not compensate for low annual yield in the energy balance. When
substantial firming infrastructure is also required, poorly oriented
deployment further reduces an already low system-level energy return.

The values in Table~\ref{tab:eroi_orientation_sensitivity} are
photovoltaic-attributed quantities within the accounting framework of
this work. In the representative 80\% wind--20\% photovoltaic mix,
orientation directly affects the photovoltaic contribution rather than
the output of the wind fleet. The table therefore isolates the
orientation sensitivity of the photovoltaic share and should not be
interpreted as an equal fractional reduction of total wind--photovoltaic
generation.

\section{Lifecycle CO$_2$ intensity and net-energy implications}
\label{sec:co2}

The system-level EROI results derived in the preceding sections
quantify the primary-energy investment required to deliver usable
electricity. Translating this energetic investment into lifecycle
greenhouse-gas emissions requires a distinction between emissions
associated with industrial infrastructure supply chains and additional
emissions arising from the operation of a particular firming pathway.
This distinction provides a common carbon-accounting framework for
battery storage, seasonal hydrogen storage, and dispatchable gas
backup.

\subsection{Carbon-accounting framework for firmed electricity systems}
\label{sec:co2_framework}

By definition, the system-level energy return is
\[
\mathrm{EROI}_{\mathrm{sys}}
=
\frac{E_{\mathrm{out}}}
     {E_{\mathrm{in,sys}}},
\]
where $E_{\mathrm{in,sys}}$ denotes the total primary-energy input
represented within the system boundary and $E_{\mathrm{out}}$ the
corresponding usable electricity delivered over the system lifetime.

The lifecycle carbon intensity is separated into two contributions,
\[
I_{\mathrm{sys}}
=
I_{\mathrm{infra}}
+
I_{\mathrm{pathway}},
\]
where $I_{\mathrm{infra}}$ represents emissions associated with
constructing, maintaining, and replacing the required generation,
storage, conversion, and network infrastructure. The term
$I_{\mathrm{pathway}}$ accounts for additional greenhouse-gas
emissions specific to the firming pathway that are not represented by
the industrial energy investment alone.

If $E_{\mathrm{in,infra}}$ denotes the primary-energy investment in
the complete infrastructure, its carbon-intensity contribution is
\[
I_{\mathrm{infra}}
=
I_{\mathrm{industrial}}
\frac{E_{\mathrm{in,infra}}}
     {E_{\mathrm{out}}},
\]
where $I_{\mathrm{industrial}}$ is the effective carbon intensity of
the industrial energy supplying the corresponding manufacturing,
construction, maintenance, and replacement processes.

For non-combustion pathways whose system-level energy investment is
dominated by infrastructure supply chains,
$E_{\mathrm{in,infra}}=E_{\mathrm{in,sys}}$, and therefore
\[
I_{\mathrm{infra}}
=
\frac{I_{\mathrm{industrial}}}
     {\mathrm{EROI}_{\mathrm{sys}}}.
\]
Lifecycle infrastructure emissions consequently scale inversely with
system-level EROI. Every reduction in energy return requires a larger
industrial energy investment for each unit of delivered electricity
and therefore increases the corresponding lifecycle carbon intensity
for a given industrial supply-chain carbon intensity.

The present calculations adopt
\[
I_{\mathrm{industrial}}
=
600\text{--}800\,\mathrm{gCO_2eq/kWh_{primary}},
\]
representing the predominantly fossil-based industrial supply chains
currently used for photovoltaic modules, wind turbines, batteries,
electrical equipment, mining, refining, transport, and
high-temperature industrial processes. This range is consistent with
present manufacturing conditions dominated by China and other
comparatively carbon-intensive industrial regions
\cite{OWID2025:ChinaElectricityCO2}.

The pathway-specific contribution depends on the firming technology.
No separate operational greenhouse-gas term is assigned to battery
storage within the present system boundary. Seasonal hydrogen storage
introduces the indirect climate effect of hydrogen leakage, whereas
dispatchable gas generation introduces fuel-cycle emissions from gas
production, processing, transport, leakage, and combustion. These
contributions must be evaluated independently and added to the
infrastructure-related carbon intensity.

Unless stated otherwise, the principal CO$_2$-equivalent values
reported in this work use the 100-year global-warming-potential
horizon, GWP$_{100}$, following the convention commonly used for
greenhouse-gas inventories and lifecycle
comparisons~\cite{IPCC:2021:WGI}. Because hydrogen and methane exert substantial
climate effects on shorter timescales, the corresponding values are
additionally evaluated using GWP$_{20}$ and reported separately as a
near-term climate sensitivity. Values calculated using different time
horizons are not combined within the same lifecycle total.

For gas-backed systems, the continuous primary-fuel input is already
included in the denominator of the corresponding system-level EROI.
It must not also be assigned the industrial supply-chain carbon
intensity. The infrastructure and fuel-cycle contributions are
therefore separated explicitly in the gas-specific calculation below.

As established in Section~\ref{sec:EROI_sys}, photovoltaic energy is
used as the accounting reference throughout this work, whereas the
derived system-level EROI represents the net energetic performance of
the complete renewable electricity system.
The resulting relationship between EROI and lifecycle infrastructure
emissions therefore applies approximately to electricity systems
dominated by combinations of photovoltaics and wind power,
irrespective of the precise photovoltaic-to-wind ratio adopted.

The following subsections apply this common framework first to
seasonal hydrogen firming, then to combined battery--hydrogen storage,
and finally to dispatchable gas backup.

\subsection{Carbon intensity of seasonal hydrogen firming}
\label{sec:co2_h2}

For seasonal hydrogen firming, the lifecycle greenhouse-gas intensity
contains the infrastructure-related contribution derived from the
system-level EROI and the indirect climate contribution of hydrogen
leakage,
\[
I_{\mathrm{sys,H2}}^{(T)}
=
I_{\mathrm{infra,H2}}
+
I_{\mathrm{leak,H2}}^{(T)},
\]
where $T$ denotes the global-warming-potential time horizon.

Within the benchmark boundary adopted in
Section~\ref{sec:h2_storage}, the system-level energy investment is
dominated by the renewable-generation infrastructure required to
compensate hydrogen-cycle losses. The infrastructure-related carbon
intensity is therefore
\[
I_{\mathrm{infra,H2}}
=
\frac{I_{\mathrm{industrial}}}
     {\mathrm{EROI}_{\mathrm{PV,sys}}},
\]
using the full-precision EROI values reported in
Table~\ref{tab:eroi_h2_storage}.

The hydrogen-leakage contribution depends on the mass of hydrogen
handled for each unit of electricity delivered through the seasonal
storage pathway. Let $\lambda$ denote the fraction of produced
hydrogen that escapes before reconversion, $\eta_{\mathrm{reconv}}$
the hydrogen-to-electricity reconversion efficiency, and
$H_{\mathrm{H2,LHV}}$ the lower heating value of hydrogen. The leaked
hydrogen mass per unit of total system electricity is then
\[
\frac{m_{\mathrm{leak,H2}}}
     {E_{\mathrm{out}}}
=
s
\frac{\lambda}
     {1-\lambda}
\frac{1}
     {\eta_{\mathrm{reconv}}
      H_{\mathrm{H2,LHV}}}.
\]
The factor $\lambda/(1-\lambda)$ converts the hydrogen mass reaching
the reconversion plant into the corresponding mass lost throughout
the preceding supply chain.

For a hydrogen global-warming potential
$\mathrm{GWP}_{T,\mathrm{H2}}$, the resulting indirect climate
contribution becomes
\[
I_{\mathrm{leak,H2}}^{(T)}
=
s
\frac{\lambda}
     {1-\lambda}
\frac{\mathrm{GWP}_{T,\mathrm{H2}}}
     {\eta_{\mathrm{reconv}}
      H_{\mathrm{H2,LHV}}},
\]

The calculation adopts
\[
H_{\mathrm{H2,LHV}}
=
33.33\,\mathrm{kWh/kg}
\]
and the leakage range
$\lambda=0.02$--$0.07$ introduced in
Section~\ref{sec:hydrogen_storage}. For the large-scale
firm-electricity application considered here, electrical reconversion
is represented by combined-cycle turbine generation with
$\eta_{\mathrm{reconv}}=0.50$--$0.60$, consistent with the
fleet-average conversion-efficiency range adopted for dispatchable
generation in Section~\ref{sec:dispatchable_backup}.

The central hydrogen global-warming potentials are
\[
\mathrm{GWP}_{100,\mathrm{H2}}=11.6
\qquad\text{and}\qquad
\mathrm{GWP}_{20,\mathrm{H2}}=37.3,
\]
with reported one-standard-deviation uncertainties of $\pm2.8$ and
$\pm15.1$, respectively~\cite{Sand2023:HydrogenGWP}. The benchmark
ranges below use the central GWP values and combine the lower and upper
endpoints of the leakage and reconversion-efficiency ranges. The
atmospheric-model uncertainty is not folded into these intervals and
would scale the leakage contribution proportionally.

On a GWP$_{100}$ basis, hydrogen leakage adds approximately
1.8--7.9~gCO$_2$eq/kWh at $s=15\%$ and
3.6--15.7~gCO$_2$eq/kWh at $s=30\%$. On a GWP$_{20}$ basis, the
corresponding near-term contributions increase to approximately
5.7--25.3 and 11.4--50.5~gCO$_2$eq/kWh, respectively. These values are
averaged over the total electricity delivered by the complete system,
not only over the electricity supplied through hydrogen storage.

Table~\ref{tab:co2_h2} combines the leakage contribution with the
infrastructure-related intensity derived from the system-level EROI.
GWP$_{100}$ provides the principal lifecycle comparison, while
GWP$_{20}$ is reported separately as the near-term climate
sensitivity.

\begin{table}[htbp]
\centering
\caption{Lifecycle greenhouse-gas intensity of seasonal hydrogen
firming for the rooftop, utility-scale, and Alpine photovoltaic
accounting configurations. Infrastructure-related emissions are
derived from the full-precision system-level EROI values using
$I_{\mathrm{industrial}}=
600\text{--}800\,\mathrm{gCO_2eq/kWh_{primary}}$.
The hydrogen-leakage contribution assumes
$\lambda=2$--$7\%$,
$\eta_{\mathrm{reconv}}=0.50$--$0.60$, and the central hydrogen
global-warming potentials reported for the 100-year and 20-year time
horizons. Values are averaged over total delivered electricity.}
\label{tab:co2_h2}

\renewcommand{\arraystretch}{1.1}

\begin{tabular}{cccc}
\toprule
$s$ &
Rooftop PV &
Utility-scale PV &
Alpine PV \\
&
\multicolumn{3}{c}{(gCO$_2$eq/kWh)} \\
\midrule

\multicolumn{4}{l}{\textit{GWP$_{100}$: principal comparison}} \\
\addlinespace[2pt]

15\% & 80--121 & 88--155 & 92--163 \\

20\% & 94--144 & 104--183 & 108--193 \\

25\% & 108--166 & 119--211 & 124--222 \\

30\% & 122--188 & 135--239 & 140--252 \\

\midrule
\multicolumn{4}{l}{\textit{GWP$_{20}$: near-term sensitivity}} \\
\addlinespace[2pt]

15\% & 84--139 & 92--172 & 96--181 \\

20\% & 99--167 & 109--206 & 113--216 \\

25\% & 114--195 & 126--240 & 131--251 \\

30\% & 129--223 & 142--274 & 148--287 \\

\bottomrule
\end{tabular}
\end{table}

The GWP$_{100}$ results show that the infrastructure contribution
remains dominant under the adopted benchmark assumptions, reflecting
the low system-level EROI caused by renewable overbuild. Hydrogen
leakage nevertheless raises the lifecycle intensity in every case and
becomes increasingly important as the seasonal-storage fraction
increases. Its influence is substantially larger over the 20-year
horizon because the indirect atmospheric effects of hydrogen are
concentrated on shorter timescales.

These values remain optimistic. The EROI calculation neglects the
embodied-energy investments associated with electrolysers,
compressors, storage caverns, pipelines, and reconversion equipment,
and retains the benchmark output factors at unity. Including these
contributions would increase the infrastructure-related carbon
intensity, while leakage above the adopted range would increase the
pathway-specific contribution independently.

\subsection{Carbon intensity of combined battery--hydrogen systems}
\label{sec:co2_batt_h2}

Combined battery--hydrogen systems contain the same two
carbon-intensity contributions as seasonal hydrogen firming. The
infrastructure contribution includes both the battery investment and
the renewable overbuild required by the hydrogen pathway, while the
pathway-specific contribution accounts for the indirect climate effect
of hydrogen leakage:
\[
I_{\mathrm{sys,batt+H2}}^{(T)}
=
I_{\mathrm{infra,batt+H2}}
+
I_{\mathrm{leak,H2}}^{(T)}.
\]

The system-level EROI derived in
Section~\ref{sec:combined_storage} already includes the battery
embodied-energy investment and the renewable overbuild required for
seasonal hydrogen storage. The infrastructure contribution is therefore
\[
I_{\mathrm{infra,batt+H2}}
=
\frac{I_{\mathrm{industrial}}}
     {\mathrm{EROI}_{\mathrm{PV,sys}}}.
\]
The battery embodied energy is assigned the same industrial
supply-chain carbon intensity as the remaining infrastructure and must
not be added again as a separate carbon term.

Adding a battery does not reduce the prescribed seasonal-storage
fraction $s_0$. The hydrogen-leakage contribution consequently remains
\[
I_{\mathrm{leak,H2}}^{(T)}(s_0)
=
s_0
\frac{\lambda}
     {1-\lambda}
\frac{\mathrm{GWP}_{T,\mathrm{H2}}}
     {\eta_{\mathrm{reconv}}H_{\mathrm{H2,LHV}}},
\]
identical to the contribution of the hydrogen-only architecture at the
same seasonal-storage fraction. Battery capacity, annual cycling, and
cumulative battery throughput do not enter this expression.

The calculation retains
$I_{\mathrm{industrial}}=
600\text{--}800\,\mathrm{gCO_2eq/kWh_{primary}}$,
$\lambda=0.02$--$0.07$, and
$\eta_{\mathrm{reconv}}=0.50$--$0.60$.
The central hydrogen global-warming potentials and the lower heating
value are taken from Section~\ref{sec:co2_h2}. The infrastructure term
uses the full-precision EROI values derived for the
\SIlist{0.25;1;4}{h} combined systems in
Table~\ref{tab:eroi_batt_h2}.

Table~\ref{tab:co2_batt_h2} gives the principal GWP$_{100}$ lifecycle
comparison. Table~\ref{tab:co2_batt_h2_20} reports the corresponding
GWP$_{20}$ near-term sensitivity. The atmospheric-model uncertainty of
the hydrogen global-warming potentials is not folded into the benchmark
intervals.

\begin{table}[htbp]
\centering
\caption{Lifecycle greenhouse-gas intensity of representative combined
battery--hydrogen systems on a GWP$_{100}$ basis. Each battery duration
is combined with the full seasonal-storage fraction $s_0$.
Infrastructure emissions use
$I_{\mathrm{industrial}}=
600\text{--}800\,\mathrm{gCO_2eq/kWh_{primary}}$.
Hydrogen leakage uses $\lambda=2$--$7\%$,
$\eta_{\mathrm{reconv}}=0.50$--$0.60$, and the central 100-year
hydrogen global-warming potential. Values are averaged over total
delivered electricity.}
\label{tab:co2_batt_h2}

\renewcommand{\arraystretch}{1.08}

\begin{tabular}{lccc}
\toprule
$s_0$ &
Rooftop PV &
Utility-scale PV &
Alpine PV \\
&
\multicolumn{3}{c}{(gCO$_2$eq/kWh)} \\
\midrule

\multicolumn{4}{l}{\textit{\SI{0.25}{h} battery}} \\
\addlinespace[2pt]
15\% & 82--133 & 91--166 & 94--173 \\
20\% & 96--155 & 106--194 & 110--202 \\
25\% & 110--177 & 122--222 & 127--232 \\
30\% & 124--199 & 137--250 & 143--261 \\

\addlinespace[4pt]
\multicolumn{4}{l}{\textit{\SI{1}{h} battery}} \\
\addlinespace[2pt]
15\% & 90--166 & 98--197 & 101--201 \\
20\% & 104--188 & 114--225 & 117--230 \\
25\% & 118--210 & 129--253 & 133--260 \\
30\% & 132--233 & 145--281 & 149--289 \\

\addlinespace[4pt]
\multicolumn{4}{l}{\textit{\SI{4}{h} battery}} \\
\addlinespace[2pt]
15\% & 121--299 & 128--325 & 128--313 \\
20\% & 135--321 & 144--353 & 144--342 \\
25\% & 149--344 & 159--381 & 160--372 \\
30\% & 163--366 & 175--409 & 176--401 \\
\bottomrule
\end{tabular}
\end{table}

\begin{table}[htbp]
\centering
\caption{Near-term lifecycle greenhouse-gas sensitivity of the
combined battery--hydrogen systems in
Table~\ref{tab:co2_batt_h2}, evaluated using the central GWP$_{20}$ of
hydrogen. All other assumptions are unchanged.}
\label{tab:co2_batt_h2_20}

\renewcommand{\arraystretch}{1.08}

\begin{tabular}{lccc}
\toprule
$s_0$ &
Rooftop PV &
Utility-scale PV &
Alpine PV \\
&
\multicolumn{3}{c}{(gCO$_2$eq/kWh)} \\
\midrule

\multicolumn{4}{l}{\textit{\SI{0.25}{h} battery}} \\
\addlinespace[2pt]
15\% & 86--150 & 95--183 & 98--190 \\
20\% & 101--178 & 111--217 & 116--225 \\
25\% & 117--206 & 128--251 & 133--261 \\
30\% & 132--234 & 145--284 & 151--296 \\

\addlinespace[4pt]
\multicolumn{4}{l}{\textit{\SI{1}{h} battery}} \\
\addlinespace[2pt]
15\% & 94--183 & 102--215 & 105--218 \\
20\% & 109--211 & 119--249 & 122--253 \\
25\% & 125--239 & 136--282 & 140--289 \\
30\% & 140--267 & 152--316 & 157--324 \\

\addlinespace[4pt]
\multicolumn{4}{l}{\textit{\SI{4}{h} battery}} \\
\addlinespace[2pt]
15\% & 125--317 & 132--342 & 131--330 \\
20\% & 141--345 & 149--376 & 149--365 \\
25\% & 156--373 & 166--410 & 166--401 \\
30\% & 171--401 & 182--443 & 184--436 \\
\bottomrule
\end{tabular}
\end{table}

For a fixed seasonal-storage fraction, hydrogen leakage is unchanged by
the addition of battery capacity. The battery therefore raises the
total lifecycle intensity through its embodied-energy contribution,
while providing no compensating reduction in the leakage term. The
increase becomes progressively larger from the \SI{0.25}{h} to the
\SI{4}{h} cases as the system-level EROI declines.

This result does not imply that batteries lack operational value. It
shows that their load-balancing and grid-support functions require
additional infrastructure whose supply-chain emissions must be included
without treating repeated cycling as seasonal energy. Under the present
benchmark, adding battery capacity cannot improve lifecycle carbon
intensity relative to the corresponding hydrogen-only system at the
same $s_0$.

The combined-system values remain optimistic. Operational battery
losses are excluded, substantial parts of the hydrogen infrastructure
carry no embodied-energy contribution, and the orientation, site,
curtailment, and delivery factors remain at unity. Including these
effects would lower system-level EROI and increase the corresponding
infrastructure-related carbon intensity.

\subsection{Carbon intensity of dispatchable gas-backed systems}
\label{sec:co2_gas}

Dispatchable gas backup differs fundamentally from the storage-based
firming pathways considered previously because natural gas is consumed
continuously whenever the dispatchable fleet generates electricity.
The resulting lifecycle greenhouse-gas intensity contains two distinct
contributions,
\[
I_{\mathrm{sys,gas}}^{(T)}
=
I_{\mathrm{emb}}
+
I_{\mathrm{gas}}^{(T)},
\]
where $I_{\mathrm{emb}}$ represents the emissions associated with
constructing and maintaining the renewable and dispatchable
infrastructure, and $I_{\mathrm{gas}}^{(T)}$ represents the complete
fuel-cycle emissions associated with supplying and burning the gas.
The superscript $T$ denotes the global-warming-potential time horizon.

The embodied contribution must be evaluated from the infrastructure
energy investment alone. Operational fuel consumption is treated
separately through $I_{\mathrm{gas}}^{(T)}$ and must not also be assigned
the industrial supply-chain carbon intensity. Using the quantities
introduced in Section~\ref{sec:disp_infrastructure},
\[
I_{\mathrm{emb}}
=
I_{\mathrm{industrial}}
\frac{E_{\mathrm{PV}}+E_{\mathrm{disp}}}
     {E_{\mathrm{gross}}}
=
I_{\mathrm{industrial}}
\frac{1+f_{\mathrm{disp}}}
     {\mathrm{EROI}_{\mathrm{fleet}}}.
\]
The first form is used for the numerical calculation so that the
underlying full-precision energy inventories remain paired
consistently.

With
\[
I_{\mathrm{industrial}}
=
600\text{--}800\,\mathrm{gCO_2eq/kWh_{primary}},
\]
the resulting embodied contributions are
43--70~gCO$_2$eq/kWh for rooftop PV,
47--86~gCO$_2$eq/kWh for utility-scale PV, and
48--88~gCO$_2$eq/kWh for Alpine PV.

The gas-related contribution scales directly with the fraction $s$ of
total electricity supplied by dispatchable generation:
\[
I_{\mathrm{gas}}^{(T)}
=
\frac{s}
     {\eta_{\mathrm{disp}}}
I_{\mathrm{gas,fuel}}^{(T)},
\]
where $\eta_{\mathrm{disp}}=0.50$--$0.60$ is the fleet-average
electrical conversion efficiency and
$I_{\mathrm{gas,fuel}}^{(T)}$ is the lifecycle greenhouse-gas
intensity per unit of gas energy consumed at the power plant.

The fuel-cycle factors are derived from the carbon-dioxide and methane
flows reported in the cited lifecycle assessment
\cite{Howarth2024:LNG}. For pipeline gas, the reported emissions are
67.6~gCO$_2$/MJ and 0.64~gCH$_4$/MJ. For LNG supplied over the
representative mean transport distance, the corresponding values are
83.1~gCO$_2$/MJ and 0.93~gCH$_4$/MJ. Applying the methane global-warming
potentials used in that assessment gives
\[
I_{\mathrm{gas,fuel}}^{(100)}
=
\begin{cases}
312\,\mathrm{gCO_2eq/kWh_{fuel}}, & \text{pipeline gas},\\
399\,\mathrm{gCO_2eq/kWh_{fuel}}, & \text{LNG},
\end{cases}
\]
for GWP$_{100}=29.8$, and
\[
I_{\mathrm{gas,fuel}}^{(20)}
=
\begin{cases}
434\,\mathrm{gCO_2eq/kWh_{fuel}}, & \text{pipeline gas},\\
575\,\mathrm{gCO_2eq/kWh_{fuel}}, & \text{LNG},
\end{cases}
\]
for GWP$_{20}=82.5$.

These factors include direct combustion emissions and the upstream
carbon-dioxide and methane emissions associated with production,
processing, transport, and delivery. LNG additionally includes
liquefaction, tanker transport, and regasification. The values therefore
represent the complete fuel-cycle contribution per unit of chemical
energy consumed by the generating plant.

The complete lifecycle intensity is
\[
I_{\mathrm{sys,gas}}^{(T)}
=
I_{\mathrm{industrial}}
\frac{E_{\mathrm{PV}}+E_{\mathrm{disp}}}
     {E_{\mathrm{gross}}}
+
\frac{s}
     {\eta_{\mathrm{disp}}}
I_{\mathrm{gas,fuel}}^{(T)}.
\]
Table~\ref{tab:co2_dispatchable_100} reports the principal
GWP$_{100}$ comparison, while
Table~\ref{tab:co2_dispatchable_20} gives the corresponding
GWP$_{20}$ near-term sensitivity.

\begin{table}[htbp]
\centering
\caption{Lifecycle greenhouse-gas intensity of renewable electricity
systems firmed by dispatchable gas backup on a GWP$_{100}$ basis.
Values combine the infrastructure contribution with the complete
pipeline-gas or LNG fuel-cycle contribution. Ranges include
$I_{\mathrm{industrial}}=
600\text{--}800\,\mathrm{gCO_2eq/kWh_{primary}}$ and
$\eta_{\mathrm{disp}}=0.50$--$0.60$.}
\label{tab:co2_dispatchable_100}

\begin{tabular}{llccc}
\toprule
Fuel &
$s$ &
Rooftop PV &
Utility-scale PV &
Alpine PV \\
&
&
\multicolumn{3}{c}{(gCO$_2$eq/kWh)} \\
\midrule

Pipeline
& 15\% & 121--164 & 125--179 & 126--182 \\
& 20\% & 147--195 & 151--211 & 152--213 \\
& 25\% & 173--226 & 177--242 & 178--244 \\
& 30\% & 199--258 & 203--273 & 204--275 \\

\addlinespace[3pt]
LNG
& 15\% & 143--190 & 146--205 & 148--208 \\
& 20\% & 176--230 & 180--245 & 181--247 \\
& 25\% & 209--270 & 213--285 & 214--287 \\
& 30\% & 242--309 & 246--325 & 247--327 \\

\bottomrule
\end{tabular}
\end{table}

\begin{table}[htbp]
\centering
\caption{Near-term lifecycle greenhouse-gas sensitivity of the
dispatchable gas-backed systems in
Table~\ref{tab:co2_dispatchable_100}, evaluated using GWP$_{20}$ for
methane. All other assumptions are unchanged.}
\label{tab:co2_dispatchable_20}

\begin{tabular}{llccc}
\toprule
Fuel &
$s$ &
Rooftop PV &
Utility-scale PV &
Alpine PV \\
&
&
\multicolumn{3}{c}{(gCO$_2$eq/kWh)} \\
\midrule

Pipeline
& 15\% & 151--200 & 155--216 & 157--218 \\
& 20\% & 188--244 & 191--259 & 193--262 \\
& 25\% & 224--287 & 228--303 & 229--305 \\
& 30\% & 260--331 & 264--346 & 265--349 \\

\addlinespace[3pt]
LNG
& 15\% & 187--243 & 191--258 & 192--261 \\
& 20\% & 235--300 & 238--316 & 240--318 \\
& 25\% & 283--358 & 286--373 & 288--376 \\
& 30\% & 330--415 & 334--431 & 336--433 \\

\bottomrule
\end{tabular}
\end{table}

On the principal GWP$_{100}$ basis, pipeline-gas backup reaches
approximately 120--180~gCO$_2$eq/kWh already at $s=15\%$ and
approximately 200--280~gCO$_2$eq/kWh at $s=30\%$, depending on the
photovoltaic accounting configuration and benchmark endpoint. LNG is
higher, reaching approximately 140--210~gCO$_2$eq/kWh at $s=15\%$ and
approximately 240--330~gCO$_2$eq/kWh at $s=30\%$.

The near-term GWP$_{20}$ values are substantially higher because
methane emissions contribute more strongly over the shorter time
horizon. The difference is especially pronounced for pipeline gas,
for which methane constitutes a larger fraction of the lifecycle
footprint. LNG remains the higher-emission pathway because its
additional processing and transport stages increase both carbon-dioxide
and methane emissions.

In both time horizons, the fuel-cycle contribution increasingly
dominates as the firm-electricity fraction rises. Dispatchable gas can
maintain electrical adequacy, but it does not preserve a low lifecycle
carbon intensity. The GWP$_{100}$ values are used in the principal
summary comparison of Section~\ref{sec:co2_summary}; GWP$_{20}$ is
retained as a separate near-term sensitivity.

\subsection{Lifecycle carbon intensity of benchmark electricity systems}
\label{sec:co2_summary}

The preceding derivations provide the quantities required to compare
the energetic and lifecycle greenhouse-gas performance of the benchmark
electricity systems considered in this work. For non-combustion
pathways, the infrastructure contribution follows from the inverse
relationship between system-level EROI and lifecycle carbon intensity.
Hydrogen-based systems additionally include the indirect climate effect
of hydrogen leakage. For dispatchable gas backup, infrastructure and
fuel-cycle emissions are calculated separately and then added.

Table~\ref{tab:co2_summary} summarizes the principal GWP$_{100}$ results
for the three photovoltaic accounting configurations. Carbon
intensities are calculated from the underlying full-precision values,
not from the rounded EROI values displayed. The GWP$_{20}$ results for
hydrogen leakage and gas supply are retained as separate near-term
sensitivities in the pathway-specific subsections and are not mixed
with the principal comparison.

\begin{sidewaystable*}[p]
\centering
\caption{Summary of system-level energetic and lifecycle greenhouse-gas
performance for the benchmark electricity architectures investigated
in this work. Results are reported separately for rooftop,
utility-scale, and Alpine photovoltaic accounting configurations. The
upper block contains reference configurations that do not independently
provide year-round firm electricity. The middle block contains complete
firm-electricity systems. Carbon intensities use GWP$_{100}$. For
non-combustion pathways, the infrastructure contribution is derived
using
$I_{\mathrm{industrial}}=
600\text{--}800\,\mathrm{gCO_2eq/kWh_{primary}}$;
hydrogen leakage is added where applicable. For gas-backed systems,
infrastructure and fuel-cycle contributions are calculated separately.
The \SI{1}{h} battery in the combined systems is a representative
load-balancing case, not an optimized duration.}
\label{tab:co2_summary}

\renewcommand{\arraystretch}{1.10}
\setlength{\tabcolsep}{6pt}
\small

\begin{tabular}{@{}ll *{3}{cc}@{}}
\toprule
&
&
\multicolumn{2}{c}{Rooftop PV} &
\multicolumn{2}{c}{Utility-scale PV} &
\multicolumn{2}{c}{Alpine PV} \\
\cmidrule(lr){3-4}
\cmidrule(lr){5-6}
\cmidrule(l){7-8}
Architecture &
Benchmark case &
EROI & CO$_2$eq &
EROI & CO$_2$eq &
EROI & CO$_2$eq \\
&
&
& (g/kWh) &
& (g/kWh) &
& (g/kWh) \\
\midrule

\multicolumn{8}{@{}l}{\textit{Reference configurations}} \\
\addlinespace[2pt]

PV installation
& Installation boundary
& 18.6--20.4 & 29--43
& 15.1--17.9 & 34--53
& 13.8--16.7 & 36--58 \\

Renewable fleet
& No firming
& 14.6--16.0 & 38--55
& 11.3--14.4 & 42--71
& 10.7--13.8 & 44--75 \\

Battery load balancing
& \SI{1}{h}
& 8.1--12.5 & 48--99
& 7.1--11.6 & 52--113
& 7.1--11.4 & 52--112 \\

& \SI{4}{h}
& 3.4--7.6 & 79--233
& 3.3--7.3 & 82--241
& 3.6--7.6 & 79--224 \\

& \SI{24}{h}
& 0.7--2.1 & 288--1122
& 0.7--2.1 & 282--1089
& 0.8--2.3 & 257--971 \\

\midrule
\multicolumn{8}{@{}l}{\textit{Complete firm-electricity systems}} \\
\addlinespace[2pt]

Seasonal hydrogen
& $s=15\%$
& 7.0--7.7 & 80--121
& 5.4--7.0 & 88--155
& 5.1--6.7 & 92--163 \\

& $s=20\%$
& 6.0--6.6 & 94--144
& 4.6--5.9 & 104--183
& 4.4--5.7 & 108--193 \\

& $s=25\%$
& 5.2--5.7 & 108--166
& 4.0--5.2 & 119--211
& 3.8--4.9 & 124--222 \\

& $s=30\%$
& 4.6--5.1 & 122--188
& 3.6--4.6 & 135--239
& 3.4--4.4 & 140--252 \\

\addlinespace[2pt]

Battery--hydrogen
& $s_0=15\%$, \SI{1}{h} battery
& 5.1--6.8 & 90--166
& 4.2--6.2 & 98--197
& 4.1--6.1 & 101--201 \\

& $s_0=20\%$, \SI{1}{h} battery
& 4.5--5.9 & 104--188
& 3.7--5.4 & 114--225
& 3.6--5.2 & 117--230 \\

& $s_0=25\%$, \SI{1}{h} battery
& 4.1--5.2 & 118--210
& 3.3--4.8 & 129--253
& 3.2--4.6 & 133--260 \\

& $s_0=30\%$, \SI{1}{h} battery
& 3.7--4.7 & 132--233
& 3.0--4.3 & 145--281
& 2.9--4.1 & 149--289 \\

\addlinespace[2pt]

Pipeline-gas backup
& $s=15\%$
& 2.2--2.6 & 121--164
& 2.1--2.6 & 125--179
& 2.1--2.5 & 126--182 \\

& $s=20\%$
& 1.7--2.0 & 147--195
& 1.6--2.0 & 151--211
& 1.6--2.0 & 152--213 \\

& $s=25\%$
& 1.4--1.7 & 173--226
& 1.4--1.7 & 177--242
& 1.4--1.7 & 178--244 \\

& $s=30\%$
& 1.2--1.4 & 199--258
& 1.2--1.4 & 203--273
& 1.2--1.4 & 204--275 \\

\addlinespace[2pt]

LNG backup
& $s=15\%$
& 1.8--2.2 & 143--190
& 1.8--2.2 & 146--205
& 1.8--2.2 & 148--208 \\

& $s=20\%$
& 1.4--1.7 & 176--230
& 1.4--1.7 & 180--245
& 1.4--1.7 & 181--247 \\

& $s=25\%$
& 1.2--1.4 & 209--270
& 1.1--1.4 & 213--285
& 1.1--1.4 & 214--287 \\

& $s=30\%$
& 1.0--1.2 & 242--309
& 1.0--1.2 & 246--325
& 1.0--1.2 & 247--327 \\

\midrule
\multicolumn{8}{@{}l}{\textit{Dispatchable low-carbon references}} \\
\addlinespace[2pt]

Hydroelectricity
& Literature range
& \multicolumn{6}{c}{EROI: 50--100;\qquad
  CO$_2$eq: 1--5~g/kWh} \\

Nuclear Gen~III+
& Literature range
& \multicolumn{6}{c}{EROI: 60--100;\qquad
  CO$_2$eq: 5--15~g/kWh} \\

Nuclear Gen~IV
& Projected
& \multicolumn{6}{c}{EROI: $>200$;\qquad
  CO$_2$eq: 2--5~g/kWh} \\

\bottomrule
\end{tabular}
\end{sidewaystable*}

The reference renewable fleet retains system-level EROI values of
approximately 11--16 and lifecycle intensities of approximately
38--75~gCO$_2$eq/kWh under the adopted industrial supply-chain
assumptions. It does not, however, independently provide year-round
firm electricity. Imposing progressively larger load-balancing or
firming requirements increases the energy investment per unit of
delivered electricity and reduces the corresponding net-energy return.

The battery-only reference cases show the energetic cost of prescribed
load-balancing capacity. A \SI{1}{h} battery retains EROI values of
approximately 7--13, whereas the \SI{4}{h} cases fall to approximately
3--8. At \SI{24}{h}, EROI reaches approximately 0.7--2.3 and lifecycle
intensity rises to approximately 260--1120~gCO$_2$eq/kWh. These
configurations still provide only daily-scale load balancing and do not
constitute complete firm-electricity systems.

Seasonal hydrogen provides long-duration adequacy but requires a large
renewable overbuild to compensate conversion losses, limited refill
availability, and the adopted reserve requirement. Across the
investigated seasonal fractions, EROI lies between approximately
3.4 and 7.7, while the GWP$_{100}$ lifecycle intensity reaches
approximately 80--250~gCO$_2$eq/kWh. Hydrogen leakage raises these
values independently of the infrastructure contribution and becomes
more consequential on the GWP$_{20}$ time horizon.

Adding a representative \SI{1}{h} battery to the seasonal-hydrogen
architecture reduces EROI to approximately 2.9--6.8 and raises the
GWP$_{100}$ lifecycle range to approximately
90--290~gCO$_2$eq/kWh. The battery can provide valuable
load-balancing and grid-support services, but it does not reduce the
prescribed seasonal fraction or the associated hydrogen leakage.
Consequently, its embodied-energy investment is additive, and the
combined architecture does not improve EROI or lifecycle carbon
intensity relative to hydrogen alone at the same $s_0$.

Gas-backed systems are dominated by continuing primary-fuel
consumption. Pipeline-gas backup yields EROI values of approximately
1.2--2.6 and GWP$_{100}$ lifecycle intensities of approximately
120--280~gCO$_2$eq/kWh across the investigated firm fractions. LNG
backup performs still worse, with EROI values of approximately
1.0--2.2 and lifecycle intensities of approximately
140--330~gCO$_2$eq/kWh. Their GWP$_{20}$ values are higher because of
the stronger near-term contribution of methane leakage.

The configuration-resolved comparison provides no systematic energetic
advantage for Alpine PV. Its higher irradiation and improved winter
yield are offset by the embodied-energy requirements of mounting,
access, grid connection, and other mountain-specific infrastructure.
That penalty propagates through the seasonal-hydrogen and combined
battery--hydrogen architectures, where Alpine PV generally remains
among the lower-EROI and higher-carbon configurations. The expectation
that Alpine deployment alone can overcome the energetic penalty of
seasonal firming is therefore not supported by the benchmark results.

All system-level values remain optimistic. The output-side factors are
set to
\[
f_{\mathrm{orientation}}
=
f_{\mathrm{site}}
=
f_{\mathrm{curtailment}}
=
f_{\mathrm{delivery}}
=1.
\]
The calculations therefore omit reductions associated with suboptimal
orientation, site limitations, soiling, snow cover, residual
curtailment, and electricity delivery. Operational battery losses are
not included in the battery benchmarks, and substantial parts of the
hydrogen production, storage, transport, and reconversion
infrastructure carry no embodied-energy contribution. Each omitted
effect would reduce EROI further and, for a given industrial
supply-chain carbon intensity, increase lifecycle emissions.

The technologies differ in physical implementation, but expose the
same system-level constraint. Batteries add embodied energy in storage
capacity, seasonal hydrogen requires renewable overbuild, and gas
backup consumes primary fuel continuously. Each pathway reduces the
net-energy surplus remaining after the electricity system has sustained
itself. Under present industrial supply chains, the same increase in
energy investment also raises lifecycle greenhouse-gas emissions.

\subsection{Renewable supply chains and the effective net-energy
constraint}

A frequently proposed response to the lifecycle emissions reported in
Table~\ref{tab:co2_summary} is that future photovoltaic modules,
batteries, inverters, transmission equipment, and firming
infrastructure could themselves be manufactured using low-carbon
electricity. In the limiting case, one may imagine that all energy
required for mining, material processing, manufacturing, construction,
transport, maintenance, and replacement is supplied by the renewable
electricity system itself, thereby avoiding the fossil-dominated
industrial carbon intensity adopted in the preceding analysis.

Such a transition would substantially reduce the energy-related
lifecycle CO$_2$ emissions of non-combustion pathways. The carbon
intensities reported in Table~\ref{tab:co2_summary} should therefore not
be interpreted as immutable technological properties. They reflect the
present assumption of industrial supply chains with an effective
primary-energy carbon intensity of
\SIrange{600}{800}{gCO2eq/kWh_{primary}}. If those supply chains were
progressively electrified and supplied by low-carbon generation, the
carbon intensity per unit of energy invested would decline.

Even in this limiting case, lifecycle emissions would not necessarily
become zero. Some emissions arise from chemical processes rather than
energy supply, including process emissions from cement and other
material production. Residual emissions may also remain associated with
mining, land-use change, transport, refrigerants, hydrogen or methane
leakage, and activities that cannot be fully electrified. Most
importantly, renewable manufacturing cannot eliminate the combustion
and fuel-cycle emissions of gas-backed systems for as long as natural
gas continues to be consumed.

The thought experiment nevertheless provides a useful limiting case.
Suppose that the effective carbon intensity of all energy invested in
non-combustion infrastructure approaches zero. The energy-related
embodied carbon contribution would then also approach zero, but the
embodied energy requirement would remain. Modules, wind turbines,
batteries, electrolysers, storage facilities, transmission networks,
and replacement components would still require energy to manufacture,
install, maintain, and reproduce. That energy would have to be
withdrawn from the gross output of the low-carbon energy system itself.

Low-carbon manufacturing therefore changes the source and carbon
intensity of the invested energy, but not its magnitude in the
system-level energy balance. Complete reliance on low-carbon supply
chains makes the net-energy constraint more explicit: the energy system
must reproduce its own infrastructure before the remaining output can
be made available to the rest of society. The relevant quantity is
consequently not gross electricity generation alone, but the net surplus
remaining after internal energetic reinvestment.

As discussed in
Section~\ref{sec:eroi_societal-relevance}, the fraction of gross
electricity production available after reproducing the energy system is
\[
\frac{E_{\mathrm{net}}}
     {E_{\mathrm{out}}}
=
1-
\frac{1}
     {\mathrm{EROI}_{\mathrm{sys}}}
=
\frac{\mathrm{EROI}_{\mathrm{sys}}-1}
     {\mathrm{EROI}_{\mathrm{sys}}}.
\]

At an EROI of 8, 12.5\% of gross output must be reinvested and 87.5\%
remains available to society. At an EROI of 5, the reinvestment share
rises to 20\% and the available surplus falls to 80\%. At an EROI of 3,
one third of gross output is required to sustain the energy system,
leaving two thirds for all other activities. An EROI of 2 requires half
of gross output to be reinvested. At unity, the complete output is
required to reproduce the energy system itself; below unity, the system
becomes a net energy sink and requires continuing external energy input.

These values span the complete firm-electricity systems in
Table~\ref{tab:co2_summary}. Seasonal-hydrogen systems reach EROI values
of approximately 3.4--7.7, corresponding to steady-state net fractions
of approximately 71--87\%. Adding the representative load-balancing
battery lowers the combined-system range to approximately 2.9--6.8,
leaving approximately 66--85\% of gross output after energetic
reinvestment. Pipeline-gas backup lies at approximately 1.2--2.6,
corresponding to only about 14--62\% net output, while the LNG cases
extend from approximately 2.2 to, and slightly below, energetic
break-even. Thus, most
complete firm-electricity architectures investigated here occupy the
single-digit EROI regime, and several approach the region in which the
net-energy surplus collapses.

The same relationship can be expressed as the gross generation required
to provide a specified net-energy supply:
\[
\frac{E_{\mathrm{out}}}
     {E_{\mathrm{net}}}
=
\frac{\mathrm{EROI}_{\mathrm{sys}}}
     {\mathrm{EROI}_{\mathrm{sys}}-1}.
\]

Providing one unit of net energy therefore requires approximately
1.14 units of gross generation at an EROI of 8, 1.25 units at an EROI
of 5, 1.5 units at an EROI of 3, and 2 units at an EROI of 2. The
remainder is required to sustain the energy system. The multiplier
increases progressively across the EROI range occupied by the benchmark
firm-electricity systems and diverges as EROI approaches unity.

This progression is central to the interpretation of
Table~\ref{tab:co2_summary}. Decarbonized manufacturing could greatly
reduce the lifecycle emissions assigned to battery- and
hydrogen-based systems, but it would not convert a low-EROI architecture
into a high-surplus energy source. An increasing share of gross
generation would circulate internally through the energy sector to
produce, maintain, and replace the generation, storage, conversion, and
network infrastructure. Only the residual output would remain available
for households, industry, transport, healthcare, agriculture, water
supply, research, and other non-energy-sector activities.

The difference between EROI values in the middle of the investigated
range is therefore societally relevant even before the sharp collapse
near unity is reached. Values between approximately 7 and 10 leave less
margin for adverse conditions, infrastructure omissions, economic
disruption, and system expansion than values above 10. Predominantly
single-digit EROI systems correspondingly provide less energetic
headroom for resilience, welfare, scientific and technological
development, and progress toward the Sustainable Development Goals.

The steady-state relationship also understates the burden during a
rapid transition. A growing energy system must supply energy not only
for replacing existing infrastructure, but also for manufacturing and
installing additional capacity. During expansion, this growth investment
is superimposed on the energy required to maintain the existing fleet.
Low-EROI technologies therefore create a particularly demanding
transition dynamic: gross generation and industrial activity must grow
before the intended net-energy supply becomes available to the rest of
the economy.

The conclusion of the low-carbon-supply-chain thought experiment is
therefore twofold. Decarbonizing industrial energy can substantially
reduce the lifecycle emissions of non-combustion electricity systems
and remains an essential objective. It cannot, however, bypass the
energetic cost of constructing and reproducing the infrastructure
itself. Carbon intensity and EROI are related but distinct constraints.
A credible low-carbon electricity architecture must achieve both low
lifecycle emissions and a sufficiently high system-level EROI to
provide a durable net-energy surplus after the energy system has
sustained itself.

\section{Discussion and Conclusions}
\label{sec:discussion_conclusions}

This study evaluates photovoltaics as a representative component of
variable renewable electricity systems using a fleet-equivalent,
system-level energy-accounting framework. The system boundary extends
beyond the individual installation to include component turnover, grid
integration, storage, renewable-generation overbuild, dispatchable
backup, fuel supply, and the infrastructure required to provide
reliable electricity. The analysis is a benchmark model of energy
requirements rather than a chronological simulation, optimization, or
prediction of a real electricity system. Its purpose is to identify the
energetic consequences of progressively expanding the boundary from
electricity generation to firm electricity supply.

This distinction is fundamental. An individual photovoltaic
installation may exhibit a favourable EROI while relying on balancing,
transmission, reserve capacity, and firm generation supplied by assets
outside the conventional installation boundary. These services remain
physically necessary and become increasingly important as the share of
variable generation rises. Installation-level performance therefore
cannot by itself establish the net-energy return of the complete
electricity system.

The representative mix of 80\% wind and 20\% photovoltaics reduces
variability relative to a photovoltaic-only system. Wind and solar
generation are partially complementary over daily and seasonal
timescales, and their combination reduces the firming burden. It does
not, however, guarantee continuous supply. Extended periods of weak
wind and low solar irradiation remain possible across Central Europe,
particularly during winter. Reliable electricity supply therefore
continues to require storage, dispatchable generation, demand
flexibility, imports, or other complementary sources of firm energy.

Although the accounting is formulated from the perspective of
photovoltaics, the principal system-level conclusions are not specific
to photovoltaic technology. Storage, overbuild, curtailment,
transmission, reserve capacity, and dispatchable backup arise because
the combined renewable fleet is weather-dependent. Changing the
wind--photovoltaic ratio may alter the magnitude of these requirements,
but does not remove them. The resulting energetic constraints therefore
apply more generally to deeply decarbonized electricity systems that
rely predominantly on variable renewable generation.

\subsection{Firming pathways and system-level EROI}

The energetic cost of managing renewable variability depends strongly
on the duration and physical implementation of the required response.
Batteries are effective for shifting electricity over short periods and
providing rapid grid-support services. The battery-only calculations,
however, are load-balancing references rather than complete
firm-electricity systems. A \SI{1}{h} battery yields system-level EROI
values of approximately 7--13, while the \SI{4}{h} cases fall to
approximately 3--8. At \SI{24}{h}, EROI declines to approximately
0.7--2.3 despite providing only daily rather than seasonal balancing.

This decline is driven by the embodied energy required to manufacture
and replace progressively larger battery capacities. The calculation
does not include operational output losses, so the reported values are
already favourable. Extending electrochemical storage duration cannot
therefore be treated as a neutral expansion of short-duration battery
services. The infrastructure investment grows directly with installed
capacity, while even the \SI{24}{h} benchmark remains unable to bridge
prolonged renewable-generation deficits.

Pumped-hydropower storage provides high efficiency, long service life,
and comparatively favourable energy return. Its principal limitation
is not energetic performance, but the geographical availability and
scale of suitable reservoirs. Even in hydro-rich Switzerland, existing
pumped-storage energy capacity remains small relative to
multi-terawatt-hour winter shortfalls. Pumped hydropower is therefore a
valuable regional balancing resource, but cannot generally provide the
seasonal inventory required by renewable-dominated systems across
Central Europe.

Seasonal hydrogen addresses the long-duration requirement, but incurs a
different energetic penalty. With an electric-to-electric round-trip
efficiency of 35\%, limited refill availability, and the adopted reserve
factor, seasonal-storage fractions of 15--30\% require renewable
overbuild equivalent to approximately 1.07--2.14 times the reference
fleet. The resulting system-level EROI lies between approximately
3.4 and 7.7. These values remain entirely within the single-digit range
even though the embodied energy of substantial parts of the hydrogen
production, transport, storage, and reconversion infrastructure is
neglected.

Batteries and seasonal hydrogen can perform complementary operational
functions, but their energetic investments remain additive in the
present benchmark. A battery can shift short-duration surpluses and
support grid operation, whereas hydrogen retains the energy inventory
required during prolonged deficits. Repeated battery cycling does not
reduce the prescribed seasonal-storage fraction and is therefore not
credited as a reduction in hydrogen overbuild or leakage.

Consequently, adding battery capacity lowers the EROI of the combined
architecture relative to hydrogen alone at the same seasonal fraction.
For the representative \SI{1}{h} battery, the combined systems yield
approximately 2.9--6.8 across seasonal fractions of 15--30\%.
Increasing battery duration reduces EROI further and no energetic
``sweet spot'' is inferred. This does not imply that batteries lack
system value; it means that their balancing and grid-support benefits
must be obtained while accounting for the additional energy required
to manufacture and replace them.

Dispatchable gas backup avoids the renewable overbuild required for
seasonal hydrogen, but replaces it with continuing primary-fuel
consumption. Pipeline-gas systems yield system-level EROI values of
approximately 1.2--2.6, while LNG systems yield approximately
1.0--2.2. Once fuel consumption dominates the energy balance,
improvements in photovoltaic manufacturing or yield have little
influence on the result. Gas backup can provide firm capacity, but it
preserves neither a high net-energy return nor deeply decarbonized
electricity.

The configuration-resolved results further show that the higher
irradiation and winter yield of Alpine PV do not translate into a
systematic system-level advantage. The additional output is offset by
the embodied-energy requirements of heavier structures, remote-site
access, maintenance, and grid connection. Alpine PV may provide useful
winter electricity at favourable sites, particularly where existing
hydroelectric flexibility is available, but it does not remove the
energetic requirements of seasonal firming.

The firming pathways therefore redistribute rather than eliminate the
energetic cost of weather-dependent generation. Batteries shift the
burden toward storage infrastructure, seasonal hydrogen toward
renewable overbuild and conversion losses, and gas backup toward
continuing fuel consumption. Under the favourable benchmark
assumptions, all complete firm-electricity architectures investigated
here remain below an EROI of 8, and many lie substantially lower.
Realistic orientation, site, curtailment, delivery, and infrastructure
requirements can only reduce these values further.

\subsection{Benchmark assumptions and real-world deployment}

The assumptions adopted in the system-level calculations are
deliberately favourable. The output factors are set to
\[
f_{\mathrm{orientation}}
=
f_{\mathrm{site}}
=
f_{\mathrm{curtailment}}
=
f_{\mathrm{delivery}}
=1.
\]
The reported EROI values therefore assume optimal photovoltaic
orientation, favourable site conditions, complete utilization of
renewable generation, and loss-free electricity delivery. These
assumptions isolate the energetic requirements of the firming pathways,
but do not represent the average performance of a large-scale operating
electricity system.

For a fixed infrastructure investment, the output factors act
multiplicatively on delivered electricity and reduce system-level EROI
in direct proportion to their product. Losses from different sources
compound rather than compensate one another. A system affected
simultaneously by non-optimal orientation, unfavourable site conditions,
curtailment, and delivery losses can consequently perform well below the
reported values even when no individual loss appears dominant.
Additional infrastructure requirements reduce EROI further by
increasing the energy investment in the denominator.

The orientation-sensitivity results quantify one part of this effect.
For the fully firm battery--hydrogen architecture with a \SI{1}{h}
battery and $s_0=25\%$, the optimally oriented rooftop-PV accounting
case yields an EROI of approximately 4.1--5.2. An east--west roof
reduces the range to approximately 3.4--4.7, while a flat roof yields
approximately 3.2--4.4. The penalty becomes much larger for vertical
deployment: approximately 2.4--3.6 for a south-facing fa\c{c}ade,
1.8--3.1 for a vertical east--west fa\c{c}ade, and 0.8--2.1 for a
north-facing fa\c{c}ade.

These values are photovoltaic-attributed within the accounting
framework of this study. In the representative 80\% wind--20\%
photovoltaic mix, poor module orientation does not reduce wind output by
the same factor. It does, however, reduce the electricity obtained from
the photovoltaic infrastructure while leaving its embodied-energy
investment largely unchanged. Compensating for the lost yield requires
additional photovoltaic capacity and associated manufacturing,
mounting, grid, maintenance, and replacement inputs.

This effect becomes increasingly important as photovoltaic deployment
expands beyond sites selected primarily for electricity yield. Proposed
applications include floating installations, reservoir dams, motorway
sound barriers, railway corridors, building fa\c{c}ades, and other
surfaces whose orientation is constrained by existing structures. Such
installations may provide land-use, architectural, or local grid
benefits, but access to an otherwise unused surface is not itself an
energetic benefit. Where annual yield is reduced, more infrastructure
is required for each unit of electricity delivered.

The site factor introduces a separate penalty. Even where modules are
geometrically well aligned, installations are affected by local
topography, horizon obstruction, partial shading, seasonal snow cover,
dust and soiling, vegetation growth, access limitations, and other
site-specific conditions. As favourable locations are progressively
occupied, deployment on increasingly constrained sites can reduce the
fleet-average $f_{\mathrm{site}}$. Expansion of nominal installed
capacity therefore does not imply a proportional expansion of delivered
electricity or net-energy return.

Unconventional placement also does not remove the firming requirement.
Electricity generated by motorway barriers, railway infrastructure,
fa\c{c}ades, floating systems, or other constrained installations
remains weather-dependent and must still be integrated, transmitted,
stored, curtailed, or supported by firm generation. Low-yield placement
therefore adds an orientation or site penalty to the storage and firming
requirements already quantified. It can increase nominal capacity
without improving the system-level energy balance.

The assumption $f_{\mathrm{curtailment}}=1$ likewise implies that all
renewable generation can be used. Operating systems experience residual
curtailment because of finite storage capacity, charging-rate limits,
transmission congestion, operational constraints, and coincident
renewable production. Geographical dispersion can reduce some local
constraints, but cannot guarantee utilization during system-wide
surplus periods. Delivery losses become especially relevant where
generation, storage, and demand are geographically separated.

The reported EROI values should consequently be interpreted as
optimistic upper benchmarks. This qualification is decisive because all
complete firm-electricity architectures remain below an EROI of 8 under
the unity assumptions. Many fall below 5, while several approach unity.
Realistic output losses move these systems further away from the
net-energy surplus required for societal resilience and industrial
development.

The effect is most severe for architectures already operating in the
low-EROI regime. A system near EROI 5 has limited margin for compounded
output losses, while a value near 2 can be driven toward energetic
break-even by comparatively modest departures from the benchmark
assumptions. Installing additional capacity to compensate low yield
increases material, manufacturing, grid, and replacement requirements
and therefore does not eliminate the underlying energetic penalty.

Demand response, sector coupling, geographical interconnection, and
alternative technology combinations may reduce some firming
requirements. Conversely, multi-year meteorological variability,
reserve margins, network congestion, storage constraints, degradation,
and omitted infrastructure may increase them. Chronological modelling
can refine the temporal allocation of these requirements, but cannot
eliminate conversion losses, infrastructure replacement, or the need
for firm supply during extended periods of low renewable generation.
Nor should a finely optimized storage duration be interpreted as robust
when future demand, weather, technology, and operating conditions remain
uncertain.

Reduced electricity yield also increases embodied carbon intensity per
kilowatt-hour because the infrastructure must still be manufactured,
installed, maintained, and replaced. An aggregate output factor of 0.5
approximately halves the corresponding EROI and doubles the
non-combustion infrastructure-related carbon intensity before other
losses or additional infrastructure are included. Real-world output
losses therefore affect both the net-energy surplus and the lifecycle
emissions of the complete system.

\subsection{Lifecycle emissions and societal implications}

The lifecycle results reinforce the energetic findings. Under the
adopted present-day industrial supply-chain intensity, the renewable
fleet without firming produces approximately
38--75~gCO$_2$eq/kWh, but does not independently provide year-round
firm electricity. Once firm supply is imposed, the additional
infrastructure, renewable overbuild, conversion losses, leakage, and
fuel consumption raise the lifecycle intensity substantially.

On the principal GWP$_{100}$ basis, seasonal hydrogen reaches
approximately 80--252~gCO$_2$eq/kWh. Adding the representative
\SI{1}{h} battery raises the combined battery--hydrogen range to
approximately 90--289~gCO$_2$eq/kWh because the battery adds embodied
energy without reducing the prescribed seasonal requirement or
hydrogen leakage. Pipeline-gas backup reaches approximately
121--275~gCO$_2$eq/kWh, while LNG reaches approximately
143--327~gCO$_2$eq/kWh. The corresponding GWP$_{20}$ values are higher
because hydrogen and methane exert stronger climate effects over the
shorter time horizon.

The battery-only cases provide an additional warning concerning the
energetic scale of storage infrastructure. They are not complete
firm-electricity systems, yet the \SI{24}{h} benchmark already reaches
approximately 257--1122~gCO$_2$eq/kWh under the adopted industrial
supply chains. Low operational emissions therefore do not guarantee
low lifecycle emissions when extensive infrastructure must be
manufactured, maintained, and repeatedly replaced.

Territorial emissions accounting can obscure this burden. Countries
that install renewable generation may report declining domestic
power-sector emissions while importing modules, batteries, inverters,
steel, cement, electrolysers, and network equipment from regions that
supply the energy-intensive manufacturing processes. The emissions then
occur primarily in the producing countries rather than where the
electricity is consumed. Emissions shifted abroad are not emissions
eliminated globally.

This displacement is not a temporary installation effect. A fleet must
be maintained continuously. Shorter-lived components require repeated
manufacture and replacement, while expansion of generation, storage,
and networks creates continuing demand for materials and industrial
energy. A country can therefore reduce emissions within its domestic
electricity boundary while remaining structurally dependent on
fossil-intensive industrial production elsewhere.

Decarbonizing those supply chains is both possible and necessary.
Manufacturing modules, turbines, batteries, electrolysers, and network
equipment using low-carbon energy would reduce their embodied emissions
and could convert part of the geographical displacement into a genuine
global reduction. It would not, however, eliminate the energy required
for mining, refining, manufacturing, construction, maintenance, and
replacement. Gas-backed systems would additionally retain fuel-cycle
and combustion emissions.

If the infrastructure were manufactured entirely using the output of
the low-carbon system itself, the invested energy would have to be
withdrawn from gross generation before the remainder became available
to society. Decarbonizing the invested energy can lower lifecycle
emissions, but cannot by itself raise EROI or increase the fraction of
gross output remaining after the energy system has reproduced itself.
Carbon intensity and net-energy return are therefore distinct and
simultaneous constraints.

The complete firm-electricity systems occupy precisely the range in
which this distinction becomes societally consequential. At EROI 8,
12.5\% of gross output must be reinvested in the energy system. At EROI
5, the reinvestment share rises to 20\%; at EROI 3, it reaches one
third; and at EROI 2, one half of gross output is required merely to
sustain energy supply. As EROI approaches unity, the net-energy surplus
collapses. Below unity, the system cannot reproduce itself without a
continuing external energy input.

All complete firm-electricity architectures investigated here remain
below EROI 8 even under favourable benchmark assumptions. Seasonal
hydrogen reaches approximately 3.4--7.7, the representative combined
battery--hydrogen systems approximately 2.9--6.8, pipeline-gas backup
approximately 1.2--2.6, and LNG approximately 1.0--2.2. Realistic
orientation, site, curtailment, delivery, and infrastructure penalties
reduce these values further. Several investigated endpoints consequently
approach, or fall below, energetic break-even.

The steady-state calculation also understates the burden during a rapid
transition. A growing energy system must supply energy for expanding
generation, storage, conversion equipment, and networks while
simultaneously maintaining and replacing the existing fleet. The claim
on industrial capacity and gross energy production therefore precedes
the intended net-energy benefit. Low EROI makes this transition dynamic
particularly demanding because a large share of new output must be
reinvested before it can support the wider economy.

These physical inputs must be paid for. Primary energy, materials,
manufacturing, construction, land, networks, maintenance, and
replacement components are not free merely because the final energy
resource is renewable. A low-EROI system requires more of these inputs
for every unit of net electricity delivered. The burden must ultimately
appear through some combination of electricity tariffs, taxation,
subsidies, public expenditure, higher industrial costs, or reduced
consumption elsewhere in the economy.

EROI does not determine a market price by itself. Financing conditions,
labour costs, regulation, market structure, fuel prices, and policy all
influence the price paid by consumers. They cannot remove the physical
requirement for additional energy and infrastructure. A high EROI does
not guarantee inexpensive electricity, but a low EROI imposes an
irreducible real-resource burden and sustained pressure on the societal
cost of energy supply.

The ability to absorb this burden depends on the available surplus. A
low-surplus society has little margin for adverse weather, fuel and
material price shocks, supply interruptions, infrastructure failures,
or unexpected replacement needs. Disturbances can then propagate
rapidly through household expenditure, industrial production, and
public budgets. A high-surplus society can maintain redundancy,
strategic reserves, essential services, and infrastructure renewal while
continuing to invest through periods of disruption.

Energy surplus also determines whether a society can invest in its
future. Research without immediate commercial returns, specialist
education, demonstration projects, competing engineering approaches,
and repeated construction all require resources beyond those needed for
short-term energy-system maintenance. These investments can improve
efficiency, reduce future costs, strengthen domestic supply chains, and
limit dependence on imported designs and components.

Technological competence requires continuity in research, education,
engineering, construction, regulation, operation, and supply-chain
development. Once lost, these capabilities cannot be restored rapidly.
Discontinuing research does not prevent a technology from advancing
elsewhere; it transfers industrial learning, technological leadership,
and future deployment capability to societies that continue its
development. Maintaining competence preserves the capacity to evaluate,
improve, regulate, and, where justified, deploy future energy options.

The Sustainable Development Goals likewise require net energy rather
than nominal generating capacity. Healthcare, water and food systems,
education, scientific research, environmental remediation, industrial
decarbonization, and climate adaptation all depend on a surplus
remaining after the energy system has maintained and replaced itself.
A transition that consumes most of this surplus in reproducing its own
infrastructure cannot provide a secure foundation for these wider
objectives.

The results therefore place a transition relying predominantly on
weather-dependent generation and extensive firming infrastructure on a
critical energetic path. They do not constitute a mathematical proof
that every future low-carbon electricity system is impossible. They do
show that the architectures investigated here cannot be assumed to
provide the abundant net-energy foundation required by a resilient
industrial society. Their already-low benchmark EROI leaves little
margin for real-world losses, transition growth, modelling uncertainty,
or omitted infrastructure.

Firm low-carbon generation with high energy return is consequently not
an optional refinement, but a central requirement of a robust
decarbonization strategy. Hydroelectricity satisfies this requirement
where geography permits, but its expansion is limited in many regions.
Among the broadly deployable technologies considered here, nuclear
power provides firm output, low lifecycle emissions, and an EROI well
above the societal-resilience threshold.

All large infrastructure projects remain exposed to technical,
economic, regulatory, and construction risk. These risks apply to
nuclear power, renewable generation, hydropower, storage, hydrogen, and
transmission alike. Scientific comparison therefore requires equivalent
system boundaries and consistent criteria for reliability, lifecycle
emissions, resource requirements, cost, and net-energy return. Energy
policy based on fear of individual technologies rather than comparative
evidence risks replacing visible project risks with less visible but
systemic energetic fragility.

A credible low-carbon transition must therefore achieve more than a
reduction in territorial electricity-sector emissions. It must reduce
global lifecycle emissions, preserve reliable supply, and retain enough
net energy to sustain industrial capability, societal resilience, and
continued development. Without that surplus, decarbonization itself
becomes progressively more difficult to finance, construct, and
maintain.

\subsection{Conclusions}
\label{sec:conclusions}

This benchmark study shows that variable renewable generation and firm
electricity supply are energetically distinct products. The representative
wind--photovoltaic fleet retains a system-level EROI of approximately
11--16 before firming, but does not independently provide continuous
electricity. Once storage, renewable overbuild, dispatchable backup, fuel
supply, and component replacement are included, the energy return is
determined increasingly by the infrastructure required to overcome
weather-dependent generation rather than by the installation-level EROI of
photovoltaics alone.

Batteries remain valuable for short-duration load balancing and rapid grid
support, but they do not replace the energy inventory required during
prolonged renewable-generation deficits. Increasing battery duration adds
embodied energy and reduces system-level EROI, while seasonal hydrogen
requires substantial renewable overbuild because of conversion losses.
Combining batteries with hydrogen preserves their complementary operational
roles, but also combines their energetic investments. Gas backup avoids
renewable overbuild only by introducing continuous primary-fuel consumption
and the associated greenhouse-gas emissions.

All complete firm-electricity architectures investigated here remain below
an EROI of 8 under favourable benchmark assumptions. Seasonal hydrogen
yields approximately 3.4--7.7, the representative combined
battery--hydrogen systems approximately 2.9--6.8, pipeline-gas backup
approximately 1.2--2.6, and LNG backup approximately 1.0--2.2. These values
are not predictions of particular electricity systems. They identify the
energetic scale and direction of the penalties introduced when reliable
supply is imposed on a predominantly weather-dependent generation fleet.

The reported values are optimistic upper benchmarks. Optimal orientation,
favourable sites, complete utilization of renewable generation, and
loss-free delivery are assumed, while several infrastructure requirements
are neglected. Real installations experience orientation and site
penalties, curtailment, delivery losses, degradation, reserve requirements,
and additional network and firming investments. Deployment on motorway
barriers, building fa\c{c}ades, railway corridors, floating structures, or
other constrained surfaces may provide local benefits, but reduced yield
requires more infrastructure per unit of electricity and does not remove
the firming requirement.

Low operational emissions likewise do not establish low global lifecycle
emissions. Manufacturing, maintaining, and continuously replacing the
required generation, storage, conversion, and network infrastructure
demands industrial energy and materials. Importing this infrastructure can
reduce territorial emissions while transferring the associated production
emissions abroad. Emissions shifted abroad are not emissions eliminated
globally. Decarbonizing supply chains can reduce this carbon burden, but it
cannot eliminate the energy investment or increase the net-energy surplus
of a system with intrinsically low EROI.

This distinction is decisive for societal resilience. Low-EROI systems must
reinvest a large fraction of gross energy production merely to sustain
energy supply. The corresponding requirements for energy, materials,
manufacturing, finance, and replacement capacity must ultimately be borne
by society. A low-surplus society has little margin for adverse weather,
price shocks, supply interruptions, infrastructure failures, or unexpected
investment needs.

High-surplus energy systems provide resilience. They support redundancy,
strategic reserves, infrastructure renewal, environmental protection, and
continuity of essential services. They also provide the capacity for
research, education, demonstration projects, industrial learning, and the
development of domestic technological competence. Societies that maintain
such capabilities can improve efficiency, reduce future construction
costs, and limit dependence on imported designs and components.

The Sustainable Development Goals require net energy rather than nominal
generating capacity or gross electricity production. Healthcare, water and
food systems, education, scientific research, environmental remediation,
industrial decarbonization, and climate adaptation all depend on an energy
surplus remaining after the energy system has maintained and replaced
itself. An electricity system operating close to energetic break-even
cannot provide this foundation.

The results therefore place a transition relying predominantly on
weather-dependent generation and extensive firming infrastructure on a
critical energetic path. Variable renewables can contribute effectively
where their output can be integrated without disproportionate storage,
overbuild, or fossil backup. They do not, however, constitute by themselves
a robust net-energy foundation for a modern industrial society. Reliable
decarbonization requires complementary firm low-carbon generation with a
high system-level EROI. Hydroelectricity provides this foundation where
geography permits; among the broadly deployable options considered here,
nuclear power provides firm output, low lifecycle emissions, and a high
net-energy surplus.

Energy policy should consequently evaluate complete electricity systems
using consistent boundaries for reliability, global lifecycle emissions,
infrastructure, resource requirements, and net-energy return. Installed
capacity and territorial emissions are insufficient measures of a
successful transition. The objective must be an abundant, reliable, and
globally low-carbon net-energy supply. Only with a durable energy surplus
can society remain resilient, sustain scientific and industrial progress,
and retain the capacity to solve future environmental and societal
challenges.

\subsubsection*{Author contributions} 
H.P. Beck is the sole author of this work and is responsible for all aspects of the study.

\subsubsection*{Competing interests}
The author declares no competing interests.

\subsubsection*{Funding statement}
This research did not receive any specific grant from funding agencies in the public, commercial, or not-for-profit sectors.

\subsubsection*{Data availability}
No datasets were generated or analysed during the current study.

\bibliography{references} 

\end{document}